\documentclass[]{aastex62}
\usepackage{amsmath,xcolor}
\pdfoutput=1

\shorttitle{Zombie Vortex Instability III}
\shortauthors{Barranco, Pei \& Marcus}

\numberwithin{equation}{section}

\newcommand{\eg}{e.g.,\ }
\newcommand{\ie}{i.e.,\ }
\newcommand{\cf}{cf.,\ }

\NewPageAfterKeywords

\begin{document}

\title{Zombie Vortex Instability. III. Persistence with Nonuniform Stratification and Radiative Damping}

\author[0000-0003-2045-677X]{Joseph A. Barranco}
\affiliation{Department of Physics \& Astronomy\\ San Francisco State University\\ 1600 Holloway Avenue, San Francisco, CA 94132}
\correspondingauthor{Joseph A. Barranco}
\email{barranco@sfsu.edu}

\author[0000-0003-2244-7236]{Suyang Pei}
\affiliation{Department of Physical \& Environmental Sciences\\ Texas A\&M University, Corpus Christi\\ 6300 Ocean Drive, Corpus Christi, TX 78412}

\author[0000-0001-5247-0643]{Philip S. Marcus}
\affiliation{Department of Mechanical Engineering\\ University of California, Berkeley\\ 6121 Etcheverry Hall, Berkeley, CA 94720-1740}

\submitjournal{The Astrophysical Journal in June 2018; revised and resubmitted in September 2018}

\begin{abstract}
The Zombie Vortex Instability (ZVI) occurs in the dead zones of protoplanetary disks (PPDs) where perturbations excite baroclinic critical layers, generating ``zombie'' vortices and turbulence.  In this work, we investigate ZVI with nonuniform vertical stratification; while ZVI is triggered in the stratified regions away from the midplane, the subsequent turbulence propagates into and fills the midplane.  ZVI turbulence alters the background Keplerian shear flow, creating a steady-state zonal flow.  Intermittency is observed, where the flow cycles through near-laminar phases of zonal flow punctuated by chaotic bursts of new vortices.  ZVI persists in the presence of radiative damping, as long as the thermal relaxation timescale is more than a few orbital periods.  We refute the premature claim by \citet{lesurlatter2016} that radiative damping inhibits ZVI for disk radii $r\gtrsim0.3$~au.  Their conclusions were based on unrealistically short cooling times using opacities with virtually no grain growth.  We explore different grain growth and vertical settling scenarios, and find that the gas and dust in off-midplane regions are not necessarily in local thermodynamic equilibrium (LTE) with each other.  In such cases, thermal relaxation timescales can be orders of magnitude longer than the optically thin cooling times assuming LTE because of the finite time for energy to be exchanged between gas and dust grains via collisions.  With minimal amounts of grain growth and dust settling, the off-midplane regions of disks are susceptible to ZVI and much of the planet-forming regions can be filled with zombie vortices and turbulence.
\end{abstract}



\section{INTRODUCTION}\label{sec:introduction}

\subsection{Background}\label{sec:background}

How turbulent are the gas motions in protoplanetary disks? From both observational and theoretical perspectives, we still are not sure.  The extent of turbulence matters for understanding the late stages of star formation (\ie how gas slowly spirals inward and onto the growing protostar) and the early stages of planet formation (\ie how sub-millimeter dust grains grow into kilometer-size planetesimals via collisional agglomeration or gravitational clumping).  Until fairly recently, the only way to estimate the magnitude of turbulent velocities in circumstellar disks came from radio observations of line emission from molecular tracers (\eg CO, CN, CS) and fitting parametric models to linewidths to tease out the turbulent contribution apart from broadening from thermal motions, line opacity and beam smear \citep{najita1996b}. This is all the more challenging when the spectra are often under-resolved both spatially and spectrally, and because thermal and turbulent broadening are very degenerate in the fitting, which can only be overcome with precise measurements of the kinetic temperature at the same location as the molecular tracers.  Despite these limitations, the consensus has been that transonic turbulence is indeed necessary to explain measured linewidths in circumstellar disks \citep{carr2004, hartmann2004, najita2009, hughes2011, guilloteau2012}.

In the age of ALMA, direct measurement of turbulent line broadening in protoplanetary disks is now possible; however, recent observations seem to indicate that the magnitude of turbulence is much lower than what would be needed to account for accretion, with turbulent velocities only a few percent of the local sound speed \citep{flaherty2015, flaherty2017, teague2016}.   Yet, even these direct measurements are still sensitive to other assumptions (e.g., assuming a spatially and temporally uniform CO/H$_2$ ratio).  A common diagnostic for turbulence is the peak-to-trough ratio in rotationally broadened CO emission lines; turbulence fills in the trough and decreases this ratio \citep{simon2015}.  However, depletion of CO into organic molecules can increase the peak-to-trough ratio, which may mask the true level of turbulence \citep{yu2016,yu2017}.  Hence, it is not clear whether the ALMA results are indicating lower levels of turbulence or depletion of CO due to organic chemistry.

Historically, theorists have asserted that the effects of turbulence could be approximated with an enhanced ``eddy'' viscosity: $\nu_{eddy}\sim\alpha c_sH$, where $c_s$ is the sound speed and $H$ is the pressure scale height -- the idea being that turbulent ``blobs'' would move at subsonic speeds over distances of order the scale height, and the parameter $\alpha$ would subsume all our ignorance of the details of the transport efficiency \citep{shakura73}.  Over time, theoretical research has progressed on two parallel tracks to elucidate the mechanisms that generate and maintain turbulence: purely hydrodynamic processes versus magnetohydrodynamic (MHD) processes.  See \citet{armitage2011review} and \citet{turner2014review} for comprehensive reviews.   \citet{balbus91} applied the magnetorotational instability (MRI) of \citet{velikhov59} and \citet{chandra60}, and demonstrated that magnetic fields can destabilize the differential shear in Keplerian rotation, leading to turbulence and outward transport of angular momentum. However, there exist relatively dense, cool and nearly neutral ``dead zones'' in PPDs that likely lack sufficient coupling between matter and magnetic fields \citep{turnerdrake2009,blaes94}, except perhaps in thin surface layers that have been ionized by cosmic rays or protostellar X-rays \citep{gammie96}.  While a review of the substantial MRI in PPD literature is beyond the scope of this work, we do note that recent work has moved far beyond ideal MHD to include effects such as the Hall term and ambipolar diffusion, both which seem to make MRI-driven turbulence less effective in dead zones \citep{simon2015,kunzlesur2013,baistone2011}.  As MRI wanes as a viable mechanism in PPDs, and older idea, magnetocentrifugal winds (MCW), has resurged \citep{blandfordpayne1982,bai2013,baistone2013b,lesur2014,bai2014,bai2015}.  In an attempt to explain the surprisingly low magnitude of turbulence in protoplanetary disks observed by ALMA, \citet{simon2018} have proposed a model in which only the inner disk is threaded by a strong magnetic field which drives a robust MCW capable of blocking far ultraviolet radiation from reaching the outer disk, resulting in very low levels of ionization and no MRI-driven turbulence.

As our work on ZVI is in the purely hydrodynamic realm, we will contrast it with other non-MHD instabilities.  In Convective Overstability, radial entropy gradients that would be stable according to the Solberg-H\o iland criterion in the adiabatic limit may yet be unstable in the limit of efficient thermal relaxation \citep{klahrhubbard2014,lyra2014}.  The chief obstacles for Convective Overstability are that the cooling time must be relatively tuned, $\Omega\tau_{cool}\sim 1$,  and the radial entropy gradient must be negative so that $-1<(N_r/\Omega)^2<0$, where $\Omega$ is the Keplerian angular speed and $N_r$ is the radial Brunt-V\"{a}is\"{a}l\"{a} frequency (frequency of buoyant oscillations in a stratified background).  The latter constraint requires a disk surface density profile that is significantly flatter than most standard models.  In Vertical Shear Instability (VSI), vertical shear induced by radial gradients of temperature (\eg a thermal wind, a baroclinic effect) that would otherwise be stable to the Kelvin-Helmholtz Instability (KHI) in the adiabatic limit may yet be unstable in the limit of rapid thermal relaxation \citep{umurhan2016,barker2015,stoll2014,nelson2013,urpin2003}.  However, the cooling times must be especially short: $\Omega\tau_{cool}\sim H/r$, where $H/r\sim0.03$ is the aspect ratio of the disk \citep{lin2015}.

\subsection{Previous Work on the Zombie Vortex Instability}\label{sec:previous_work}

In our earlier computational investigations into the stability of 3D vortices in PPDs, we serendipitously discovered ``naturally'' forming vortices in the stratified layers above and below the midplane \citep{barranco05}.  In three recent papers \citep{MPJH13,MPJBHL15,MPJB16}, we demonstrated the existence of a new, purely hydrodynamic instability that fills rotating stratified shear flows with turbulence and robust vortices.  Initial perturbations (either an initial seed vortex or noise with a power-law energy spectrum) excite baroclinic critical layers, which then generate dipolar vortex layers (two juxtaposed oppositely-signed layers of vorticity); while cyclonic vortex layers remain stable, anticyclonic vortex layers roll up into anticyclonic vortices (\ie anticyclones).  We named the instability the ``Zombie Vortex Instability'' (ZVI) not only because it may occur in the dead zones of PPDs, but also because of the way one zombie vortex ``infects'' neighboring baroclinic critical layers, spawning new zombie vortices, which ``infect'' farther critical layers, and so on, {\it ad infinitum}.

Critical layers are special locations in a shear flow where a low-amplitude wave (\ie a neutral eigenmode) has a wave speed that matches the actual fluid flow. Mathematically, in a linear stability analysis, one finds that the coefficients of the highest derivatives of the linearized equations vanish at the critical layer, indicating that the neutrally stable eigenmodes are singular \citep{maslowe86, drazinreid2004}.  They are already well-studied in incompressible flow; in a frame of reference moving at the wave speed, the streamlines of the flow in the vicinity of the critical layer form the familiar Kelvin's cat's eye pattern \citep{kelvin1880,kundu90}.   \citet{MPJH13} discovered a new kind of critical layer in stratified, rotating shear flow called a ``baroclinic critical layer'' in which the wave speed matches the shear velocity plus (or minus) the Brunt-V\"{a}is\"{a}l\"{a} frequency divided by the eigenmode's wavenumber.  A critical layer (either the traditional one or the new baroclinic one) is a very narrow structure;  at its onset, the thickness is of order $\delta_{CL}\sim(\nu/\Omega k)^{1/3}\sim Re^{-1/3}H\sim (\ell_{mfp}/H)^{1/3}H$, where $\nu$ is the kinematic viscosity, $k$ is the wavenumber of a perturbation, $Re$ is the Reynolds number\footnote{The Reynolds number is the ratio of the rate of change of momentum via advection to the magnitude of the viscous force: $Re\equiv{|(\boldsymbol{v} \cdot \boldsymbol{\nabla})\boldsymbol{v}|}/{|\nu\nabla^2\boldsymbol{v}|}\approx UL/\nu\sim H/\ell_{mfp}$, where $L$ is a characteristic length which we take to be equal to the gas pressure scale height $H$, $U$ is a characteristic velocity which we take to be equal to $H$ times the orbital frequency $\Omega$, and where $\nu$ is the molecular viscosity and $\nu\sim c_s\ell_{mfp}$, where $c_s$ is the sound speed and $\ell_{mfp}$ is the mean free path between collisions of gas molecules.} and $\ell_{mfp}$ is the mean free path for collisions of gas molecules \citep{maslowe86,wang2016}.  For conditions within PPDs, critical layers are many orders of magnitude smaller than the gas scale height: $\delta_{CL}\sim 10^{-4}H$.  As the critical layer evolves, nonlinear effects cause it to expand and its thickness becomes independent of the Reynolds number \citep{maslowe86}. In simulations, numerical diffusion (\eg finite grid effects, hyperviscosity or hyperdiffusivity) also cause critical layers to smear out.  Numerical experiments confirm that critical layers still develop even in somewhat under-resolved calculations; the effective thickness is 3-5 collocation points in a spectral simulation \citep{wang2016}.   What is so special about baroclinic critical layers that make them an effective trigger for turbulence?  Within critical layers are highly-localized regions where the gradients of the density, pressure and velocity can be extremely large.  In particular, it is the coupling of the large vertical velocity of the neutral eigenmode with the rotation of the system that generates vorticity (\ie stretching ``planetary'' vorticity)  \citep{pedlosky87,kundu90}.

ZVI is not an artifact of the numerical method as we have observed it with spectral codes and finite-volume codes (\eg Athena), with Boussinesq, anelastic, and fully compressible treatments of the continuity condition, with and without the shearing box, and with hyperviscosity and real molecular viscosity.  One may ask, if it is so robust, how was it missed in previous numerical calculations?  Prior work often lacked one or more of the crucial ingredients: ZVI requires vertical stratification, high resolution to resolve the narrow critical layers, a broad spectrum of perturbations (\ie Kolmogorov, but not Gaussian-peaked, so that the vorticity peaks on the small scales), and enough simulation time to allow the critical layers to amplify perturbations.  Two recent papers \citep{umurhanshariff2016,lesurlatter2016} have now independently confirmed the existence of ZVI and its nonlinear development into zombie vortices and turbulence.

\newpage
\subsection{Goals \& Outline}\label{sec:outline}

{\color{black}
In our previous work, we investigated ZVI with uniform stratification (uniform gravity with uniform background temperature) \citep{MPJH13,MPJBHL15,MPJB16}.  We also worked in the limit that the timescale associated with radiative damping was infinite.  In this work, we relax both of these assumptions.  We note that \citet{lesurlatter2016} raised concerns that rapid cooling or radiative diffusion may suppress the development of ZVI.  However, they prematurely ruled out ZVI (except in very optically thick regions close to the protostar where the timescale for cooling is set by radiative diffusion) because they assumed unrealistically short cooling times based on dust opacities that included virtually no grain growth (maximum size of grains of only a few microns) \citep{semenov2003}. They also completely neglected vertical settling and assumed the dust-to-gas ratio was uniform in height.  In \S\ref{sec:zvi_with_damping}, we carefully analyze the rate at which the gas can thermally relax via collisional exchange of energy with dust grains.  We explore various realistic grain growth and vertical settling scenarios and find that the gas and dust are not necessarily in local thermodynamic equilibrium (LTE) with each other and the time for the gas and dust to ``communicate'' is a significant bottleneck in radiative cooling from thermal dust emission.  In fact, the effective cooling time in off-midplane regions of PPDs can be orders of magnitude longer than when LTE is assumed, and thus, ZVI is still an robust mechanism to generate turbulence throughout much of the planet-forming regions of protoplanetary disk.
}

The outline of the paper is as follows.  In \S\ref{sec:equations}, we describe the anelastic equations and our computational algorithm.  In \S\ref{sec:stratification}, we present the results of simulations with nonuniform vertical temperature profiles which yield nonuniform stratification.  In \S\ref{sec:zvi_with_damping}, we discuss the dependence on radiative damping on the tightness of the gas and dust thermal coupling and show that the relevant thermal relaxation time is orders of magnitude longer than the conventional optically thin cooling time for perfectly thermally coupled gas and dust.  Furthermore, we demonstrate with simulations that ZVI is operative with realistic levels of radiative damping. Finally, in \S\ref{sec:discussion}, we assess the relevance of ZVI and highlight avenues for future work.


\section{EQUATIONS OF GAS DYNAMICS WITH NON-UNIFORM STRATIFICATION}\label{sec:equations}

Consider a three-dimensional box of ideal gas that orbits the protostar at cylindrical radius $r_0$ with Keplerian angular speed $\Omega_0\equiv\Omega_K(r_0)$.  The box is sufficiently small that we ignore curvature and choose Cartesian coordinates $(x,y,z)$ and corresponding unit vectors $(\boldsymbol{\hat{x}},\boldsymbol{\hat{y}},\boldsymbol{\hat{z}})$ for the local radial, azimuthal, and vertical directions \citep{hill1878, goldreich65b}.  We ignore radial gradients in the background profiles for gas density, pressure and temperature.  The equilibrium state in this rotating reference frame is a linearized Keplerian shear flow and vertical hydrostatic balance (in what follows, we will denote vectors with boldface and steady-state variables with overbars):
\begin{subequations}\label{E:equilibrium}
\begin{align}
\boldsymbol{\bar{v}}(x)  &= \bar{v}_y(x)\boldsymbol{\hat{y}} = -\tfrac{3}{2}\Omega_0 x\boldsymbol{\hat{y}}, \label{E:shear} \\
d(\ln\bar{p})/dz &= -g_z(z)\bar{\rho}(z)/\bar{p}(z) = -(\Omega_0^2z)/[\mathcal{R}\bar{T}(z)], \label{E:hydrostatic}
\end{align}
\end{subequations}
where $\boldsymbol{\bar{v}}$ is the Keplerian shear flow in the rotating frame, $\bar{p}$ , $\bar{\rho}$, and $\bar{T}$ are the steady-state gas pressure, density and temperature, $\mathcal{R}$ is the gas constant in the ideal gas law, $\bar{p}=\bar{\rho}\mathcal{R}\bar{T}$, and $g_z(z)=\Omega_0^2z$ is the vertical component of the protostellar gravity (neglecting the gravity of the gas itself).  For vertically isothermal profiles $\bar{T}=T_0$, hydrostatic balance yields a Gaussian density profile $\bar{\rho}(z)=\rho_0\exp(-z^2/2H^2)$ where the gas scale height is $H\equiv \sqrt{\mathcal{R}T_0}/\Omega_0=c_{s0}/\Omega_0$ and where $c_{s0}$ is an isothermal sound speed.  We will not necessarily assume that the temperature profiles are uniform, but it will be convenient to use this isothermal scale height $H$ as a reference unit of length.

Vertical stratification is measured by vertical profiles of potential temperature $\bar{\theta}(z)$ and the Brunt-V\"{a}is\"{a}l\"{a} frequency $\bar{N}(z)$ (the frequency associated with buoyancy oscillations) \citep{pedlosky87}:
\begin{subequations}\label{E:stratification}
\begin{align}
\bar{\theta}(z) &= \bar{T}(z)[p_0/\bar{p}(z)]^{(\gamma-1)/\gamma},\label{E:pottemp}\\
[\bar{N}(z)]^2 &= g_z(z)d(\ln\bar{\theta})/dz,\label{E:brunt}
\end{align}
\end{subequations}
where $\bar{\theta}$ is the steady-state potential temperature, which is the temperature a parcel of gas would have if adiabatically brought to reference pressure $p_0$, and $\gamma=5/3$ is the adiabatic exponent.\footnote{The adiabatic index $\gamma=7/5$ for a diatomic gas with translational and rotational degrees of freedom, but only at temperatures sufficient to excite rotational transitions. For molecular hydrogen, rotational transitions are excited around 100~K.  The bulk of a protoplanetary disk beyond a few AU is below this temperature, so we assume that the molecular hydrogen only has translational degrees of freedom, which yields $\gamma=5/3$.} Potential temperature is just another measure of entropy $s$; to wit, $s = c_p\ln\theta + s_0$.  One needs to specify only one of $\bar{T}$, $\bar{\theta}$, or $\bar{N}$, and the others, along with $\bar{p}$ and $\bar{\rho}$ are determined from equations \eqref{E:hydrostatic}, \eqref{E:pottemp}, \eqref{E:brunt}. Of course, the vertical thermal structure is not arbitrary, but is set by the overall energy balance and radiative transfer \citep{chiang1997,dalessio1998,dullemond2002,dalessio2006,aikawa2006}, which is beyond the scope of this work.

We model the temporal evolution of the flow with the Euler equations modulo the continuity equation replaced by the anelastic approximation: 
\begin{subequations}\label{E:anelastic}
\begin{align}
\partial\boldsymbol{v}/\partial t  &= -(\boldsymbol{v}\cdot \boldsymbol{\nabla})\boldsymbol{v}-2\Omega_0\boldsymbol{\hat{z}}\times\boldsymbol{v} +3 \Omega_0^2 \, x\boldsymbol{\hat{x}}-\boldsymbol{\nabla}\Pi + (\tilde{\theta}/\bar{\theta})g_z\boldsymbol{\hat{z}}, \label{E:anelastic_momentum}\\
\partial\tilde{\theta}/\partial t &= - (\boldsymbol{v} \cdot \boldsymbol{\nabla})\tilde{\theta}  - v_z(\bar{\theta}\bar{N}^2/g_z) - \mathcal{L}_{rad}\tilde{\theta}, \label{E:anelastic_entropy}\\ 
0 &= \boldsymbol{\nabla} \cdot \left[\bar{\rho}(z)\boldsymbol{v}\right], \label{E:anelastic_continuity}
\end{align}
\end{subequations}
where $\boldsymbol{v}$ is the gas velocity in the rotating frame, $\tilde{\theta}\equiv\theta-\bar{\theta}$ is the potential temperature anomaly, and $\Pi\equiv(p-\bar{p})/\bar{\rho}$ is the kinematic pressure.   Equation \ref{E:anelastic_momentum} determines the evolution of momentum; the five terms on the right-hand side represent advection, the Coriolis force, the tidal term (the difference between the inward force of protostellar gravity and the outward centrifugal force), the pressure force, and the buoyancy force.  The viscous force is entirely neglected because the Reynolds number is very large: $Re\sim 10^{14}$. Equation \ref{E:anelastic_entropy} describes the evolution of entropy or potential temperature; the linear operator $\mathcal{L}_{rad}$ represents the effects of radiative damping, and can correspond to simple Newton cooling or a wavenumber dependent diffusion operator.  This will be discussed in greater detail in section \S\ref{sec:zvi_with_damping}. 

In subsonic flow, short-wavelength acoustic waves have periods that are much shorter than the characteristic timescale of the large-scale advective motions.  In numerical simulations, the timestep for an explicit algorithm must be short enough to temporally resolve these fast waves (\ie the  CFL condition), which may be inefficient for calculating the evolution of the large-scale flow for long integration times.  One strategy is to filter sound waves from the fluid equations so that the timestep will be limited by the longer advective timescale.  The anelastic approximation does this by replacing the full continuity equation with the kinematic constraint that the mass flux be divergence-free; this is equation \ref{E:anelastic_continuity}. This approximation still allows for the effects of density stratification (\eg buoyancy in the vertical momentum equation, pressure-volume work in the energy/entropy equation) and has been employed extensively in the study of deep, subsonic convection in planetary atmospheres \citep{ogura62,gough69,bannon96} and stars \citep{gilman81,glatzmaier81a,glatzmaier81b}. We have previously used the anelastic approximation to study three-dimensional vortices in protoplanetary disks \citep{barranco00a,barranco05,barranco06}, the Kelvin-Helmholtz instability of settled dust layers in protoplanetary disks \citep{barranco09,lee2010a,lee2010b}, and turbulent dissipation in tidally perturbed stellar convective zones \citep{penev09,penev11}.  The basic idea is that there may be large variations in the background pressure and density in hydrostatic equilibrium, but that at any height in the atmosphere, the fluctuations of the pressure and density are small compared to the background values at that height.  There are many versions of the anelastic approximation with slight differences in gravity wave propagation or energy conservation; comparisons of which can be found in \citet{brown2012} and \citet{vasil2013}.

We numerically solve these equations with our well developed and benchmarked 3D spectral code that employs specially tailored algorithms to handle the computational challenges due to rapid rotation, intense shear, and strong stratification \citep{barranco06}.  The basic philosophy of spectral methods is to approximate any function of interest with a finite sum of basis functions multiplied by spectral coefficients \citep{gottlieborszag77,marcus86a,canuto88,boyd00}.  A partial differential equation (PDE) in space and time is reduced to a coupled set of ordinary differential equations (ODE) for the time evolution of the spectral coefficients.  The chief advantage of spectral methods over finite-difference methods is accuracy per degrees of freedom (\eg number of spectral modes or number of grid points).  In one dimension, the global error (\eg $L_2$ norm) for a finite-difference method with $N$ grid points scales as $(1/N)^p$, where $p$ is the (fixed) order of the method, whereas for a spectral method with $N$ spectral modes, the error scales as $(1/N)^N$.  Thus, to get the same level of accuracy, spectral methods generally require far fewer degrees of freedom.  This advantage is even more pronounced in 3D problems requiring high resolution.

Because of the linear background shear, the equations depend explicitly on the cross-stream coordinate, making it problematic to apply periodic boundary conditions in this direction.  The equations can be made autonomous in the horizontal directions by transforming to a set of Lagrangian shearing coordinates \citep{goldreich65a,marcus77,rogallo81}.  Features in the flow that are advected by the shear appear quasi-stationary in the shearing coordinates, allowing for larger timesteps to be taken in the numerical integration.  Because the background stratification generally depends on the vertical coordinate, we do not impose periodicity in the vertical direction; instead, we use finite Chebyshev polynomial series to resolve the vertical dependence of flow variables \citep{boyd00}.  {\color{black} We force the vertical velocity and the vertical component of the pressure gradient to zero at impenetrable lids at $z=\pm4H$.   Consistent with these constraints, the vertical gradients of the horizontal velocity and potential temperature also vanish at the lids (though these are not imposed, but follow naturally from the equations of motion).  In previous works, we have explored other vertical boundary conditions (\eg forced periodicity, mapping the vertical boundary to infinity) and shown that the simulations qualitatively converge to the same results \citep{barranco05,barranco06,MPJBHL15,MPJB16}.}

Spectral methods are inherently energy conserving.  However, nonlinear interactions cause a cascade of energy from low spatial wavenumbers to high spatial wavenumbers, where it will cause aliasing and subsequent destabilization of the numerical algorithms.  Hence, spectral codes often must resort to including ``hyperviscosity'' to damp energy at high spatial wavenumbers.  In practice, the spectral coefficients of the velocity and potential temperature are exponentially damped every time step: $\exp[-\Delta t(\nu_{\bot}^{hyp}k_{\bot}^8+\nu_z^{hyp}n^8)]$, where $k_{\bot}^2\equiv k_x^2 + k_y^2$ is the square of the horizontal Fourier wavenumbers and $n$ is the order of the Chebyshev polynomial. The hyperviscosity coefficients $\nu_{\bot}^{hyp}$ and $\nu_z^{hyp}$ are initially set to values such that the highest resolved Fourier or Chebyshev number has an $e$-folding time equal to one timestep.  They can be dynamically adjusted every few hundred timesteps so that the energy spectrum does not curl-up at the highest wavenumbers.

 
\section{SIMULATIONS OF ZOMBIE VORTEX INSTABILITY WITH NON-UNIFORM STRATIFICATION}\label{sec:stratification}

For numerical simulations, we choose a convenient set of reference units: length is measured in units of the gas scale height $H$, time is measured in units of the reciprocal of the Keplerian frequency $\Omega$, and velocity is measured in units of $H\Omega=c_{s}$.  We will often report velocities in terms of Mach numbers $Ma\equiv |\boldsymbol{v}|/c_{s}$, but our simulations are not fully compressible and do not admit acoustic waves.  The unit for vorticity is the Keplerian frequency, though we will often report vorticities in terms of the Rossby number $Ro\equiv|\boldsymbol{\omega}|/2\Omega$. All simulations in this section have a  domain size of $(8H)^3$ resolved with $256^3$ spectral modes.  The size of the time steps was adjusted every $\sim$0.1 orbits to keep the target CFL number $\equiv v_{max}\Delta t/\Delta x\lesssim0.0625$.  The vertical component of gravity was taken to be $g_z(z)=\Omega^2z$ and the initial background profiles for temperature, potential temperature, density and pressure were assumed to be functions of the vertical coordinate $z$ alone.  Other hydrodynamic instabilities such as convective overstability and vertical shear instability were excluded from these simulations because there were no radial derivatives of the background profiles for the thermodynamic variables.  In the simulations described in this section, radiative damping was neglected ($\mathcal{L}_{rad}=0$).  The effects of radiative damping will be explored in section \S\ref{sec:zvi_with_damping}.

\begin{figure}
\gridline{	\fig{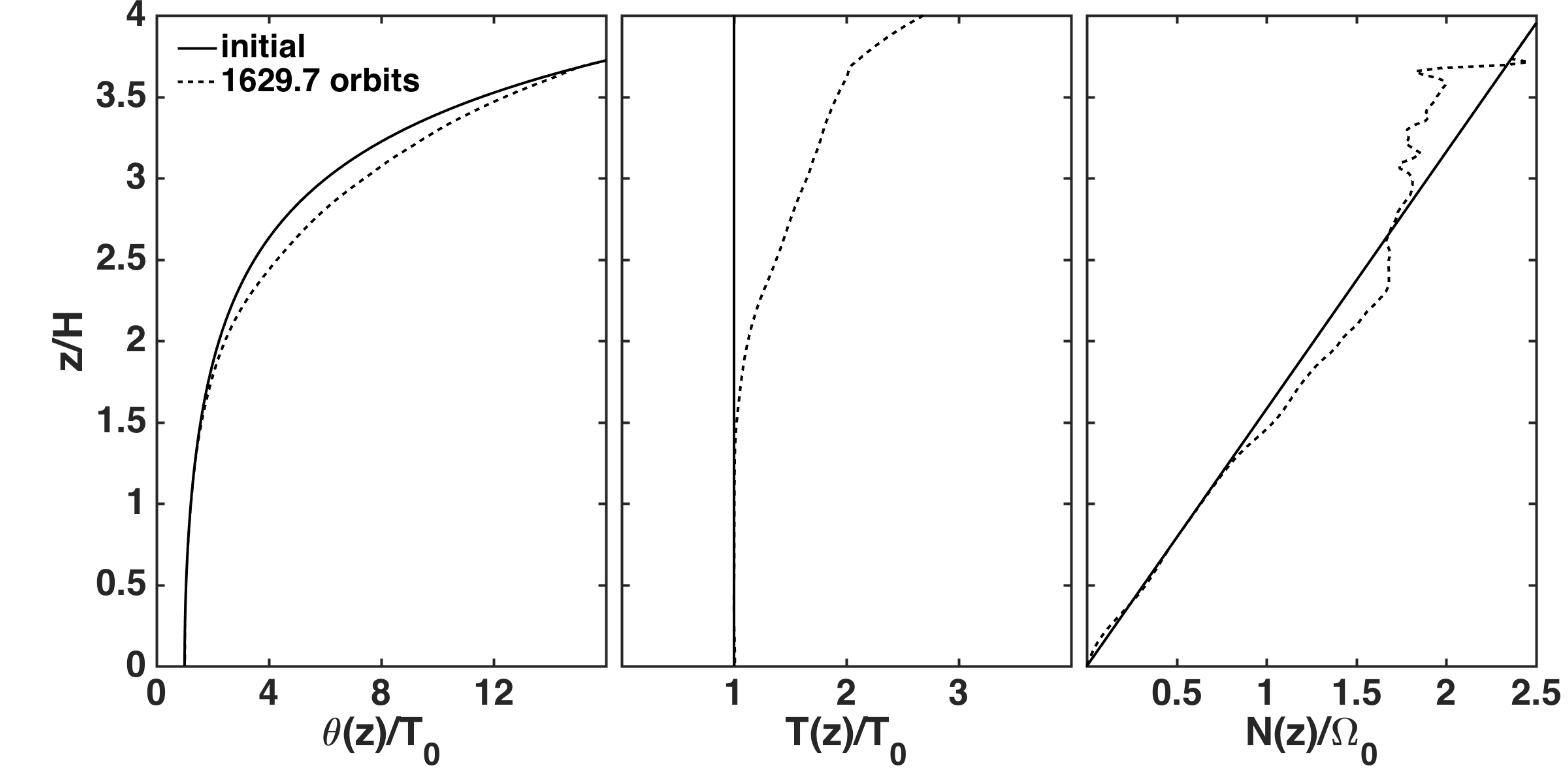}	{0.7\textwidth}{(a){\bf ``Run\_Isothermal''}}}
\gridline{	\fig{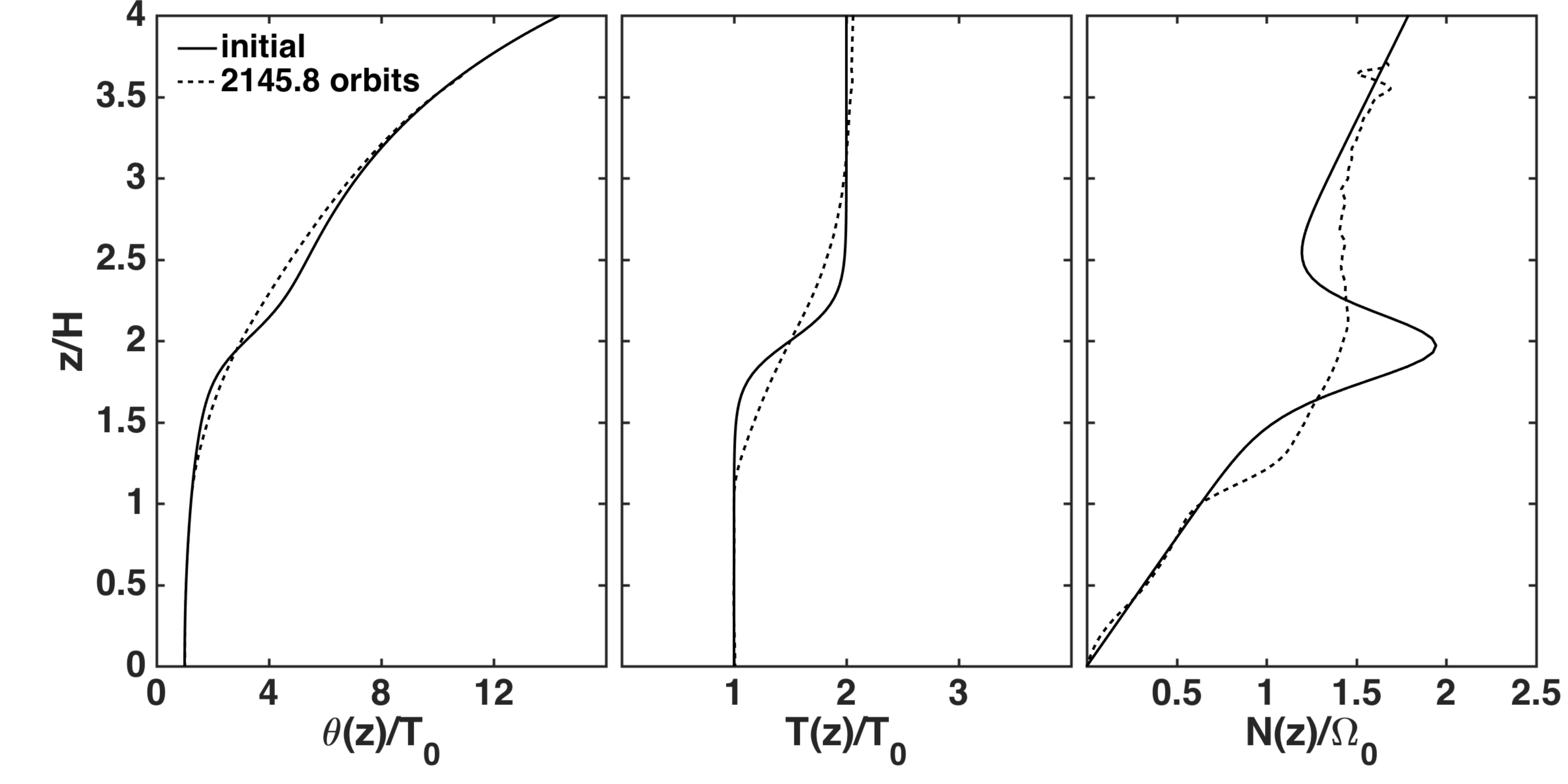}	{0.7\textwidth}{(b){\bf ``Run\_Temp\_Step''}}}
\gridline{	\fig{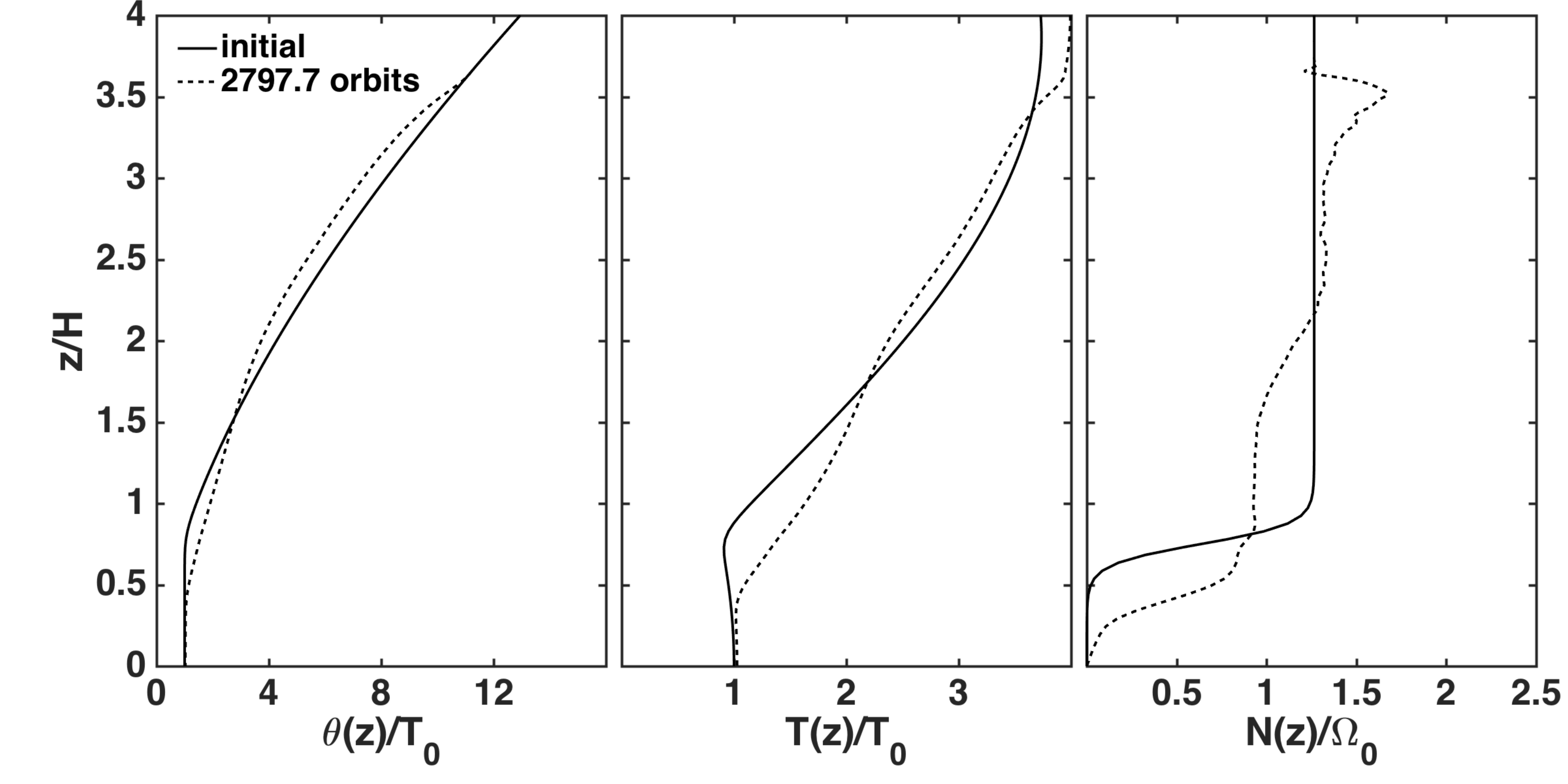}	{0.7\textwidth}{(c){\bf ``Run\_Brunt\_Step"}}}
\caption{{\bf Initial and final background profiles for potential temperature, temperature, and Brunt-V\"{a}is\"{a}l\"{a} frequency, as functions of distance from midplane.} (a) {\bf ``Run\_Isothermal''} was initialized with a uniform background temperature and a Brunt-V\"{a}is\"{a}l\"{a} frequency that was linear with height, (b)  {\bf ``Run\_Temp\_Step''} was initialized with a temperature step profile, \ie a uniform temperature $T_0$ around the midplane, which steps up to a higher uniform temperature $2T_0$ in the disk atmosphere.  This yields a Brunt-V\"{a}is\"{a}l\"{a} frequency with a local maximum.  (c) {\bf ``Run\_Brunt\_Step"} consisted of a step profile for the Brunt-V\"{a}is\"{a}l\"{a} frequency, \ie with the midplane being unstratified, and then the Brunt-V\"{a}is\"{a}l\"{a} frequency jumps up to a constant value of $1.2649\Omega$ in the atmosphere of the disk. }\label{fig:background_profiles}
\end{figure}

In order to explore how non-uniform stratification affects the development of ZVI and late-time zombie turbulence, we simulated ZVI with 3 different vertical profiles (see Figure~\ref{fig:background_profiles}):

(i) ``Run\_Isothermal'' was initialized with a uniform temperature background $\bar{T}(z)=T_0$; the corresponding Brunt-V\"{a}is\"{a}l\"{a} frequency increased linearly away from the midplane: $\bar{N}(z)=\sqrt{(1-1/\gamma)}\Omega|z|/H$. While the temperature is unlikely to be vertically uniform in realistic disks, this is the standard and simplest model for theoretical investigations.

(ii) ``Run\_Temp\_Step'' initially consisted of a uniform temperature of $T_0$ around the midplane, with a steep rise at $|z|\approx2H$ to a different uniform temperature $2T_0$; this resulted in a Brunt-V\"{a}is\"{a}l\"{a} frequency profile that has local maxima near $z=2H$.  This is a simple model for a two-temperature layered model \citep{chiang1997, dalessio1998, dalessio2006}.

(iii) ``Run\_Brunt\_Step'' had an initial Brunt-V\"{a}is\"{a}l\"{a} frequency of zero for $|z|\lesssim H$ with a steep rise at $|z|\approx H$ to a different constant value $\bar{N}=2\Omega\sqrt{(1-1/\gamma)}$.   The motivation for this model is the observation of step-like or staircase patterns in the stratification in atmospheric and oceanic flows, believed to be created by breaking internal gravity waves \citep{orlanski1969,phillips1972,pelegri1998}.   As we will see, this model is further validated by the observation that the late time evolution of ZVI-induced turbulence in Run\_Isothermal created a region of uniform Brunt-V\"{a}is\"{a}l\"{a} frequency.

\begin{figure}
\gridline{	\fig{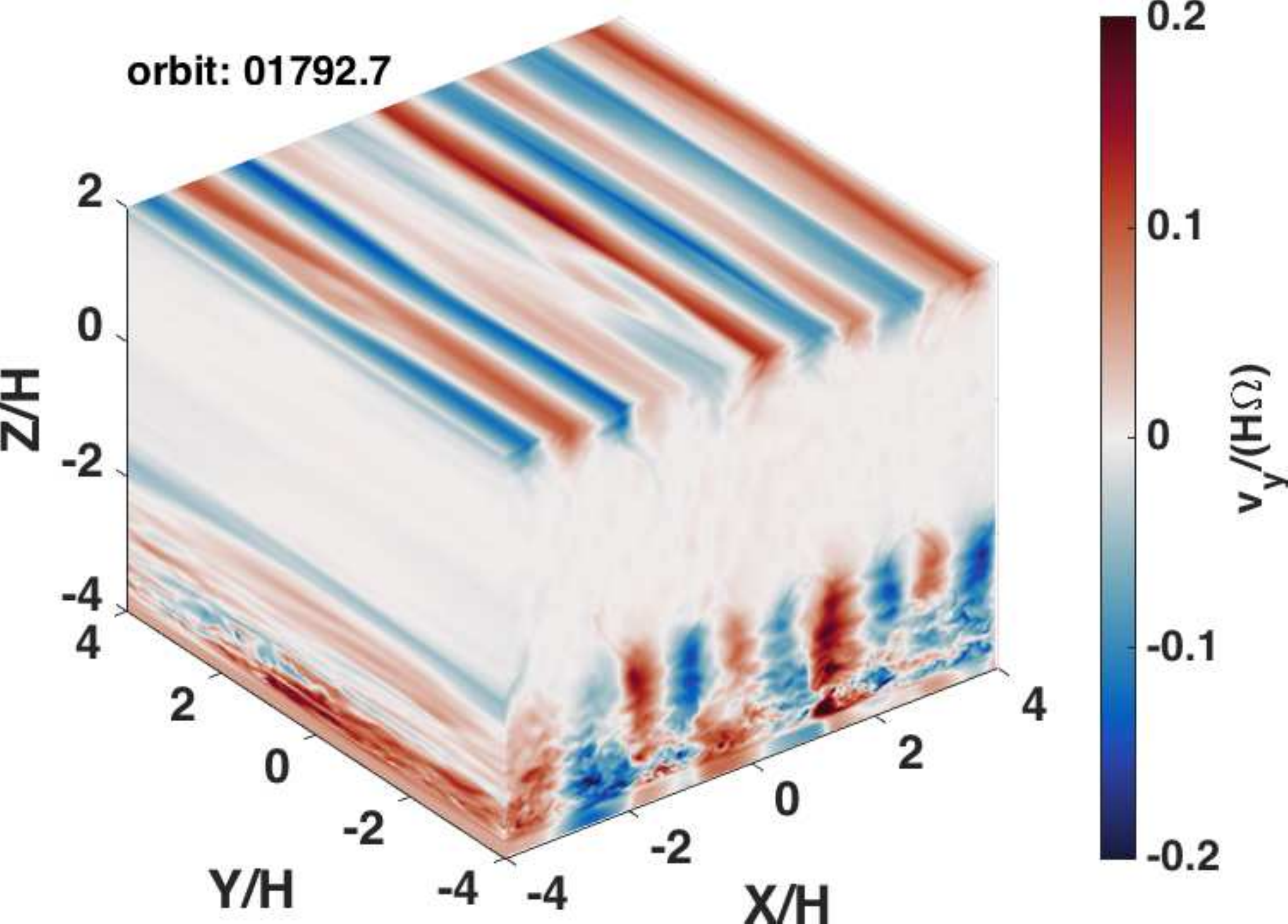}	{0.45\textwidth}{(a)}
        		\fig{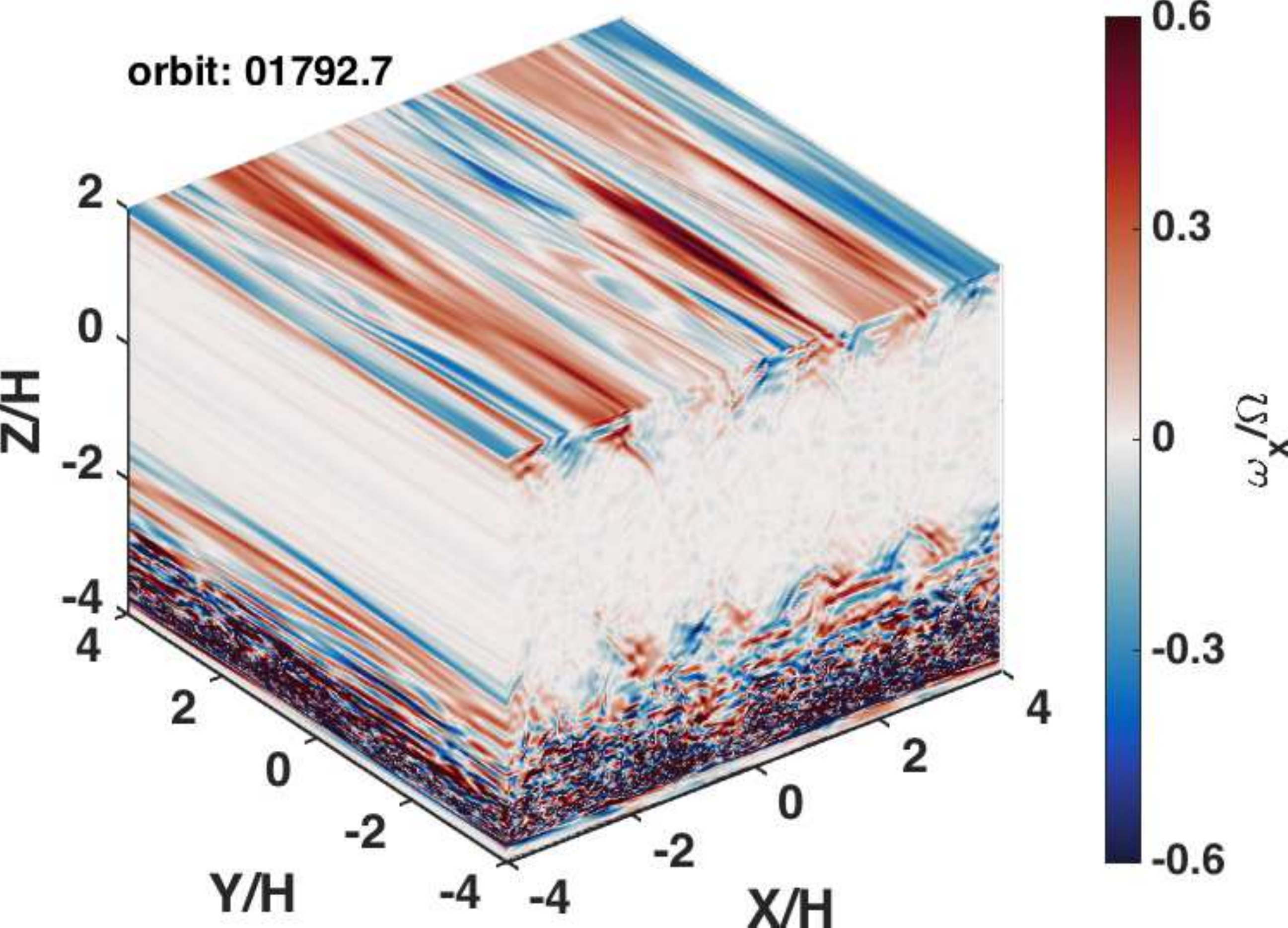}	{0.45\textwidth}{(d)}}
\gridline{	\fig{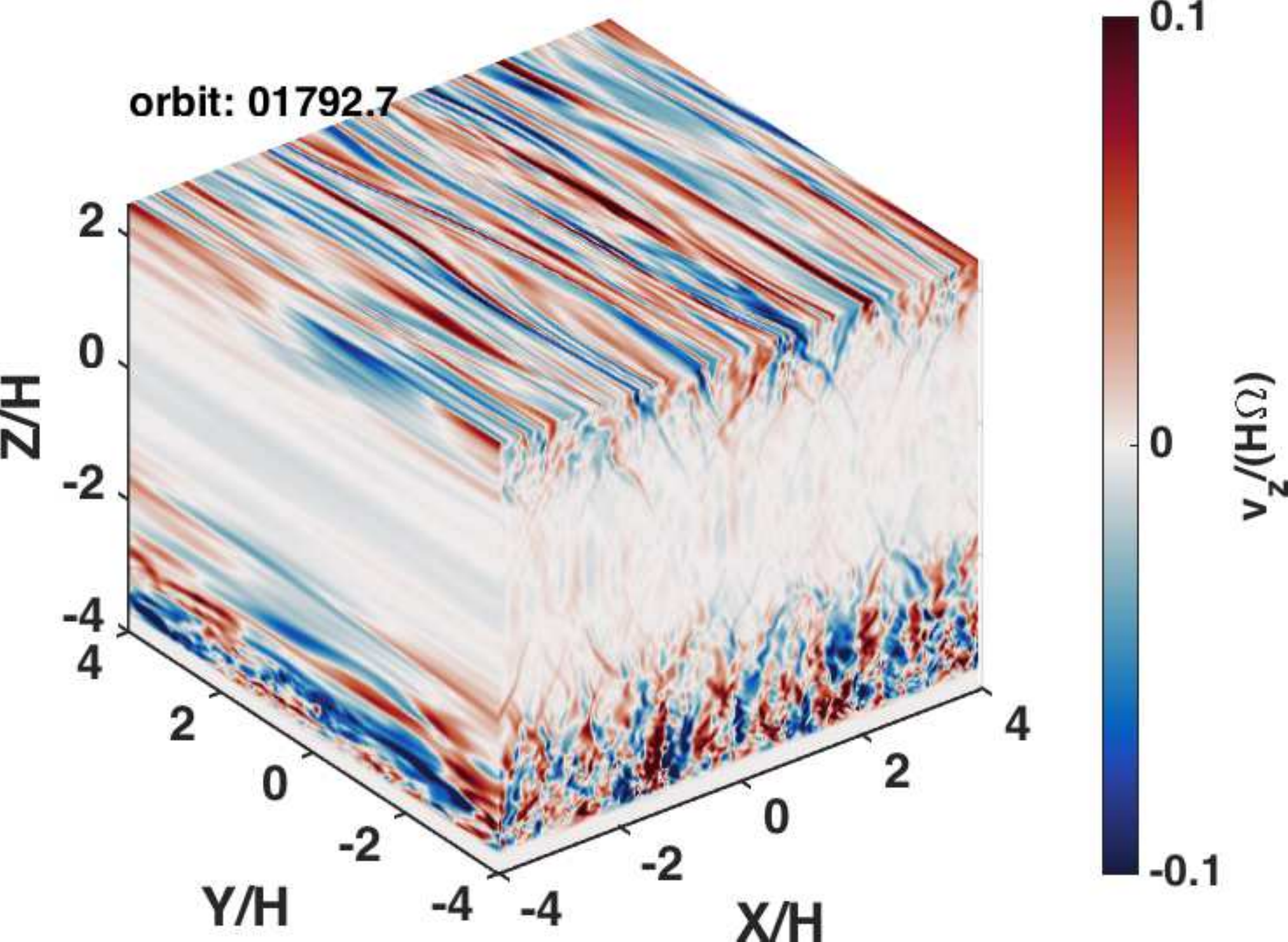}	{0.45\textwidth}{(b)}
        		\fig{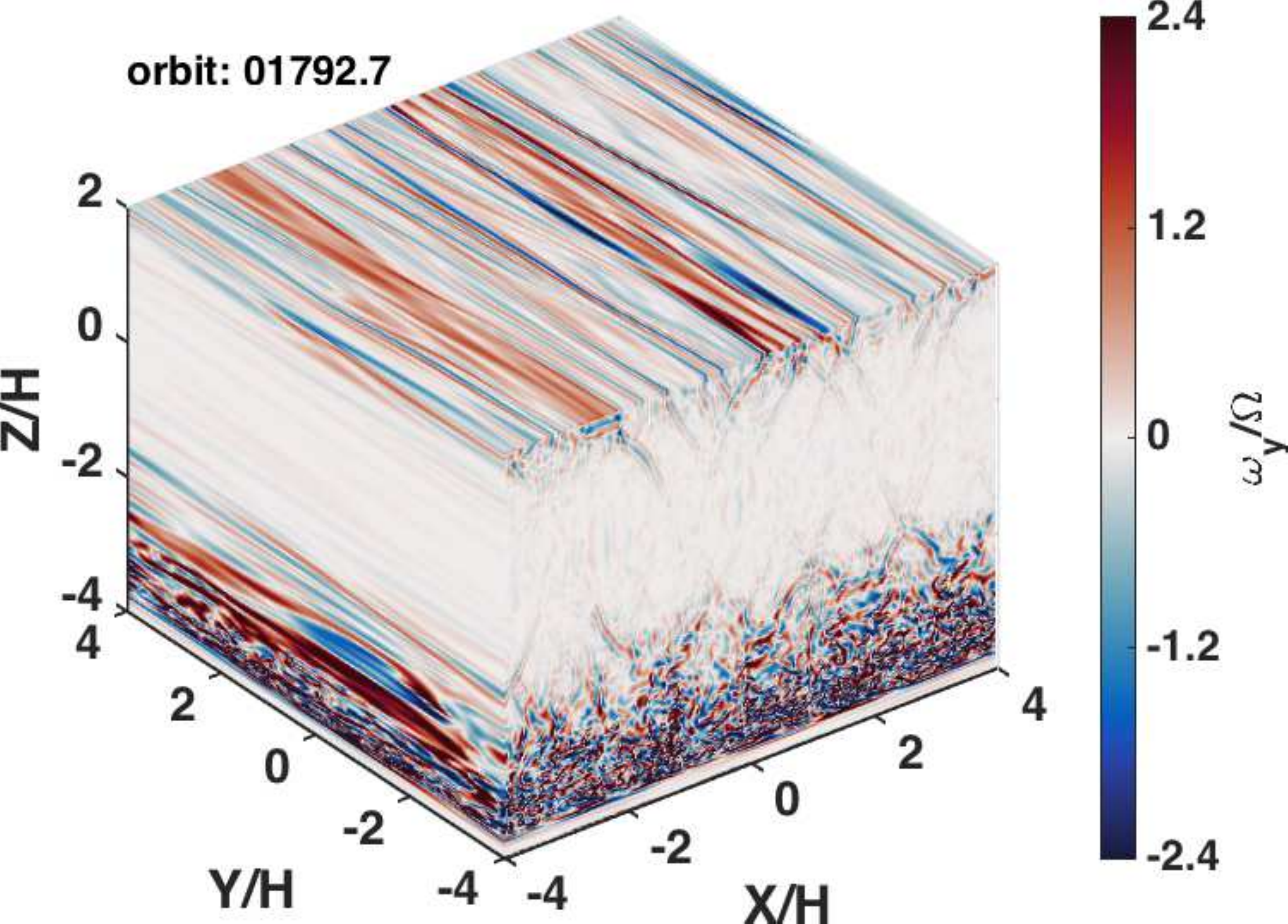}	{0.45\textwidth}{(e)}}
\gridline{	\fig{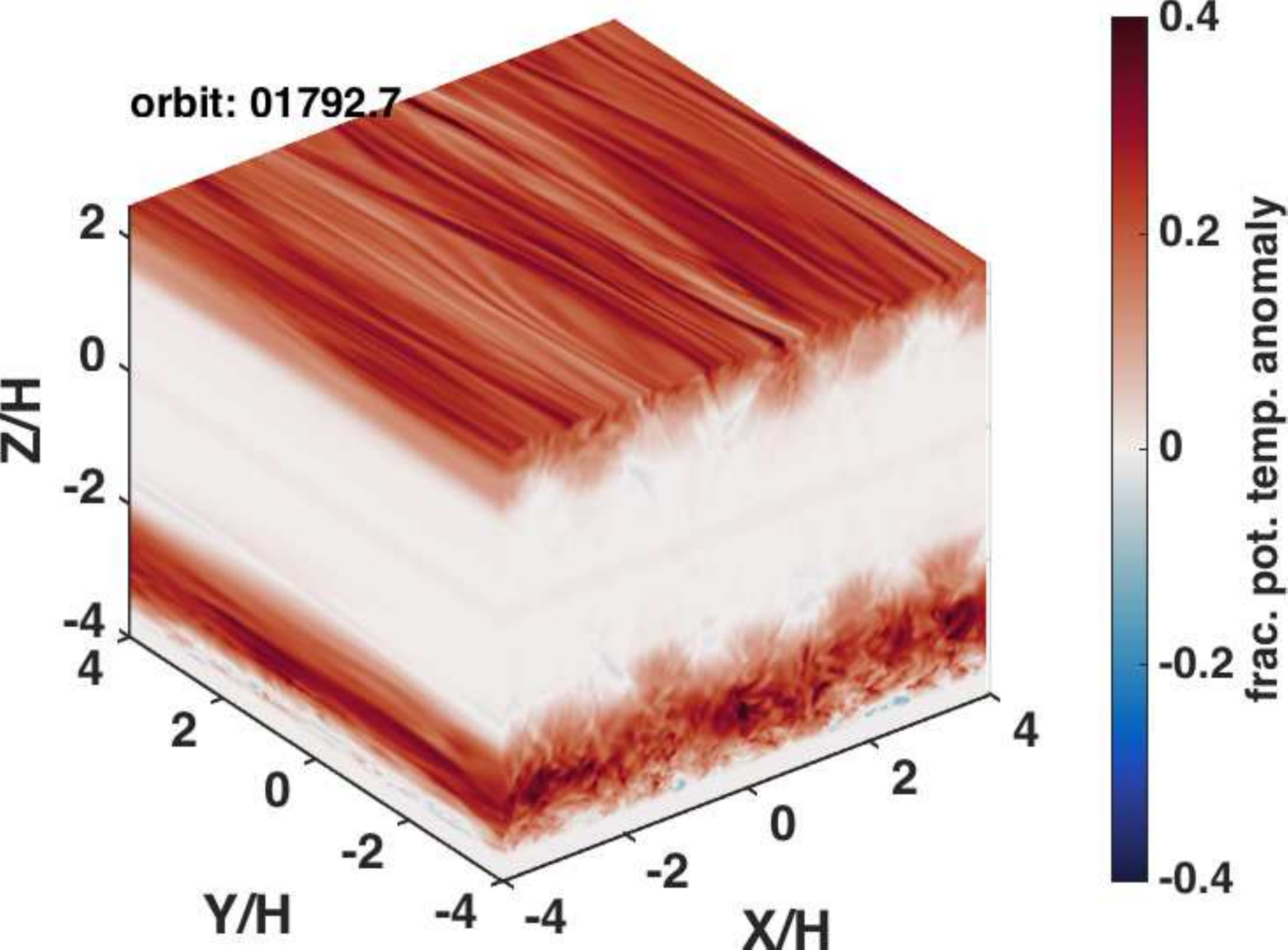}	{0.45\textwidth}{(c)}
        		\fig{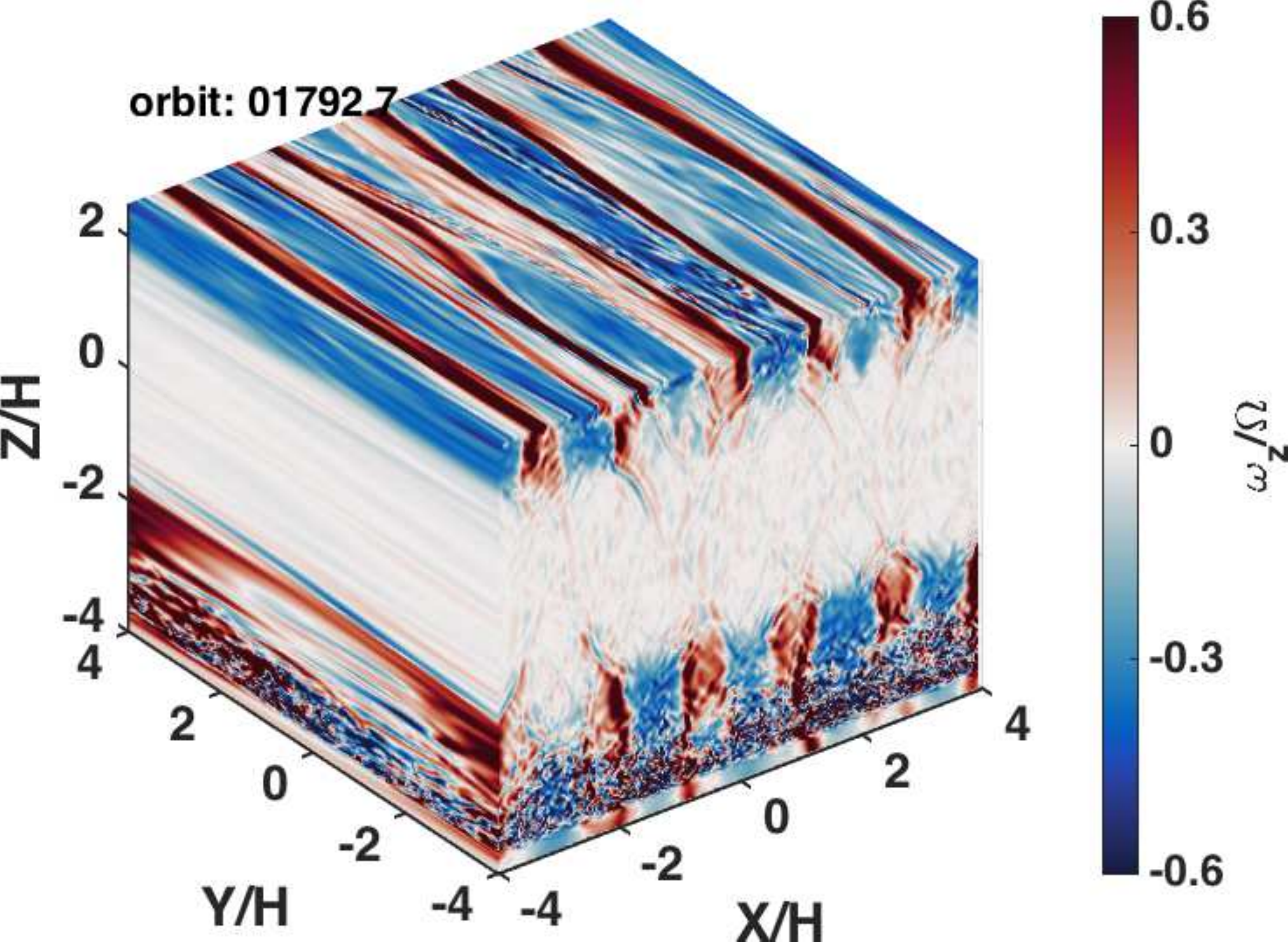}	{0.45\textwidth}{(f)}}
\caption{{\bf Late-time evolution (after 1792.7 orbits) for profile Run\_Isothermal.}  The left column (subfigures a-c) shows the azimuthal velocity (Keplerian shear subtracted), vertical velocity and fractional potential temperature anomaly.  The right column (subfigures d-f) shows the three components of the relative vorticity (Keplerian vorticity subtracted from vertical vorticity).}\label{fig:zvi_images01}
\end{figure}

\begin{figure}
\gridline{	\fig{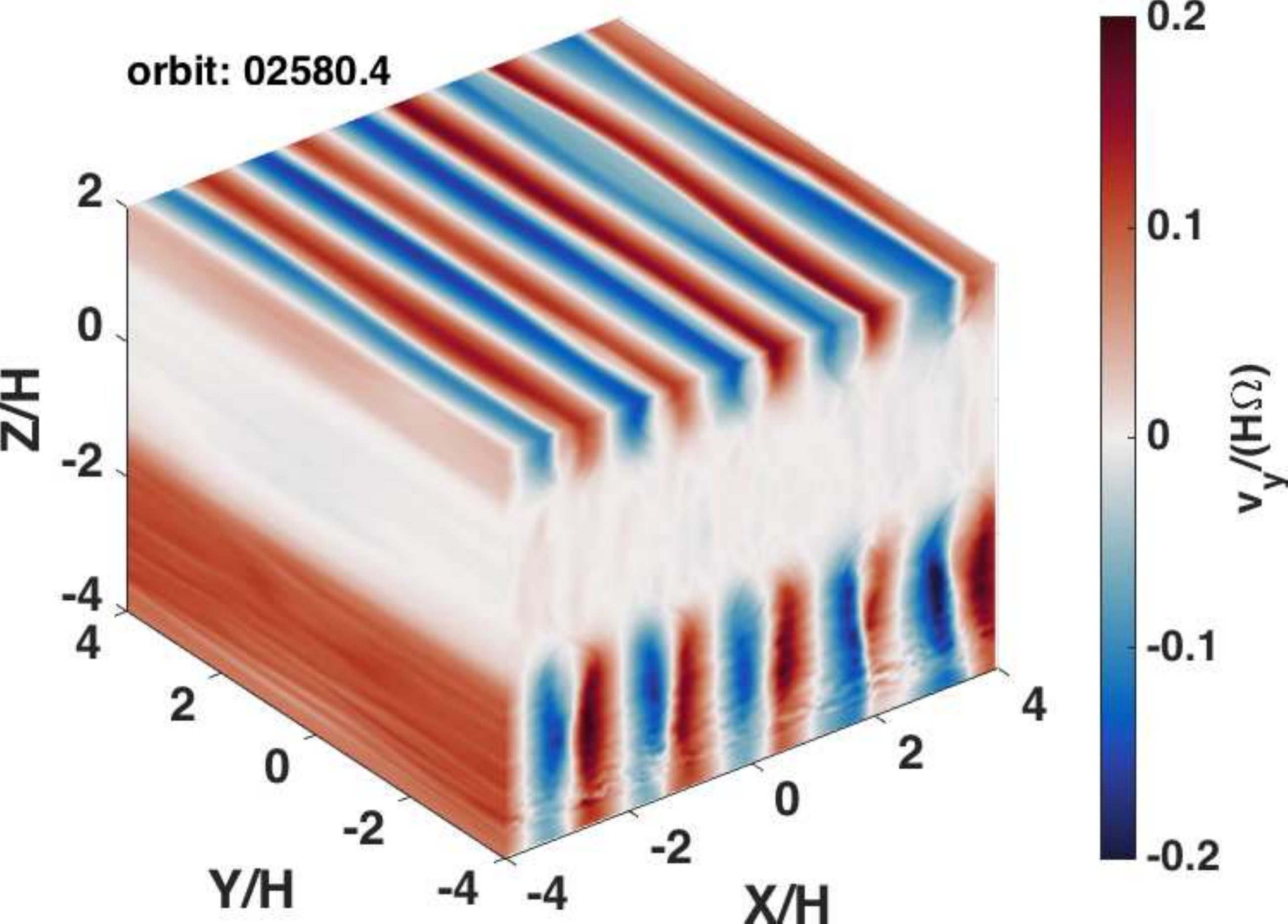}	{0.45\textwidth}{(a)}
        		\fig{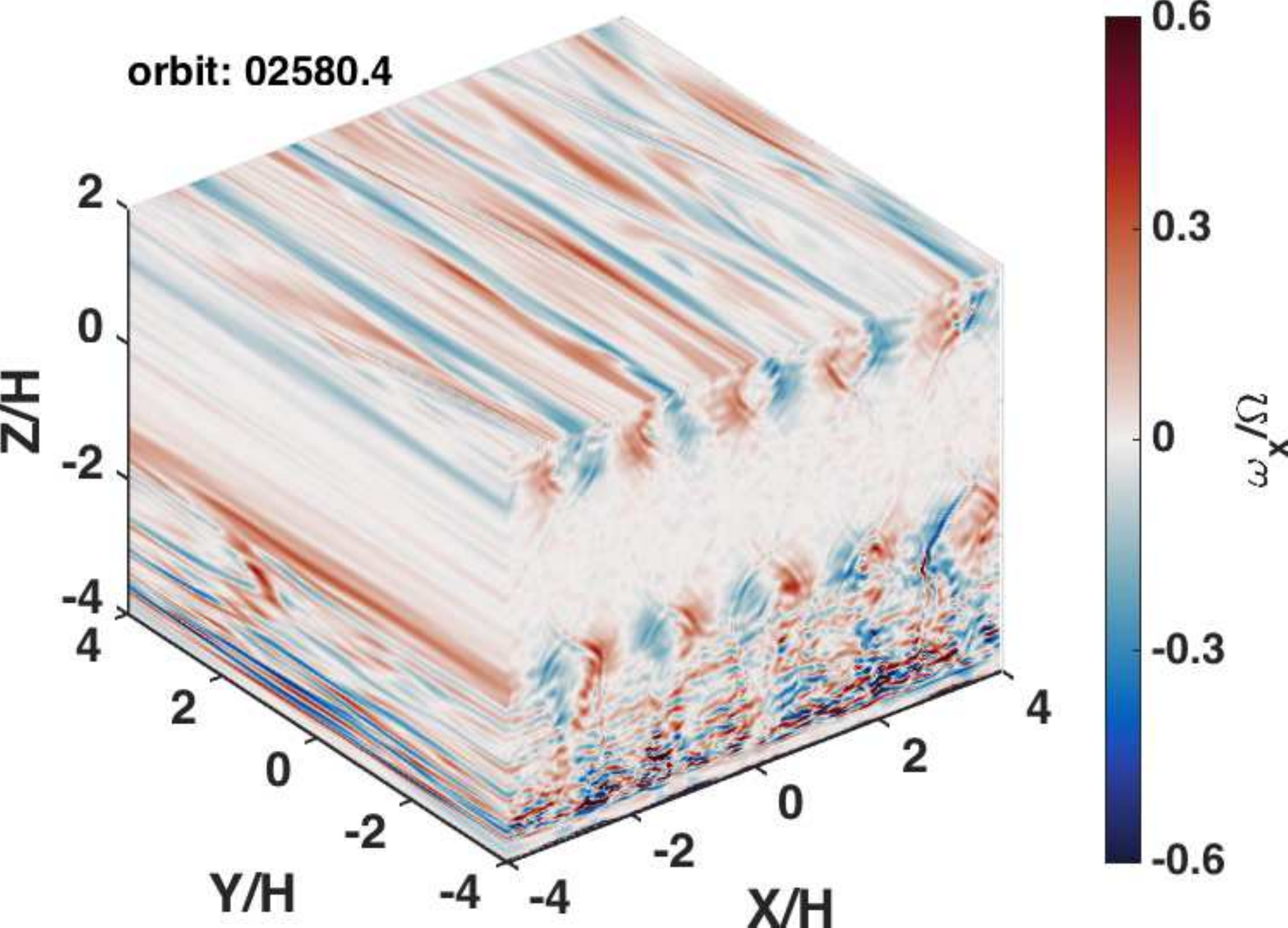}	{0.45\textwidth}{(d)}}
\gridline{	\fig{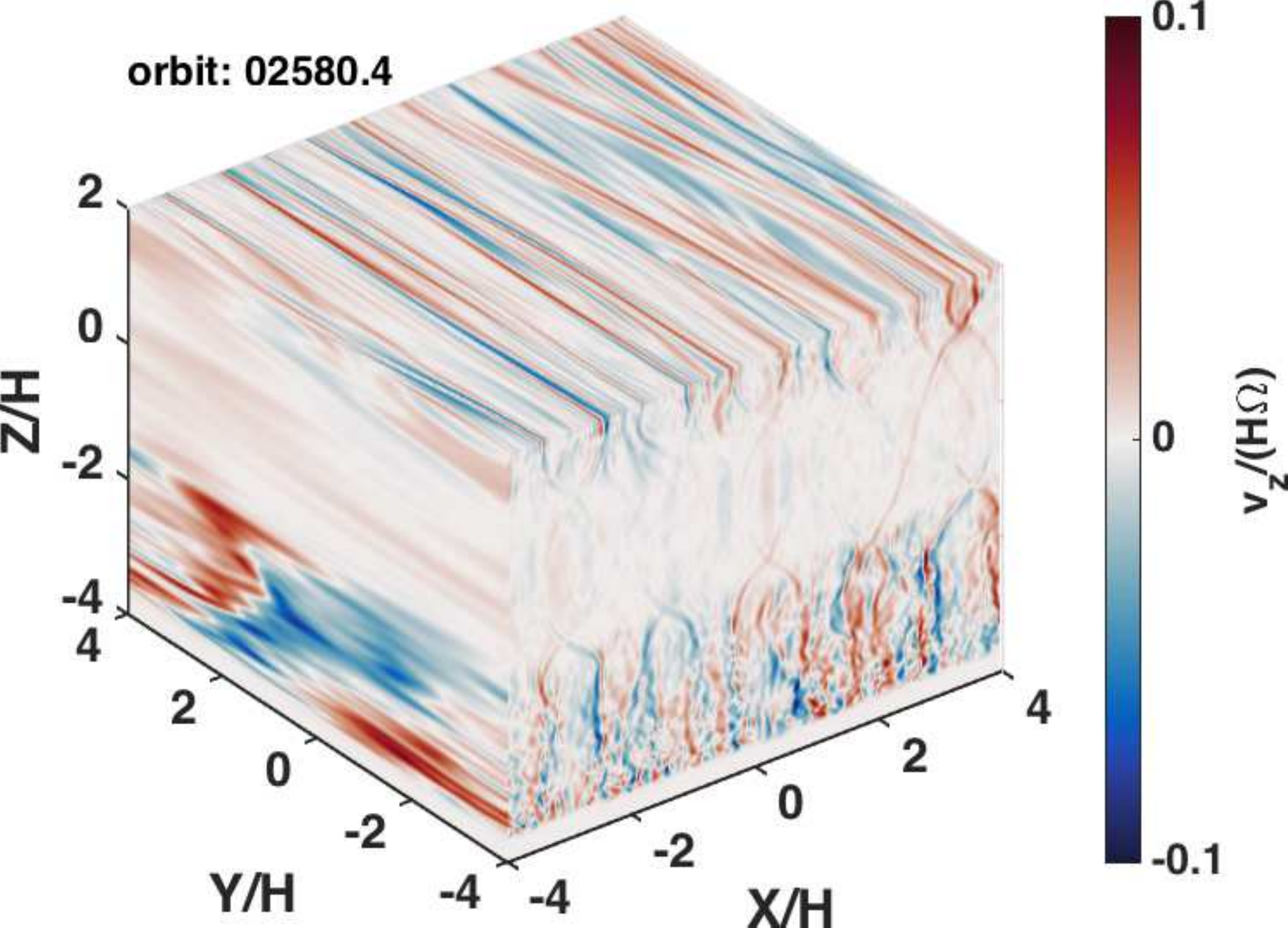}	{0.45\textwidth}{(b)}
        		\fig{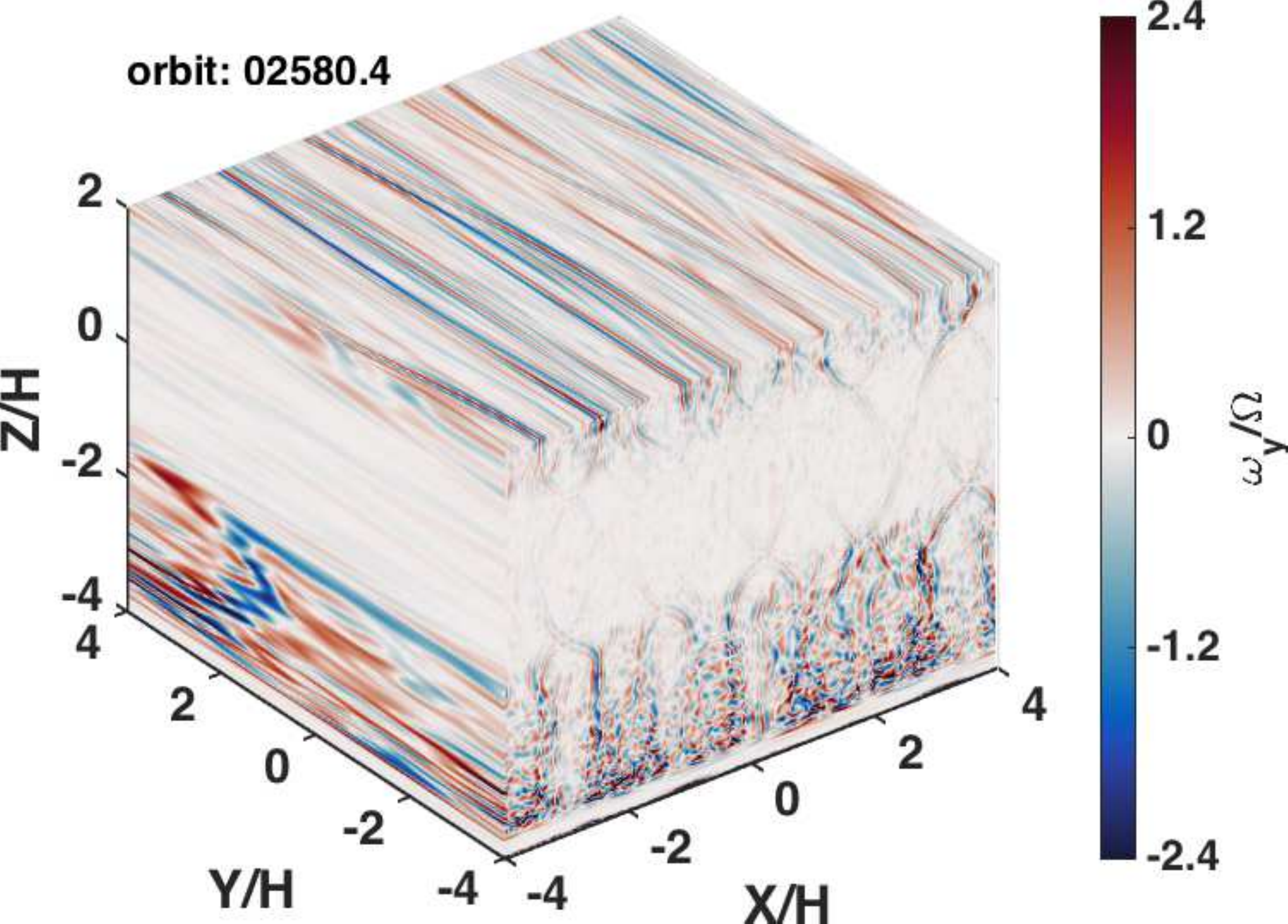}	{0.45\textwidth}{(e)}}
\gridline{	\fig{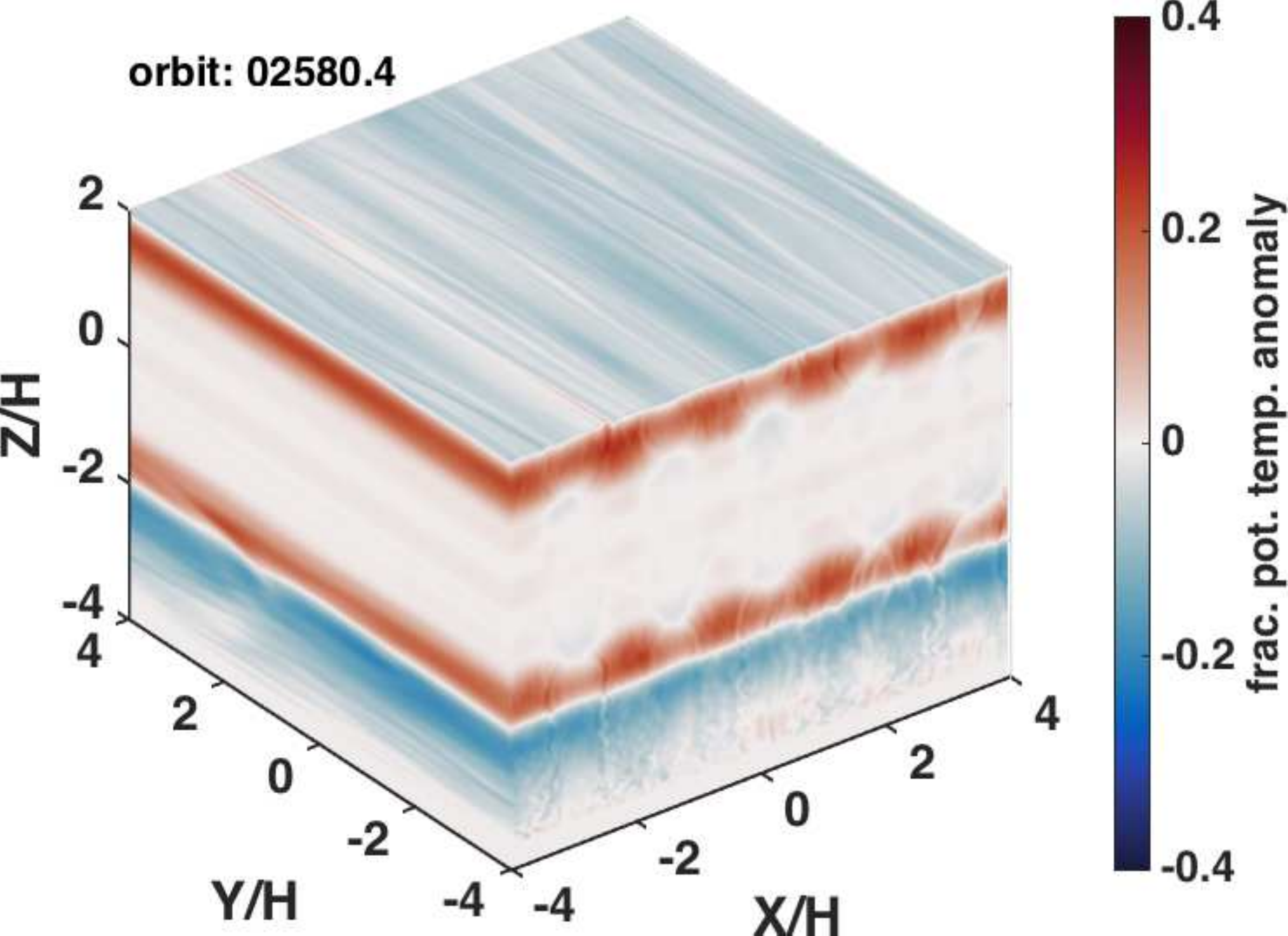}	{0.45\textwidth}{(c)}
        		\fig{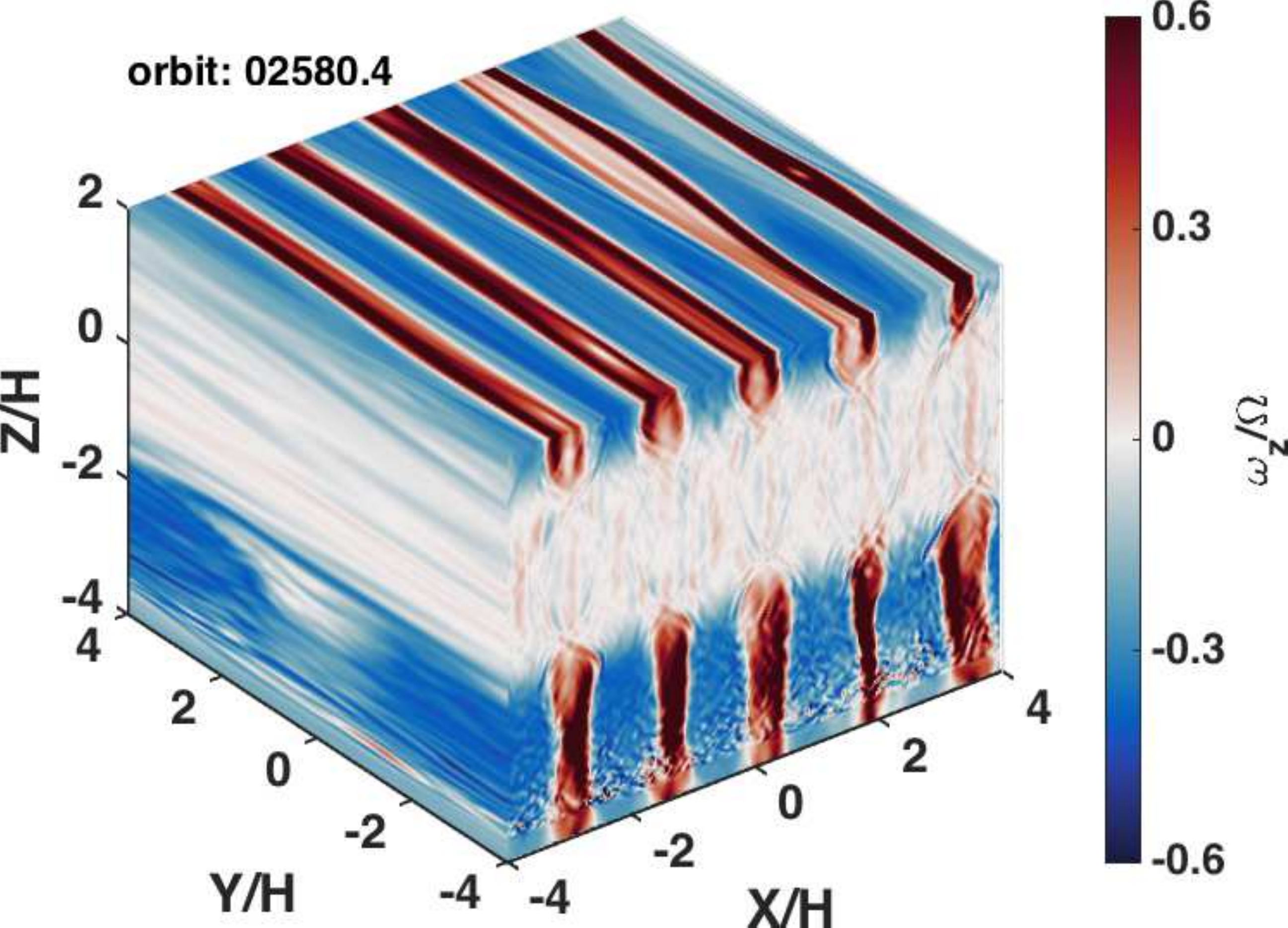}	{0.45\textwidth}{(f)}}
\caption{{\bf Late-time evolution (after 2580.4 orbits) for profile Run\_Temp\_Step.}  The left column (subfigures a-c) shows the azimuthal velocity (Keplerian shear subtracted), vertical velocity and fractional potential temperature anomaly.  The right column (subfigures d-f) shows the three components of the relative vorticity (Keplerian vorticity subtracted from vertical vorticity).}\label{fig:zvi_images03}
\end{figure}

\begin{figure}
\gridline{	\fig{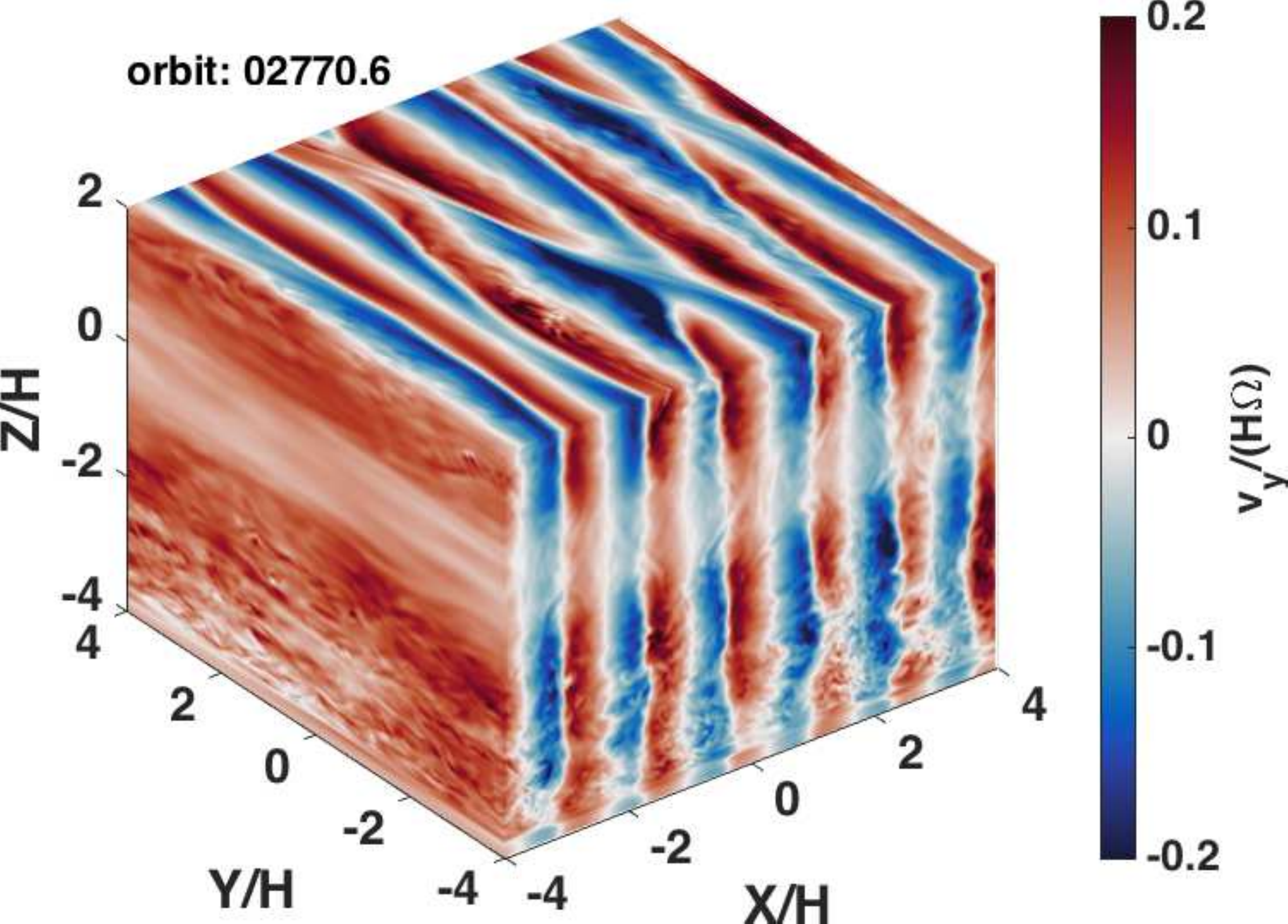}	{0.45\textwidth}{(a)}
        		\fig{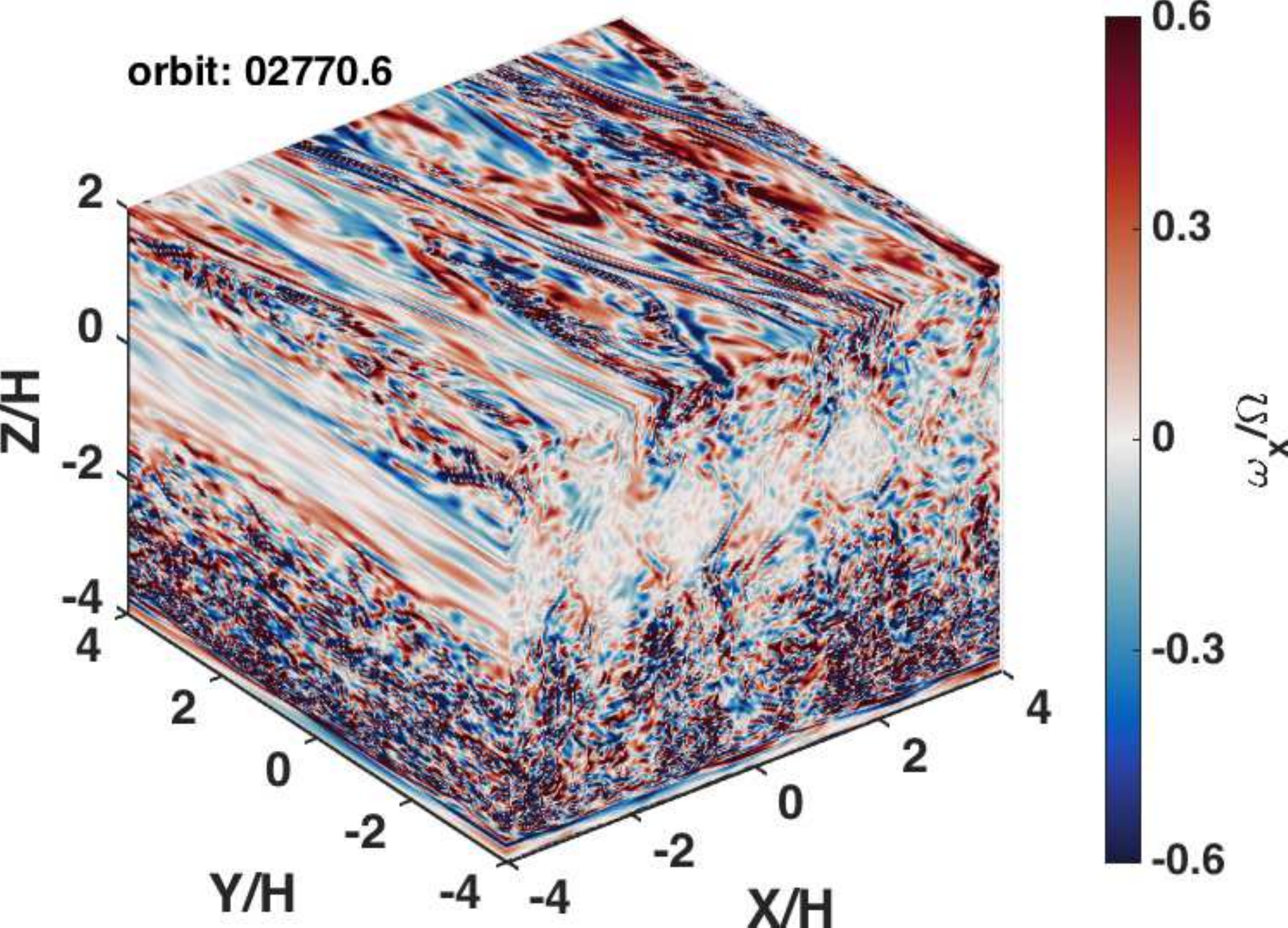}	{0.45\textwidth}{(d)}}
\gridline{	\fig{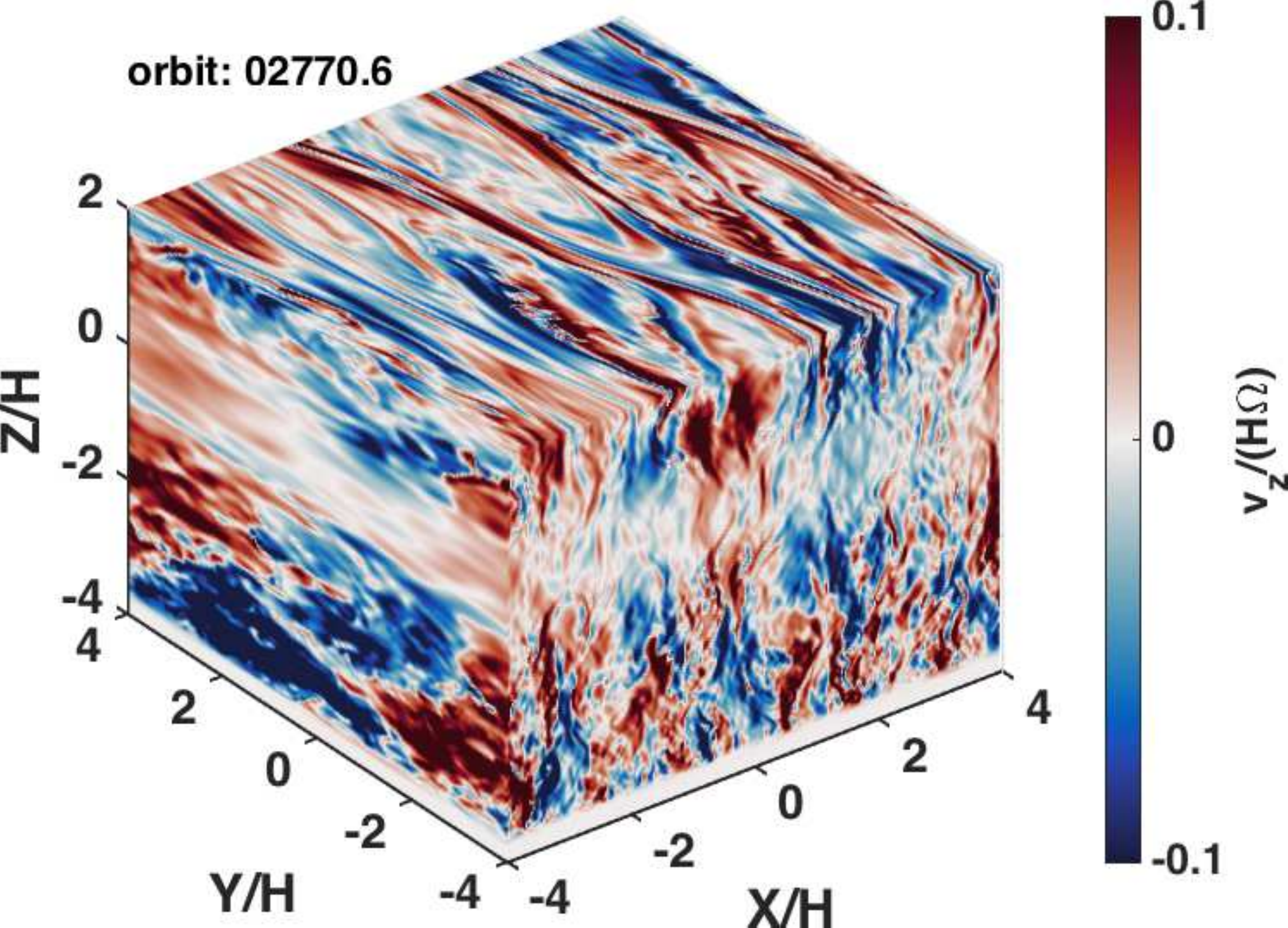}	{0.45\textwidth}{(b)}
        		\fig{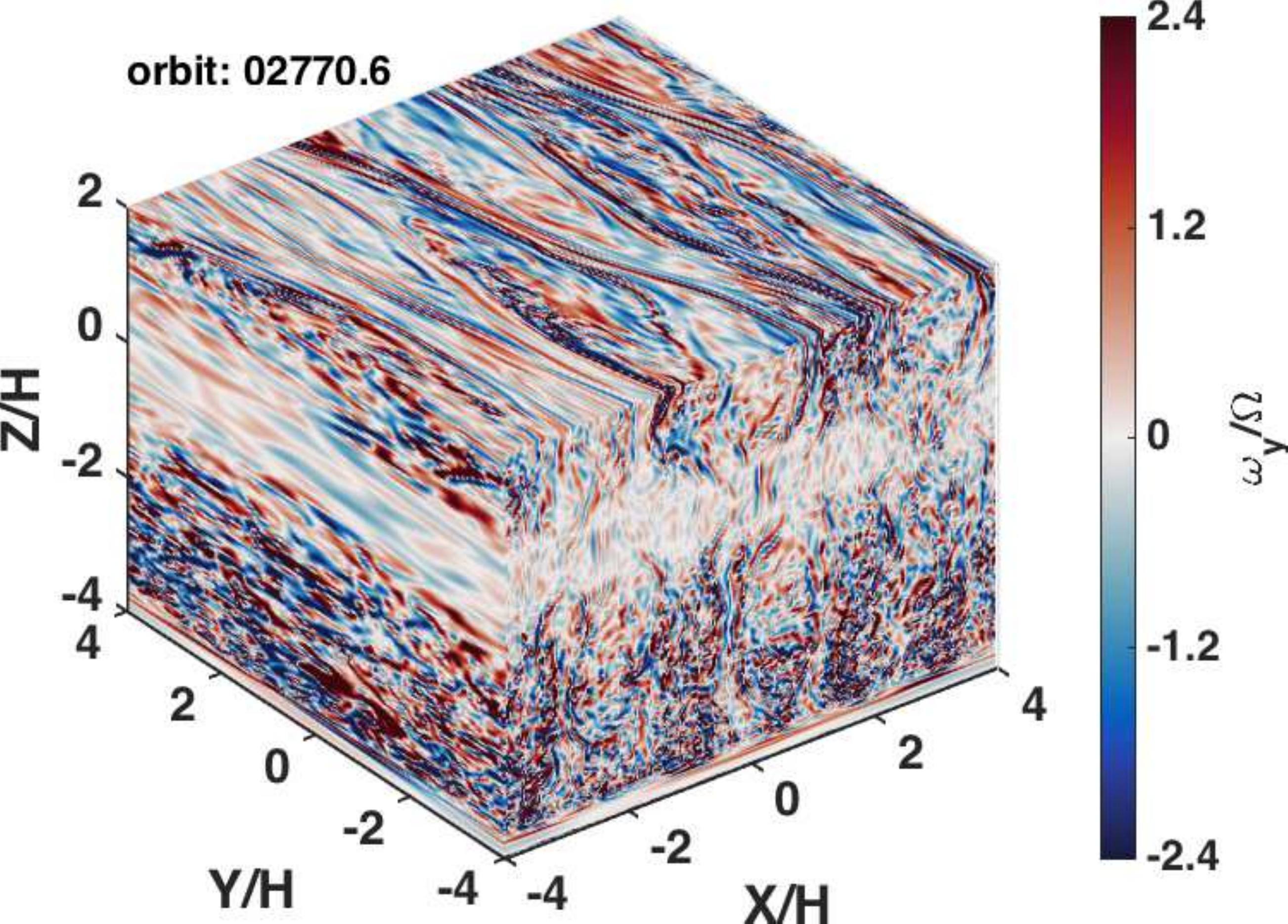}	{0.45\textwidth}{(e)}}
\gridline{	\fig{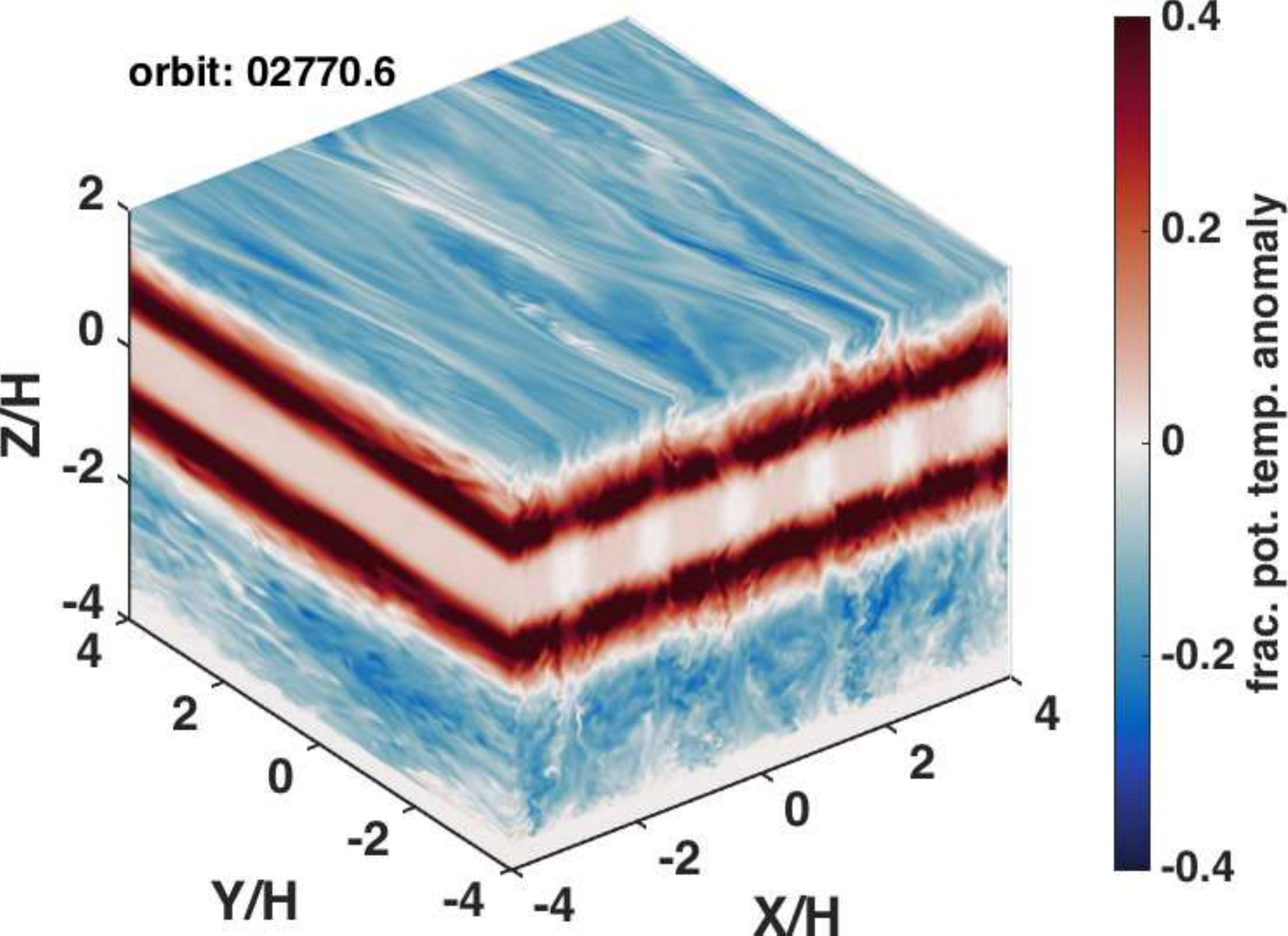}	{0.45\textwidth}{(c)}
        		\fig{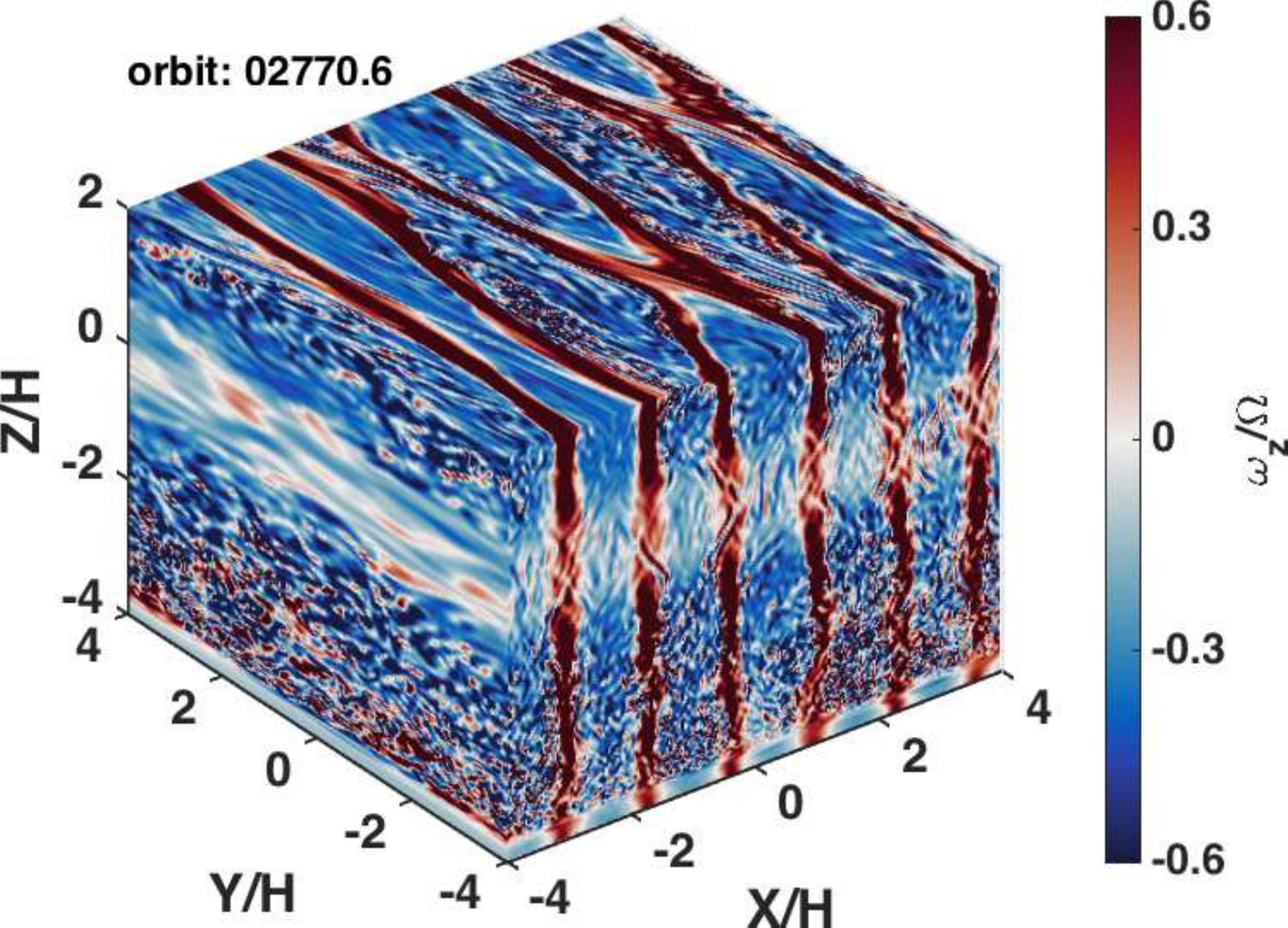}	{0.45\textwidth}{(f)}}
\caption{{\bf Late-time evolution (after 2770.6 orbits) for profile Run\_Brunt\_Step.}  The left column (subfigures a-c) shows the azimuthal velocity (Keplerian shear subtracted), vertical velocity and fractional potential temperature anomaly.  The right column (subfigures d-f) shows the three components of the relative vorticity (Keplerian vorticity subtracted from vertical vorticity).}\label{fig:zvi_images04a}
\end{figure}

\begin{figure}
\gridline{	\fig{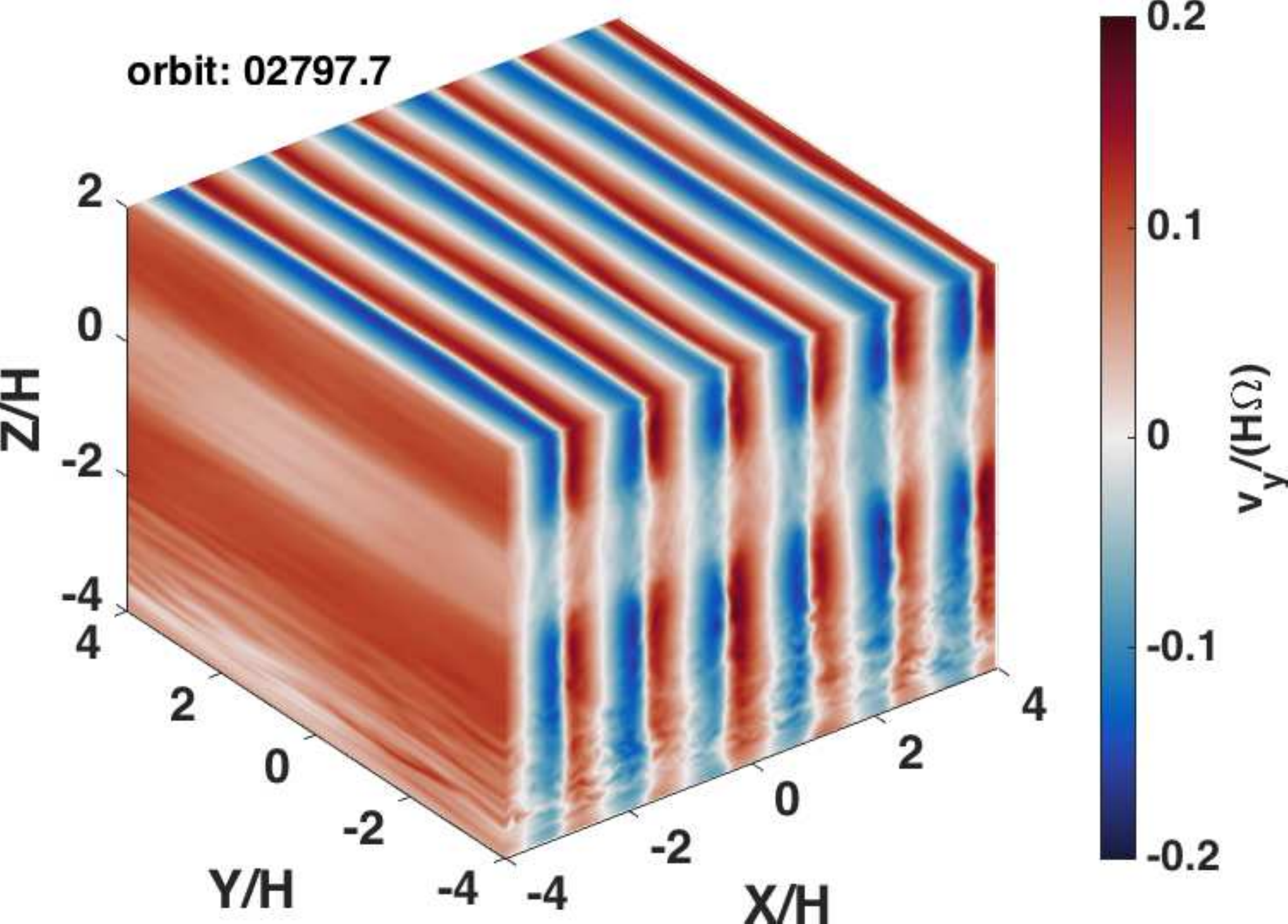}	{0.45\textwidth}{(a)}
        		\fig{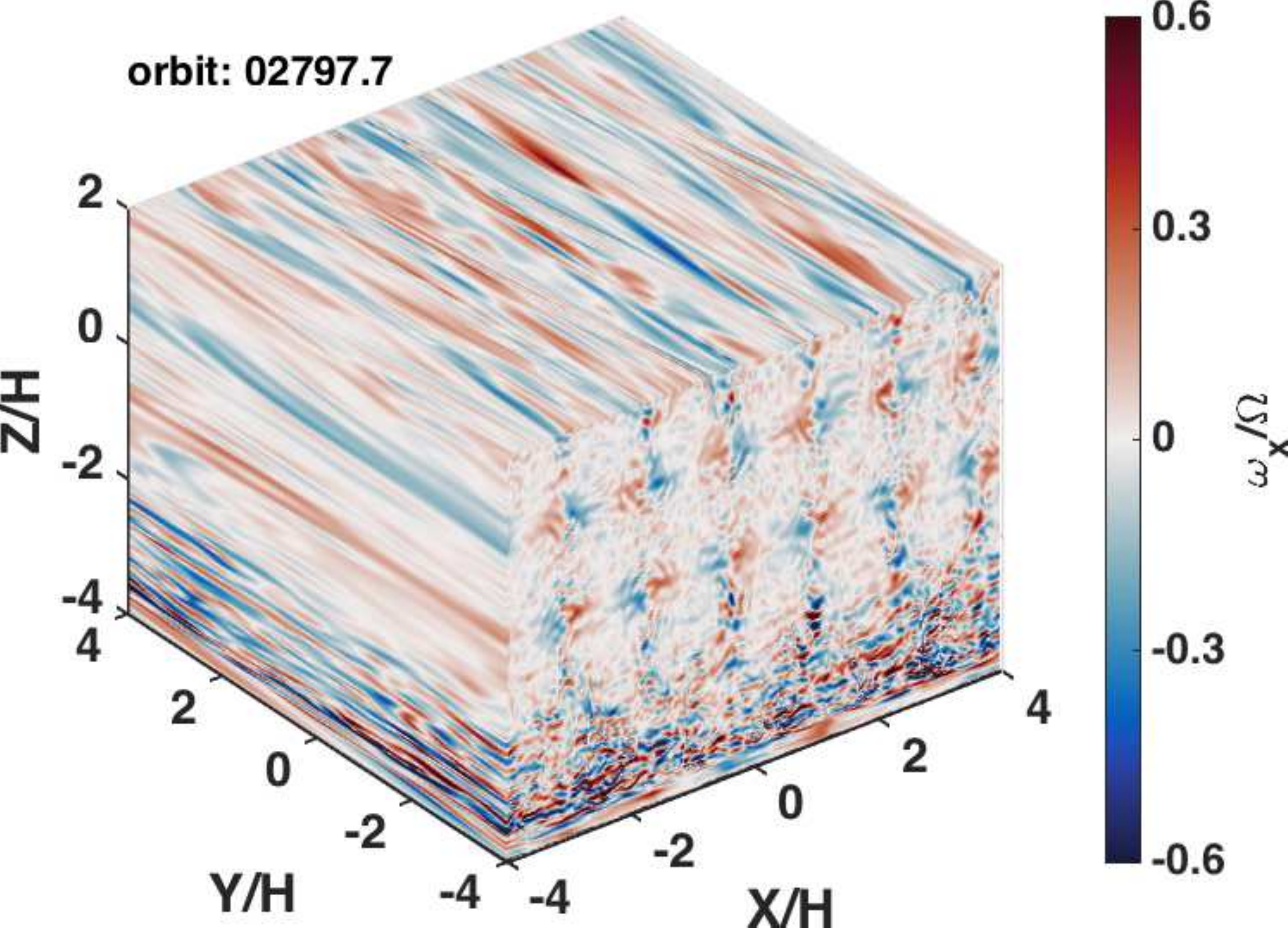}	{0.45\textwidth}{(d)}}
\gridline{	\fig{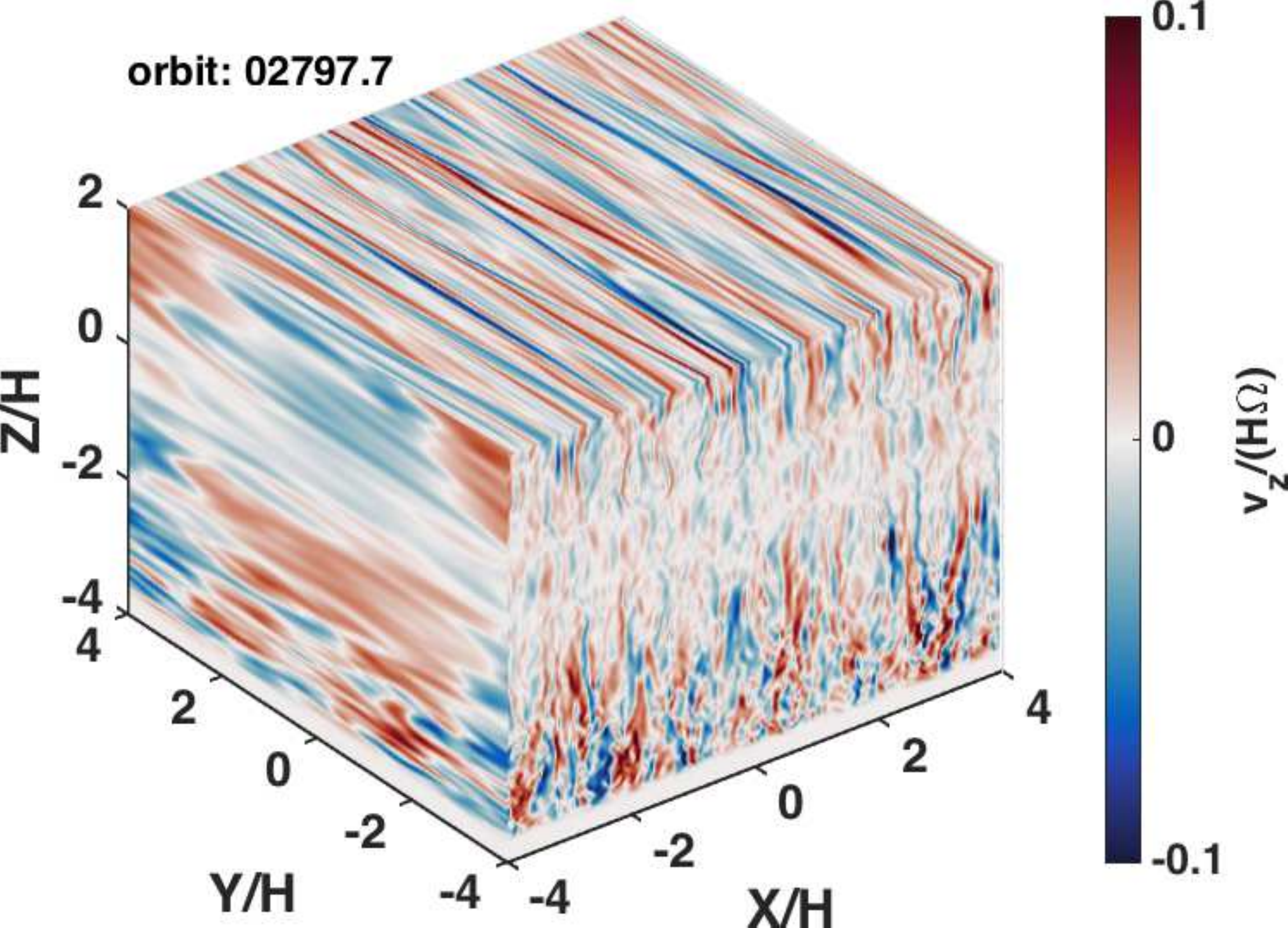}	{0.45\textwidth}{(b)}
        		\fig{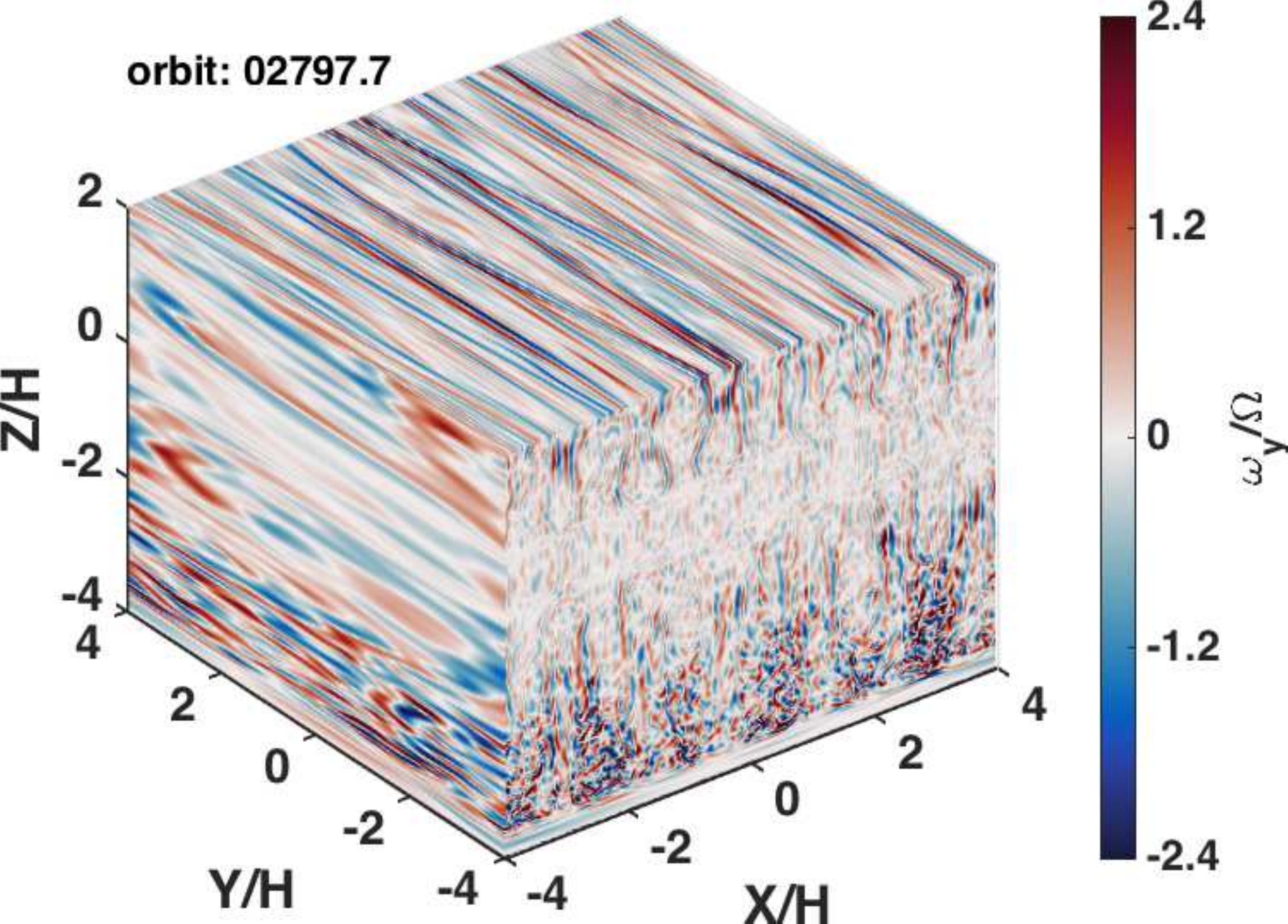}	{0.45\textwidth}{(e)}}
\gridline{	\fig{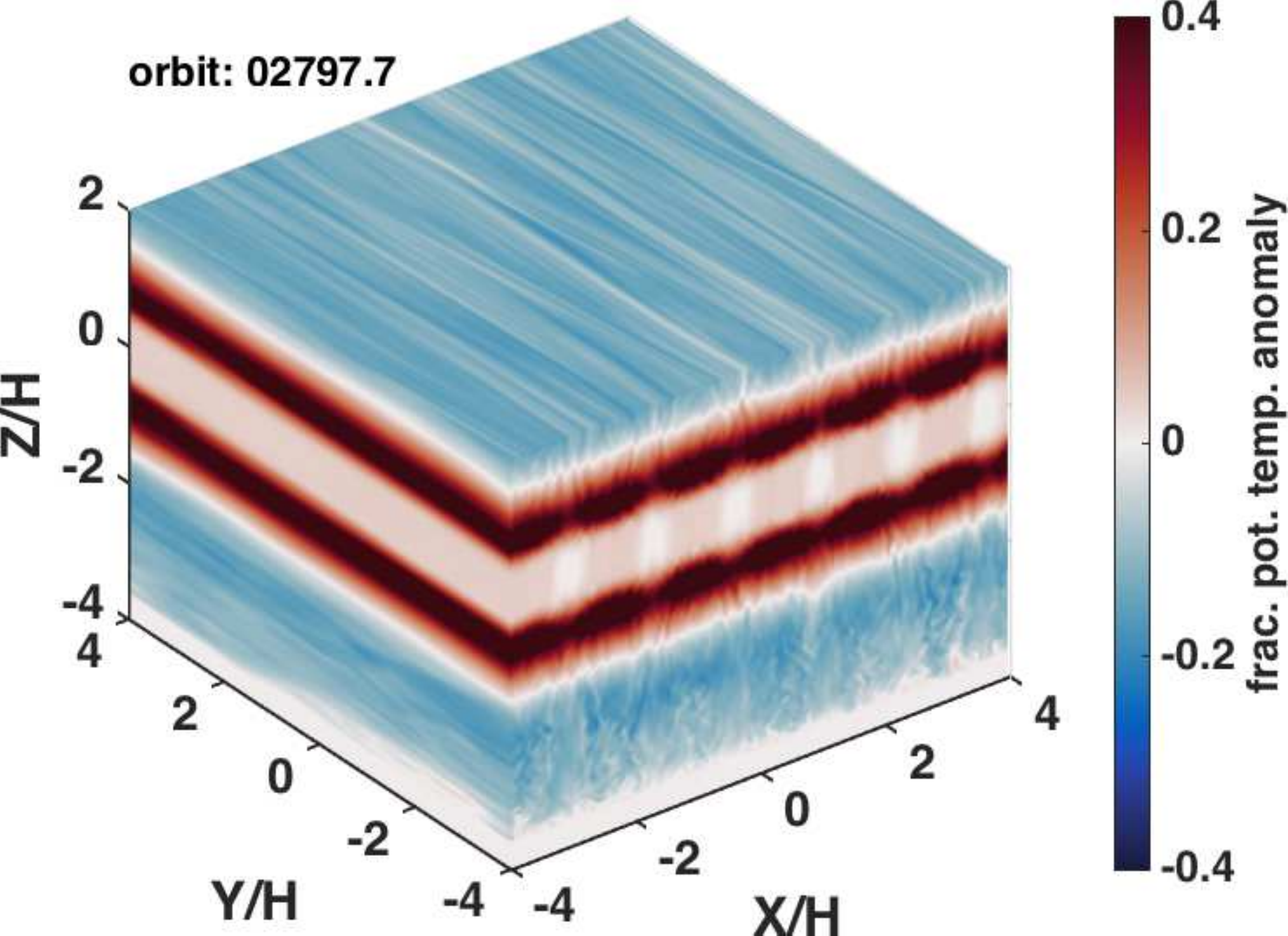}	{0.45\textwidth}{(c)}
        		\fig{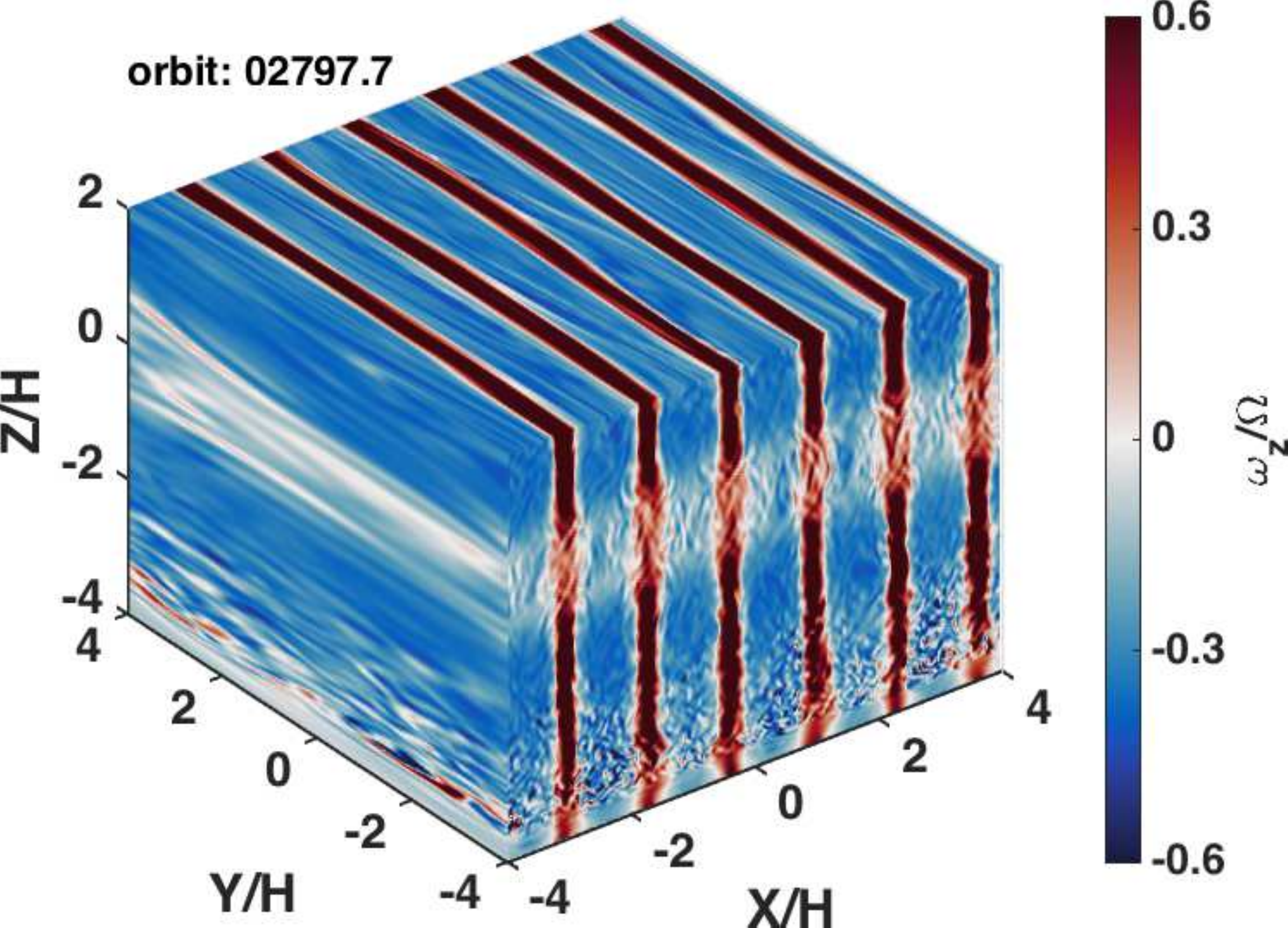}	{0.45\textwidth}{(f)}}
\caption{{\bf Late-time evolution (after 2797.7 orbits) for profile Run\_Brunt\_Step.}  The left column (subfigures a-c) shows the azimuthal velocity (Keplerian shear subtracted), vertical velocity and fractional potential temperature anomaly.  The right column (subfigures d-f) shows the three components of the relative vorticity (Keplerian vorticity subtracted from vertical vorticity).}\label{fig:zvi_images04b}
\end{figure}

\begin{figure}
\gridline{	\fig{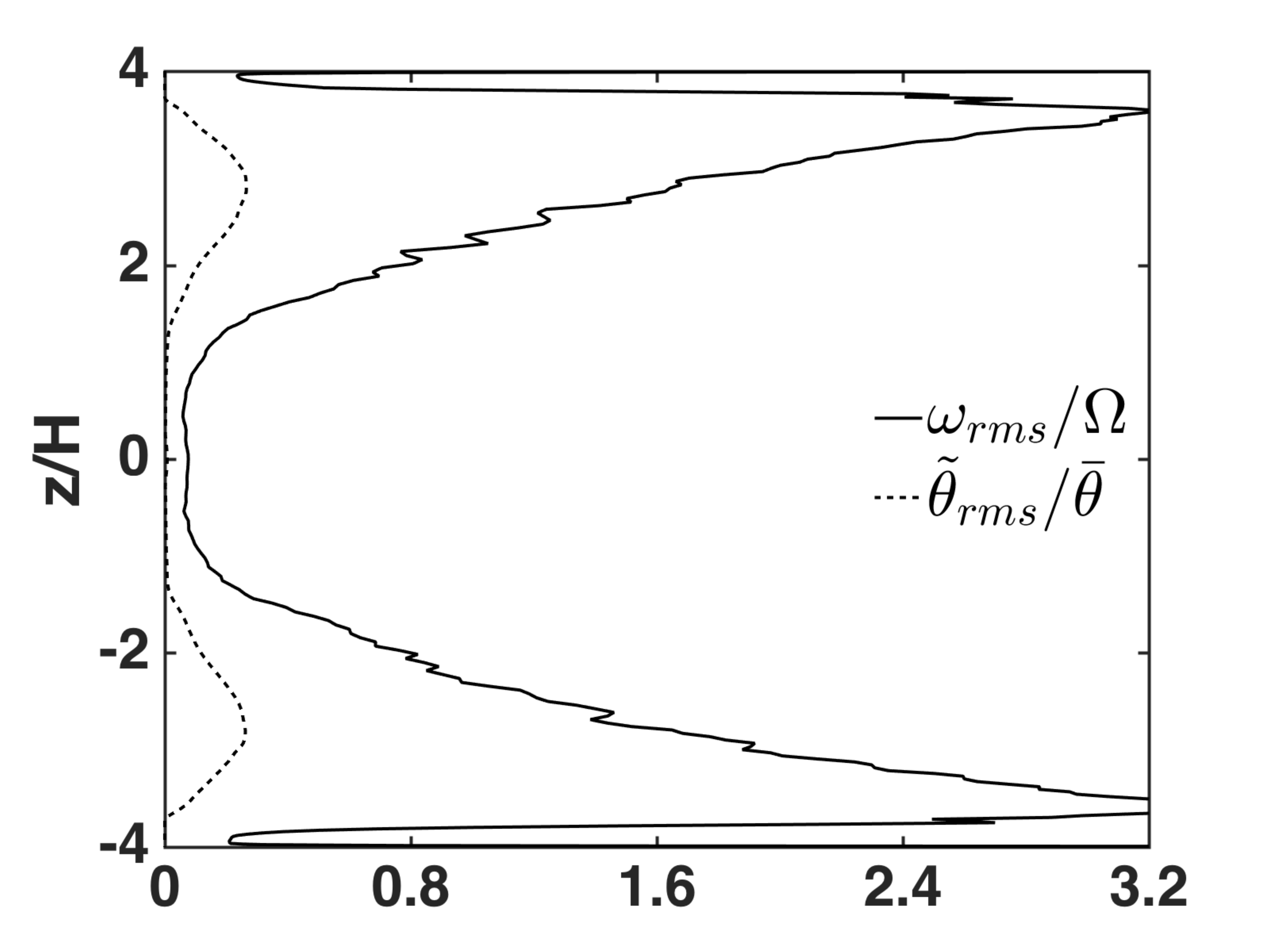}{0.33\textwidth}{(a)}
        		\fig{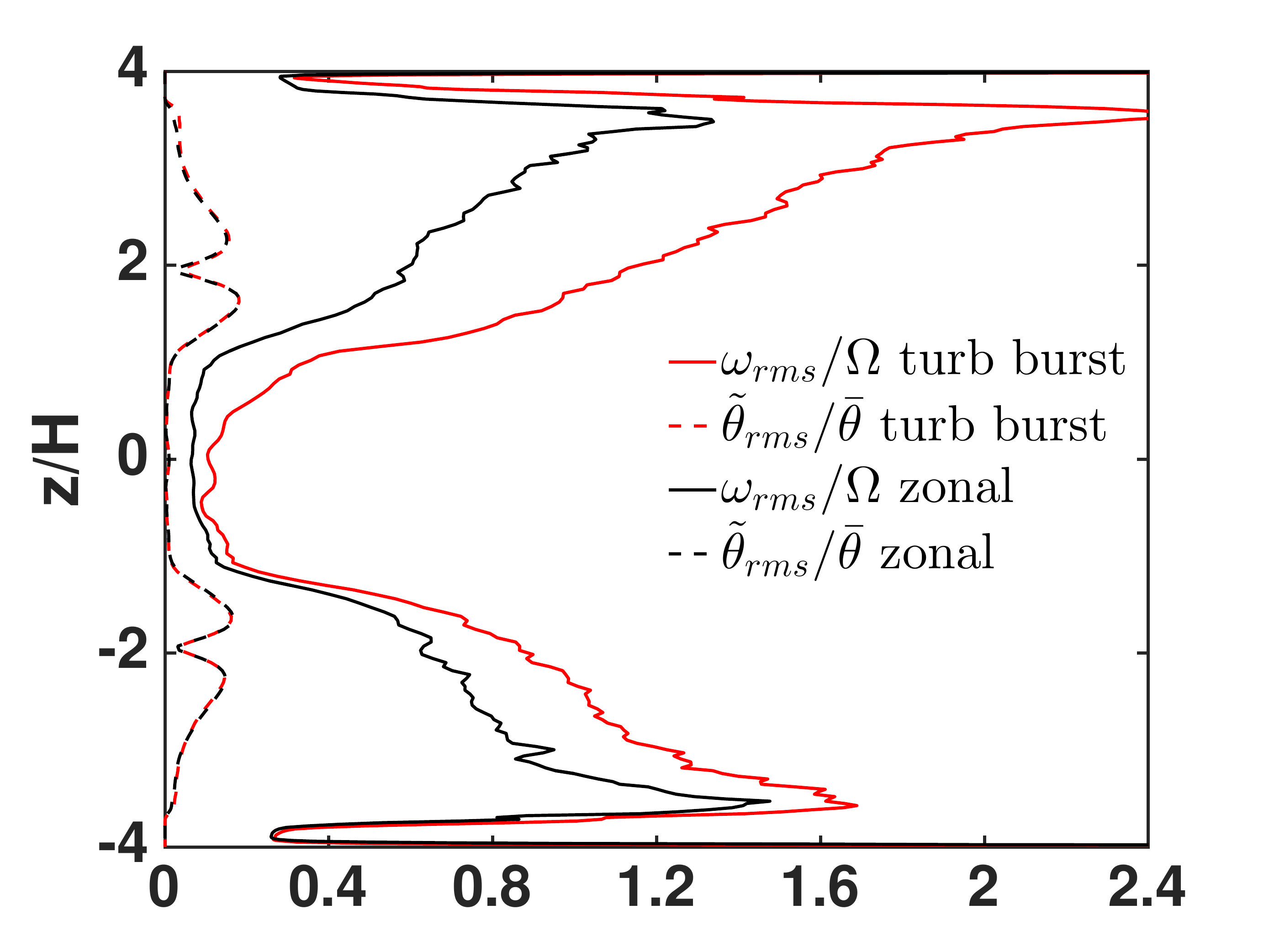}{0.33\textwidth}{(b)}
		\fig{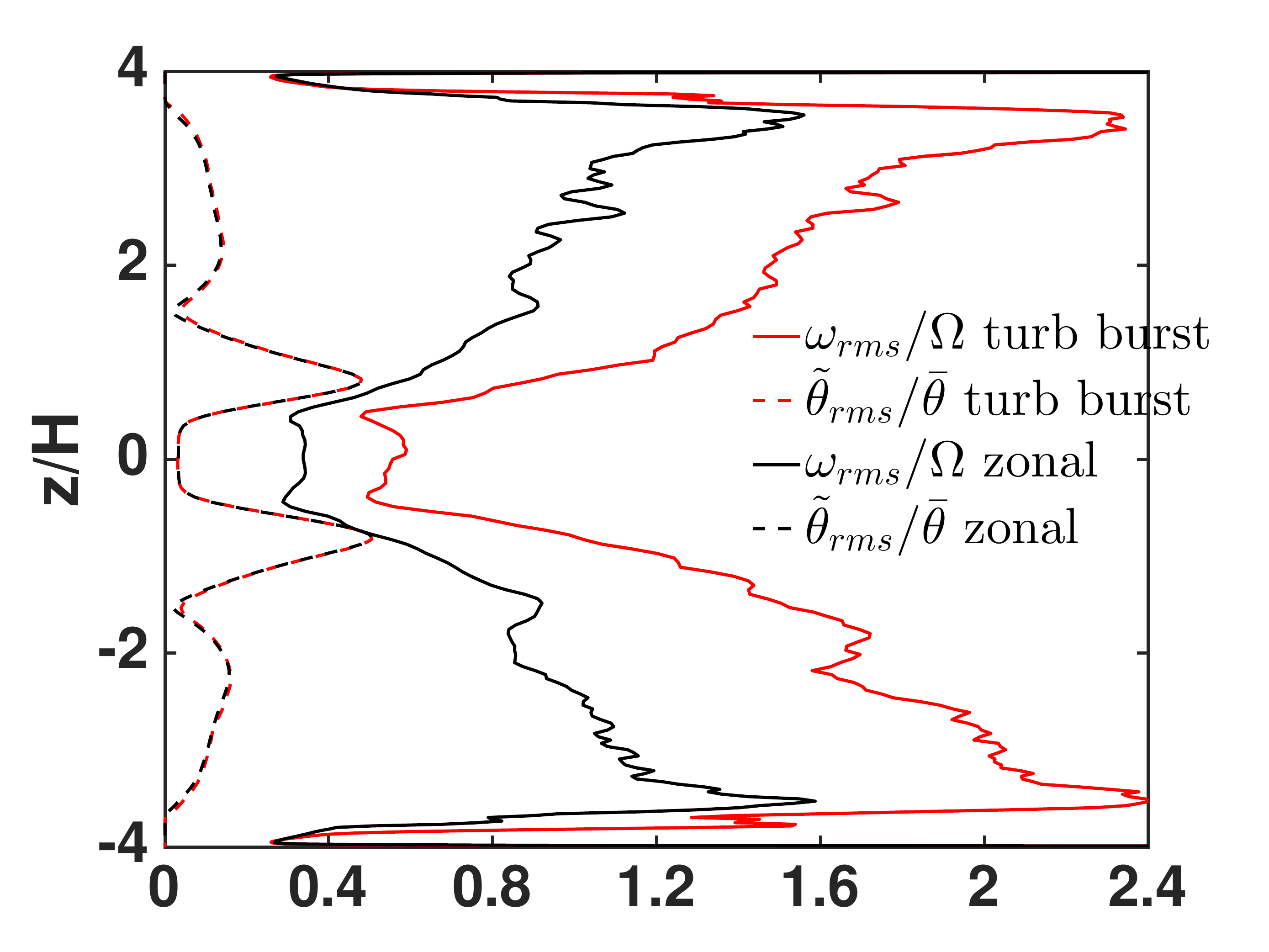}{0.33\textwidth}{(c)}}
\caption{{\bf Vertical extent of zombie turbulence.}  Root mean square (rms) of relative vorticity and fractional potential temperature anomaly (averaged over $x\!-\!y$ planes) as function of height $z$: (a) Run\_Isothermal after 1792.7 orbits, (b) Run\_Temp\_Step after 2444.6 orbits (turbulent burst phase, shown in red) and 2580.4 orbits (zonal flow phase, shown in black), (c) Run\_Brunt\_Step after 2757.2 orbits (turbulent burst phase, shown in red) and 2797.7 orbits (zonal flow phase, shown in black).}\label{fig:zombie_thickness}
\end{figure}

\begin{figure}
\gridline{	\fig{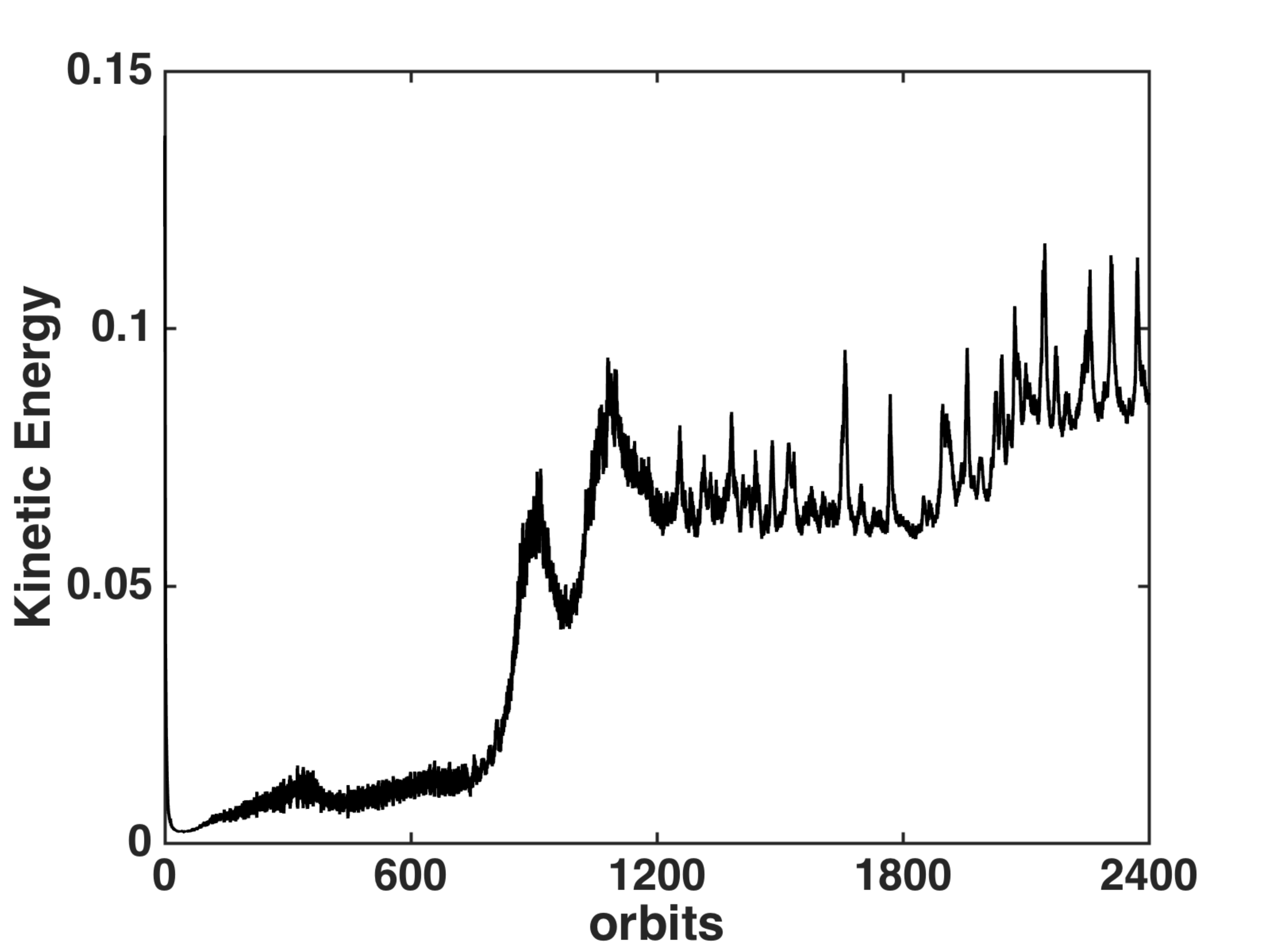}{0.5\textwidth}{(a)}
        		\fig{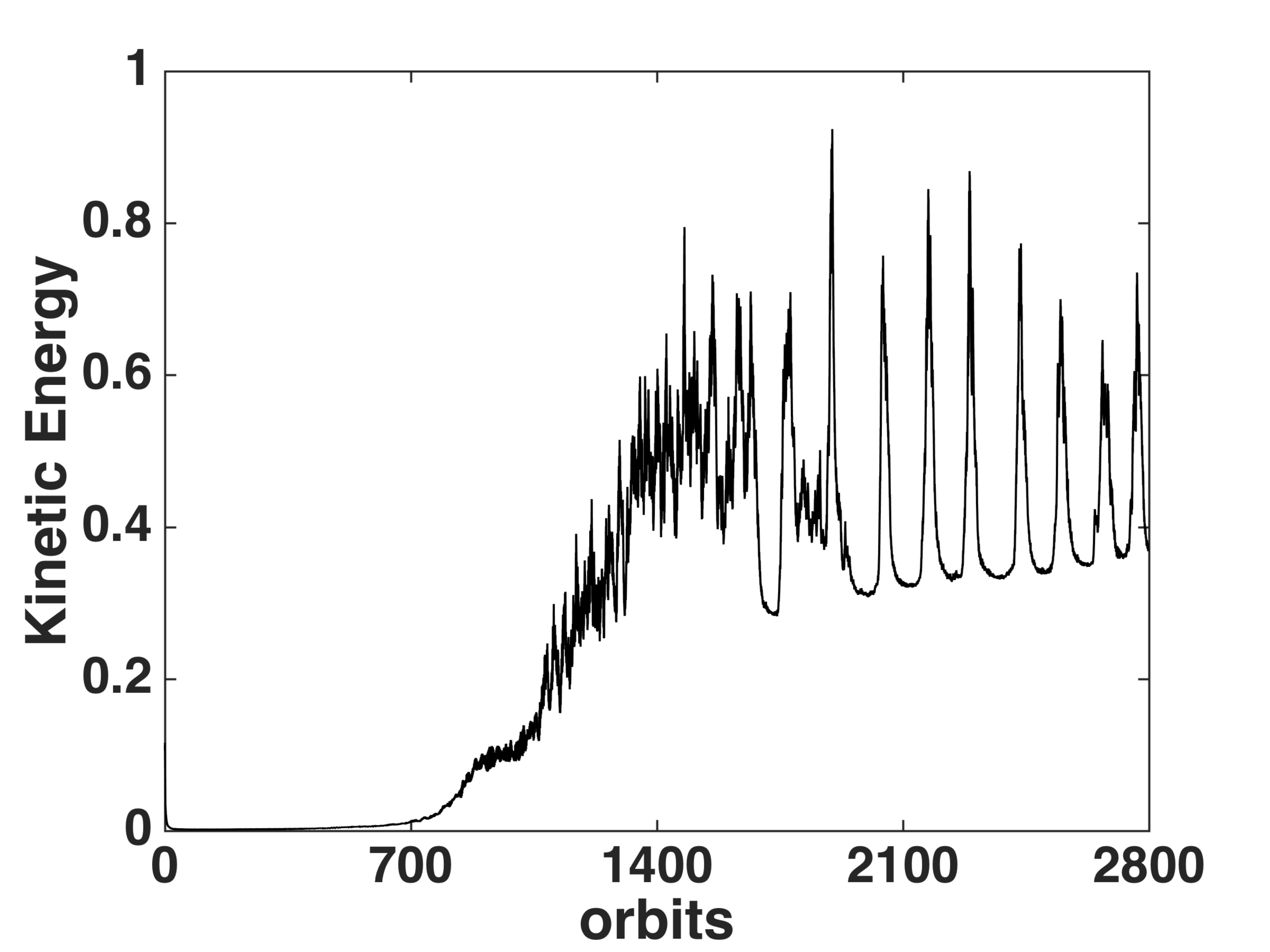}{0.5\textwidth}{(d)}}
\gridline{	\fig{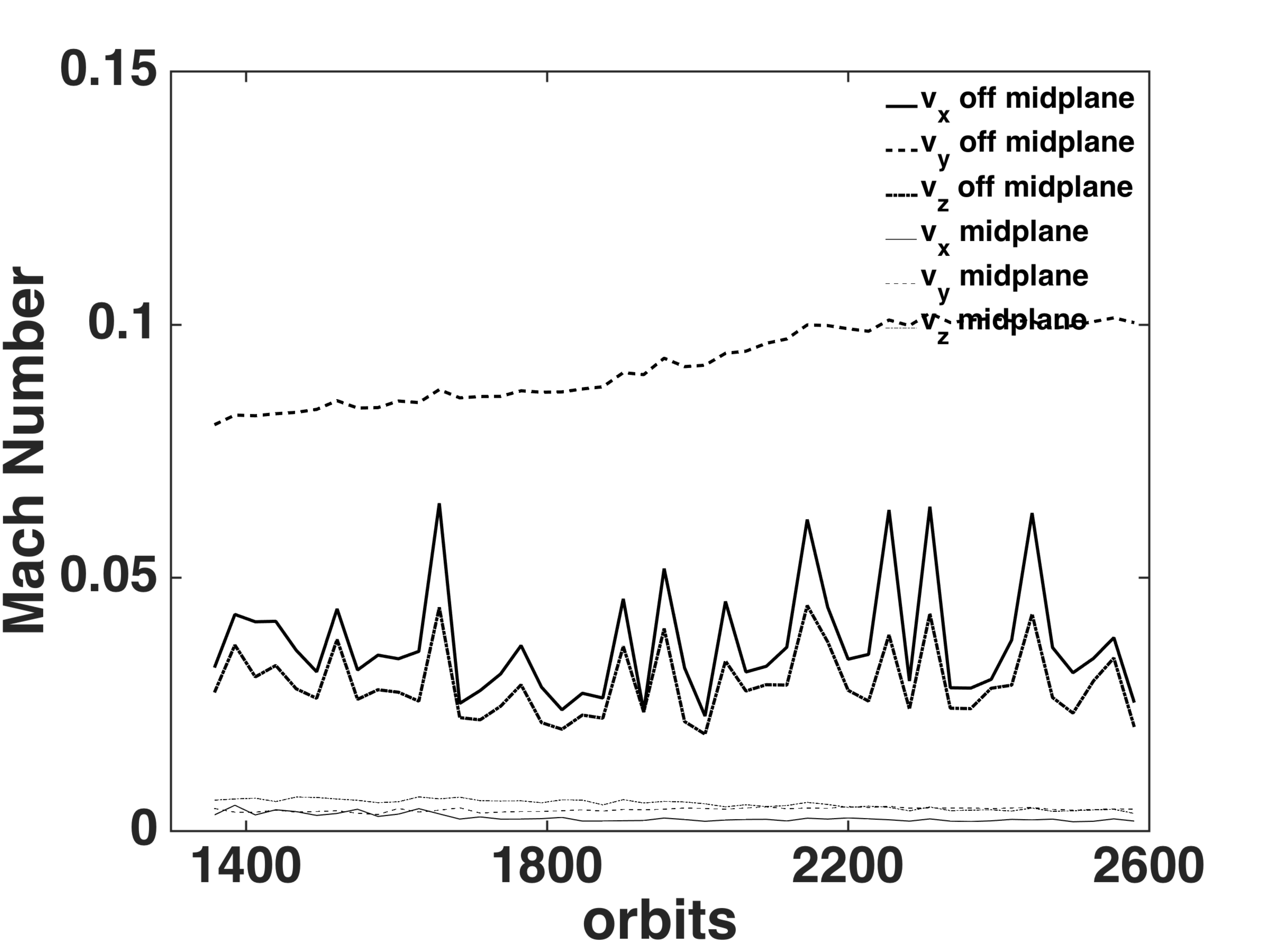}{0.5\textwidth}{(b)}
        		\fig{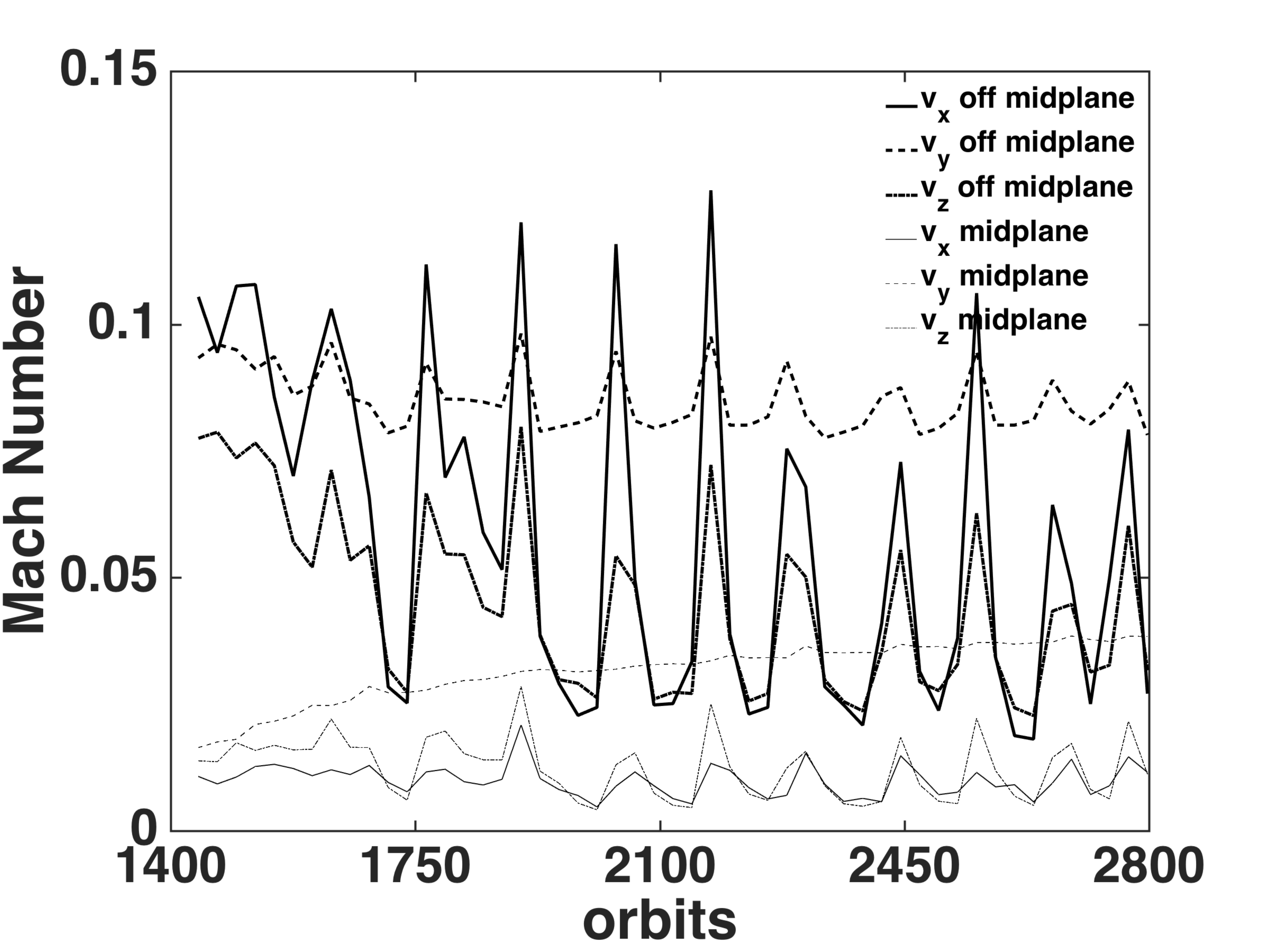}{0.5\textwidth}{(e)}}
\gridline{	\fig{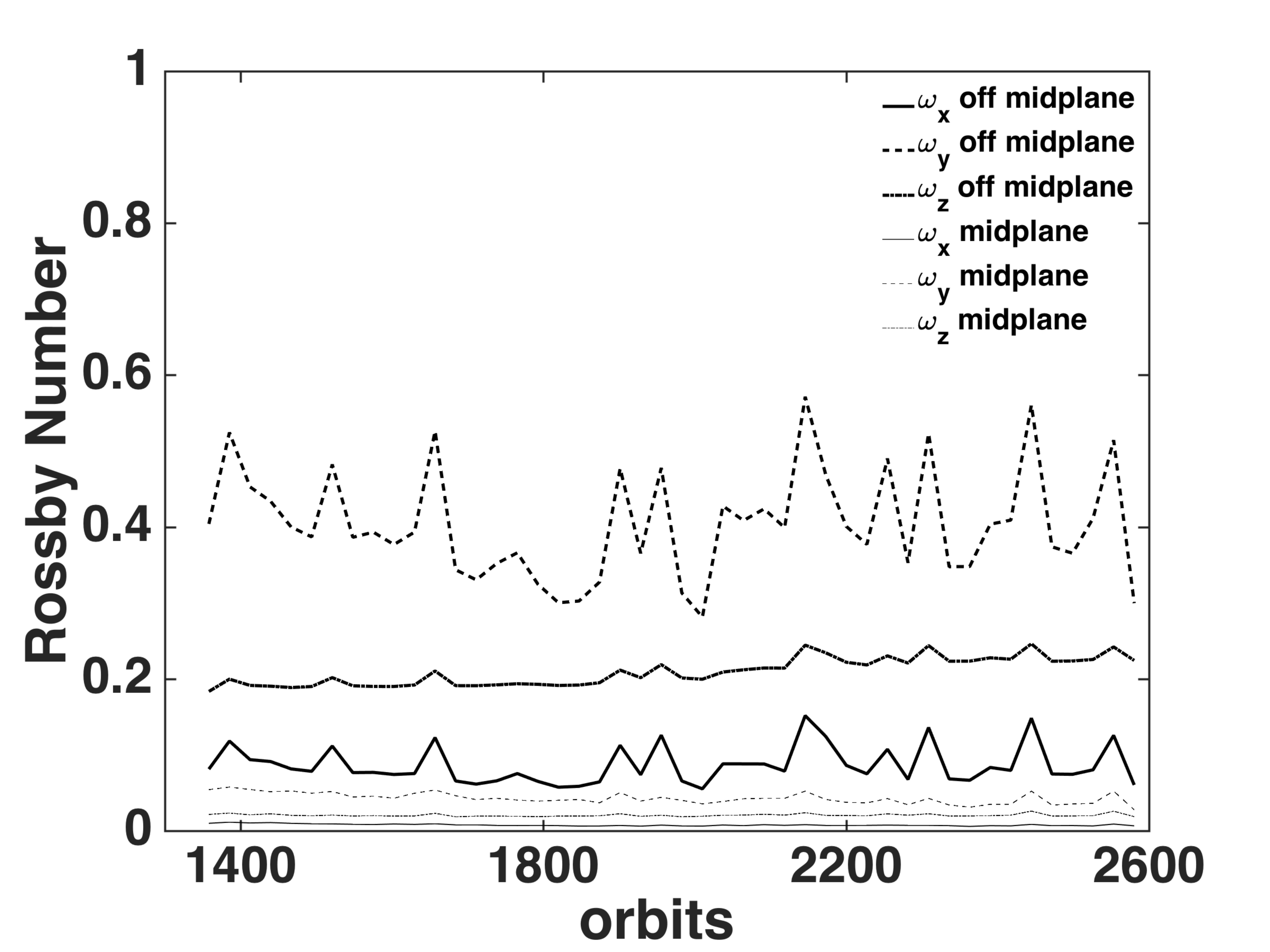}{0.5\textwidth}{(c)}
        		\fig{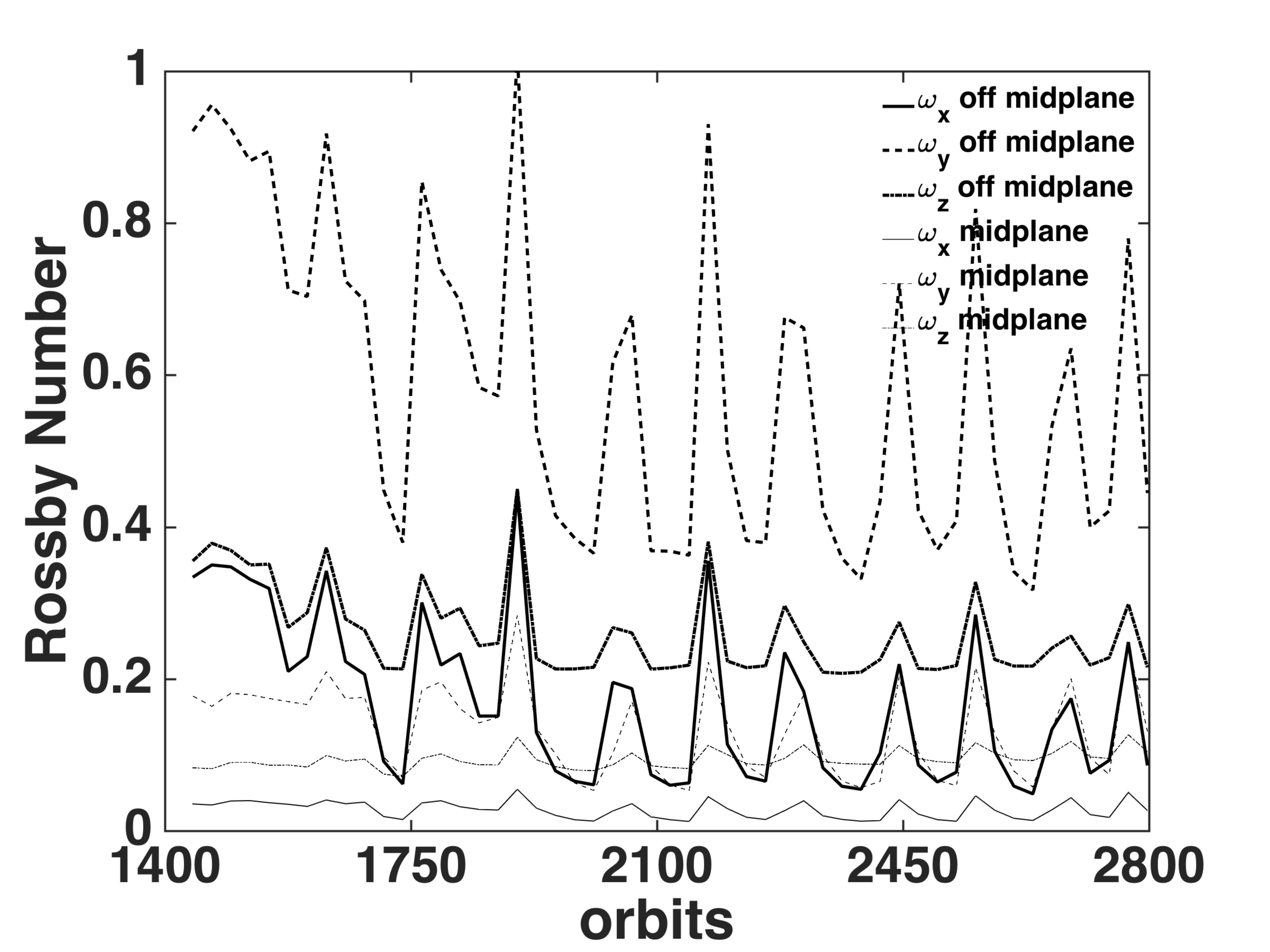}{0.5\textwidth}{(f)}}
\caption{{\bf Diagnostics for profiles Run\_Temp\_Step (a-c) and Run\_Brunt\_Step (d-f).} (a,d) Kinetic energy is $\int\tfrac{1}{2}\bar{\rho}|\boldsymbol{v}-\boldsymbol{v}_{Kep}|^2dV$ in units of $\rho_0H^5\Omega^2$. (b,e) Mach number is $Ma_{j,rms}\equiv v_{j,rms}/c_s$ for the rms of each velocity component $v_j$.  Darker lines are for averages within regions away from the midplane $2H<|z|<3H$, whereas lighter lines are for averages for regions close to the midplane $|z|<0.25H$.    (c,f) Rossby number is $Ro_{j,rms}\equiv\omega_{j,rms}/(2\Omega)$ for the rms of each component of the relative vorticity $\omega_j$.  Lighter and darker lines are for averages over the same regions as in (b,e).}\label{fig:diagnostics}
\end{figure}

The velocity fields in all runs were initialized with 3D homogeneous, isotropic Kolmogorov turbulence with rms velocity $v_{rms}=0.05H\Omega$, and the initial potential temperature anomalies were set to zero.  {\color{black}ZVI has a finite-amplitude trigger, and we previously investigated the magnitude of perturbations that are necessary to excite the baroclinic critical layers. In \citet{MPJB16}, we demonstrated that ZVI requires perturbations with Mach number as small as $Ma\gtrsim 10^{-6}$.}  
All three profiles were susceptible to ZVI and eventually evolved into fully developed ``zombie turbulence'' characterized by intense, undulating cyclonic vortex sheets parallel to the $y\!-\!z$ plane, interspersed with coherent anticyclonic zombie vortices.  Figure~\ref{fig:zvi_images01} shows the late time evolution for Run\_Isothermal after 1792.7 orbits and Figure~\ref{fig:zvi_images03} shows the late time evolution for Run\_Temp\_Step after 2580.4 orbits.  Figures~\ref{fig:zvi_images04a} and \ref{fig:zvi_images04b} shows the late time evolution for Run\_Brunt\_Step at two different time, after 2770.6 orbits and 2797.7 orbits.  In the first column of plots for Figures~\ref{fig:zvi_images01}-\ref{fig:zvi_images04b}, the displayed variables are, from top down: (a) perturbation azimuthal velocity $v_y-\bar{v}_y$, (b) vertical velocity $v_z$, (c) fractional potential temperature anomaly $\tilde{\theta}/\bar{\theta}$.  In the second column of these same plots, from top down (d)-(f) are the $x$, $y$, and $z$ components of the vorticity $\boldsymbol{\omega}\equiv\boldsymbol{\nabla}\boldsymbol{\times}\boldsymbol{v}$ (the vorticity of the Keplerian shear is subtracted from the vertical component of vorticity).  In the remainder of this section, we will highlight the main features of fully-developed zombie turbulence.

\paragraph{Vertical extent of zombie turbulence and modification of background stratification:}  

As expected, zombie turbulence is strongest in the stratified regions away from the ZVI-inactive midplane, yet it is clear that some turbulence penetrates into the unstratified midplane.  Figure~\ref{fig:zombie_thickness} show the vertical extent of the domain that is filled with turbulence:  the solid lines indicate the root mean square (rms) of the relative vorticity (averaged over horizontal planes) as functions of height, while dashed lines are for the rms fractional potential temperature anomalies (averaged over horizontal planes) as functions of height.  One can see that there is a variation in the vertical extent of zombie turbulence depending on the vertical stratification profiles: the turbulence in Run\_Isothermal is found mostly away from the midplane $|z|\gtrsim2H$, whereas the turbulence in Run\_Brunt\_Step has the greatest penetration into the midplane.  Looking more closely at Run\_Brunt\_Step, Figure~\ref{fig:diagnostics}e compares the rms Mach number for each component of the velocity in a region around the midplane $|z|<0.25H$ (lighter lines), and for regions away from the midplane $2H<|z|<3H$ (darker lines); similarly, figure~\ref{fig:diagnostics}f compares the rms Rossby number for each component of the vorticity in a region around the midplane $|z|<0.25H$ (lighter lines), and for regions away from the midplane $2H<|z|<3H$ (darker lines).  While the region around the midplane is inactive to ZVI, it is nonetheless turbulent, though with Mach numbers around a few percent, which is a factor of roughly 4 down from the more vigorous turbulence in the stratified ZVI-active regions.

At late times, we observe that zombie turbulence has altered the underlying background stratification.  Returning to Figure~\ref{fig:background_profiles}, we compare the initial and final background profiles for potential temperature, temperature, and Brunt-V\"{a}is\"{a}l\"{a} frequency, as functions of distance from midplane.  For Run\_Isothermal, the Brunt-V\"{a}is\"{a}l\"{a} frequency initially is a linear function of height, yet at late times, it appears that the Brunt-V\"{a}is\"{a}l\"{a} frequency has become roughly constant within the ZVI active region.  In Run\_Temp\_Step, it appears that the vertical mixing from zombie turbulence has somewhat flattened the initial local maximum of Brunt-V\"{a}is\"{a}l\"{a} frequency. These results suggest that perhaps mixing in zombie turbulence homogenizes the stratification.

\paragraph{Creation of quasi-steady-state zonal flow:}

\begin{figure}
\epsscale{0.85}
\plotone{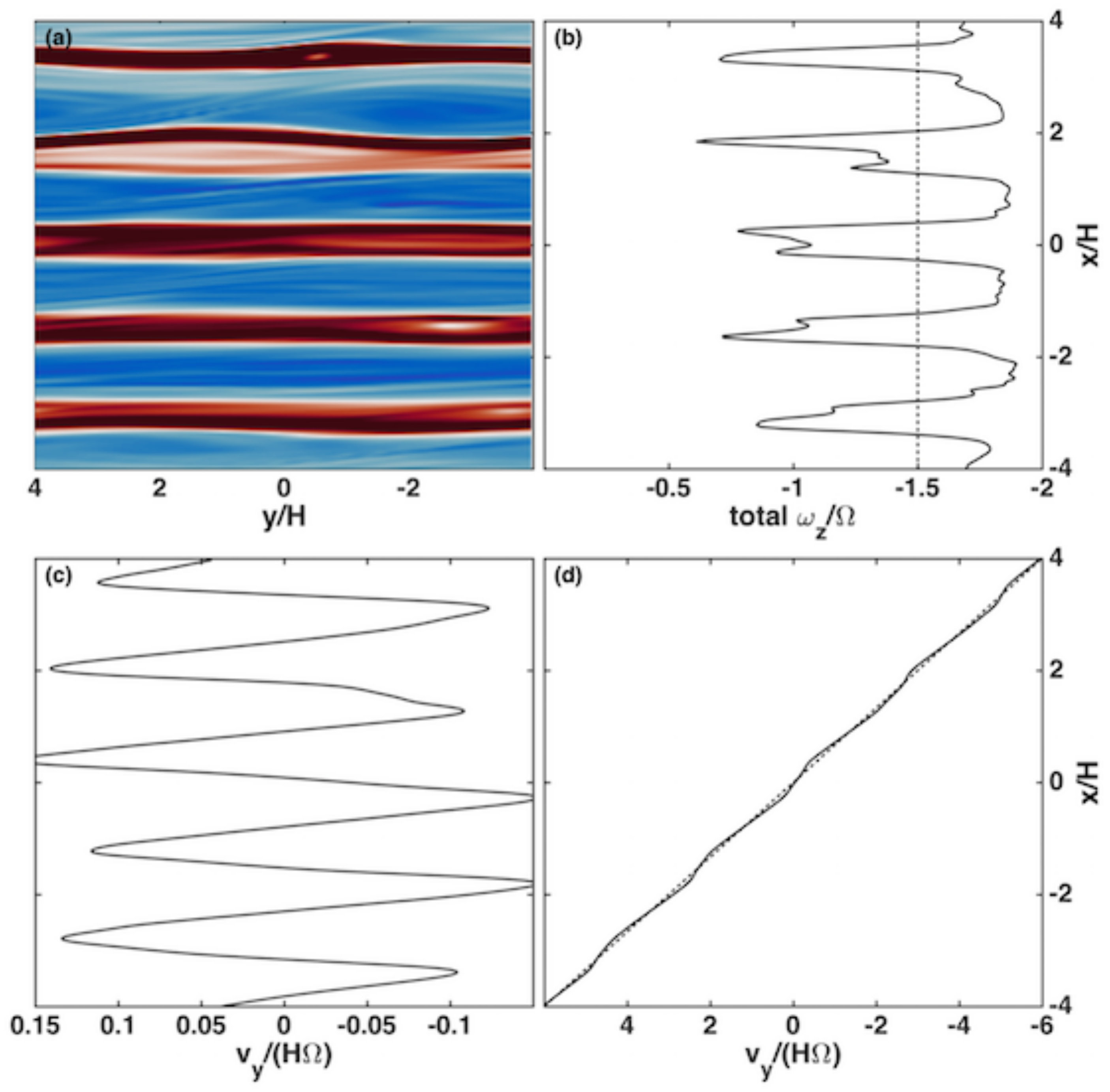}
\caption{{\bf Zonal flow in profile Run\_Temp\_Step.} (a) Vertical component of relative vorticity in a horizontal plane at height $z=2H$ after $\approx2600$ orbits.  Colormap is the same as in Figure~\ref{fig:zvi_images03}.  (b) Azimuthally averaged vertical component of total vorticity in rotating frame.  The total vorticity here is the relative vorticity plus the vorticity of the Keplerian shear, $\omega_{Kep}=-(3/2)\Omega$, indicated by the vertical dashed line.  (c) Corresponding azimuthally averaged relative azimuthal velocity $v_y-v_{Kep}$. (d)  Corresponding azimuthally averaged azimuthal velocity, including background Keplerian shear flow.  Note that the graph in panel (b) is exactly the first derivative of the graph in panel (d).}\label{fig:zonal03}
\end{figure}

\begin{figure}
\epsscale{0.85}
\plotone{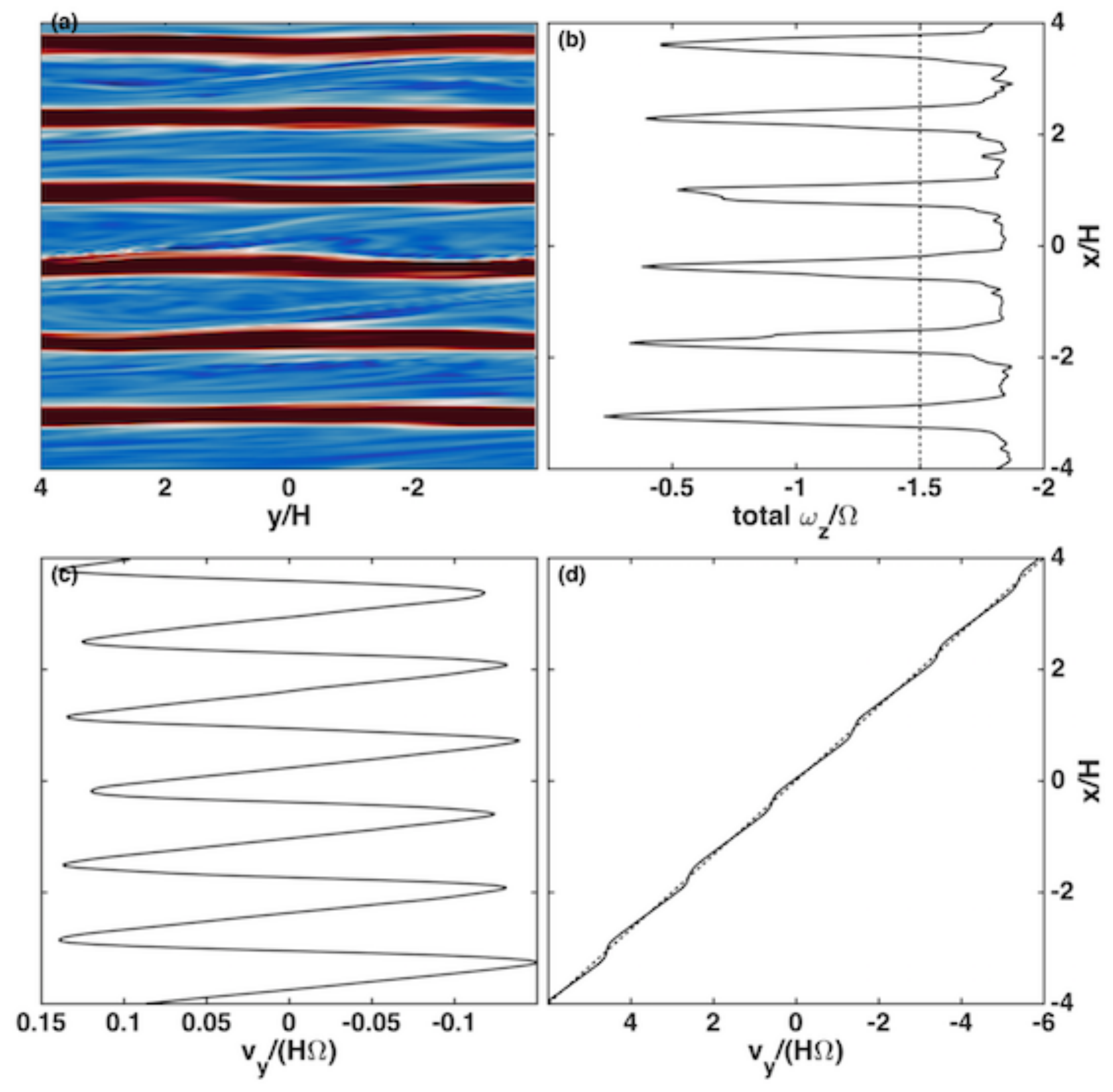}
\caption{{\bf Zonal flow in profile Run\_Brunt\_Step.} (a) Vertical component of relative vorticity in a horizontal plane at height $z=2H$ after $\approx2800$ orbits.  Colormap is the same as in Figure~\ref{fig:zvi_images04a}.  (b) Azimuthally averaged vertical component of total vorticity in rotating frame.  The total vorticity here is the relative vorticity plus the vorticity of the Keplerian shear, $\omega_{Kep}=-(3/2)\Omega$, indicated by the vertical dashed line.  (c) Corresponding azimuthally averaged relative azimuthal velocity $v_y-v_{Kep}$. (d)  Corresponding azimuthally averaged azimuthal velocity, including background Keplerian shear flow.  Note that the graph in panel (b) is exactly the first derivative of the graph in panel (d).}\label{fig:zonal04}
\end{figure}

At late times, the computational domain is segmented in the radial ($x$) direction into alternating $y\!-\!z$ slabs of vertical vorticity; there are five or six pairs of opposite-signed vortex layers over a radial extent of $8H$, for a radial separation $\Delta x=1.33H$ (for Run\_Brunt\_Step)  or $1.60H$ (for Run\_Isothermal and Run\_Temp\_Step).  The radial thickness of the cyclonic layers is approximately one third the thickness of the anticyclonic layers, but this is compensated by the fact that the cyclonic layers are roughly three times more intense as the anticyclonic layers, so that the average vertical vorticity over any $x$-$y$ plane vanishes.   What is the physics that sets the scale of the radial separation?  Most likely it is imprinted from the initial critical layer separation. In \citet{MPJH13} and \citet{MPJBHL15}, we show that the fundamental critical layer separation in Keplerian shear is $\Delta\equiv NL_y/(3\pi\Omega)$.  For Run\_Brunt\_Step, this is $\Delta\approx 1.07H$, which is a bit smaller than the observed separation of the cyclonic sheets.  However, we noted in \citet{MPJBHL15} that zombie vortices spawned in neighboring critical layers may merge to increase the separation of zones in late time zombie turbulence.

At late times, we see that there is very stable zonal azimuthal flow superposed on the Keplerian shear.  In Figures~\ref{fig:zonal03}a~and~\ref{fig:zonal04}a, we show the vertical component of the relative vorticity in a horizontal plane at $z=2H$ for the final times for Run\_Temp\_Step and Run\_Brunt\_Step. Here, we can more clearly see the radial spacing of the five or six pairs of alternating cyclonic and anticyclonic vortex layers.  We average the relative vorticity azimuthally (over the $y$ coordinate) and add the Keplerian vorticity to yield the total vorticity in the rotating frame, graphed in panel (b) of both figures.  We have not done so here, but one could add a constant $+2\Omega$ to the vorticity in the rotating frame to yield the total vorticity in an inertial reference frame.  In panel (c) of Figures~\ref{fig:zonal03}~and~\ref{fig:zonal04}, we show the azimuthally averaged relative azimuthal velocity $v_y-v_{Kep}$.  We add the Keplerian shear flow to this to yield the total azimuthal velocity in the rotating frame, graphed in panel (d).  We note that the graph in panel (b) is the exact first derivative of the graph in panel (d).

\paragraph{Turbulent burst cycles:}

\begin{figure}
\epsscale{1.0}
\plotone{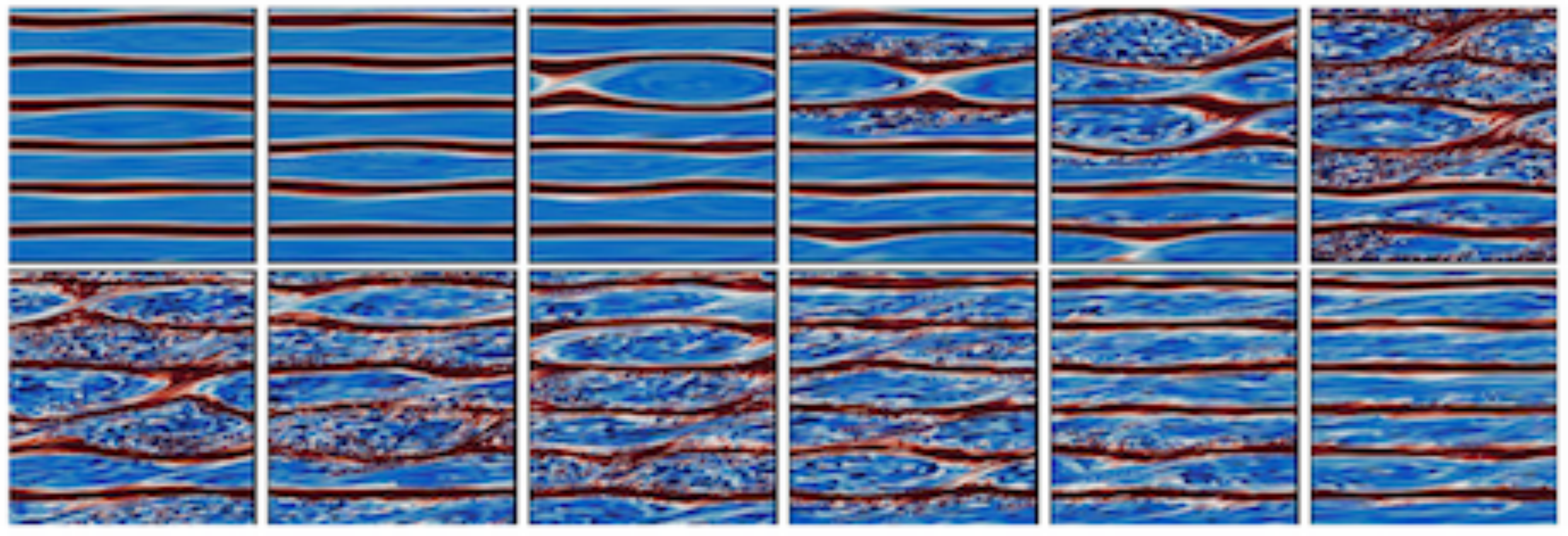}
\caption{Turbulent burst cycle in Run\_Brunt\_Step.  Shown are $x-y$ slices (in each panel, $x$ increases upwards, $y$ increases leftwards) at $z\!=\!2H$ of the vertical component of relative vorticity, with same color scale as Fig.~\ref{fig:zvi_images04a}.  First frame is at orbit 2750.2 and last frame is at orbit 2787.5 with other frames at equal intervals of 3.4 orbits.}\label{fig:turb_burst04}
\vspace{0.125in}
\end{figure}

In Figure~\ref{fig:turb_burst04}, we zoom in to show the vertical component of relative vorticity for Run\_Brunt\_Step in horizontal slices at $z\!=\!2H$ between orbits 2750.2 to 2787.6 at intervals of 3.4 orbits.  Over this period, we can see the zonal flow make a transition from a laminar state to a more turbulent, chaotic state.  The vertical sheets of cyclonic vorticity appear to wildly undulate and at times intersect neighboring cyclonic sheets, pinching off anticyclonic vortices.  The transition does not start at the same time throughout the domain, but appears to be initiated between a pair of cyclonic sheets and then progress to neighboring zones.  In Figure~\ref{fig:diagnostics}, turbulent bursts are easily identified with spikes of kinetic energy.   For Run\_Brunt\_Step, the average period between bursts is roughly 125 orbits, but varies between 100 and 150 orbits.  In Run\_Temp\_Step, the bursts are not as prominent with much more of a spread in the interval between bursts.  Turbulent bursts and intermittency are common features of rotating flows in the laboratory \citep{coughlin96, colovas1997}.  We hypothesize that one of the drivers for this turbulent bursting is the discrepancy between the fundamental separation of critical layers $\Delta$ and the  actual observed separation after mergers of neighboring critical layer regions.  We highlight this as an area for future investigations.

\newpage
\section{SIMULATIONS OF ZOMBIE VORTEX INSTABILITY WITH RADIATIVE DAMPING}\label{sec:zvi_with_damping}

We now address how radiative damping of the thermal fluctuations inside baroclinic critical layers affects the onset of ZVI.  \citet{lesurlatter2016} performed a series of simulations with optically thin cooling (with a Newton cooling prescription) and with optically thick radiative diffusion (with a Laplacian operator).  They found that ZVI required an optically thin cooling time $t_{thin}\gtrsim10\Omega_K^{-1}$ or a radiative P\'{e}clet number $Pe_{rad}\ge 10^4$ (definition in Appendix A).  In Appendix A, we show that results of our own simulations are consistent with these critical values.  Where we disagree with \citet{lesurlatter2016} is how they mapped these results onto different locations within a protoplanetary disk. {\color{black} Using opacities from \citet{semenov2003} (largest dust grains of only a few microns with no significant grain growth), they concluded that the cooling time in the optically thin parts of the disk are too short to allow ZVI, and so they claimed ZVI can only operate in the denser regions close to the protostar where the gas is optically thick and the cooling is set by radiative diffusion.  

In this section, we show that the time for thermal relaxation can be orders of magnitude longer than asserted by \citet{lesurlatter2016}.  In their choice of opacity, they assumed complete mixing of gas and dust and neglected settling into the midplane, which would deplete grains not far from the midplane where ZVI is triggered.  They also chose a grain distribution with very little grain growth, with maximum grain sizes of only a few microns.  We explore various grain growth and settling scenarios, and find that the gas and dust in the off-midplane regions of PPDs are not necessarily in local thermodynamic equilibrium (LTE).  The hydrodynamics which leads to the creation of the baroclinic critical layers generates thermal anomalies in the gas (via pressure-volume work, buoyancy).  The temperature of the dust does not instantaneously match that of the gas, and the gas and dust become out of local thermodynamic equilibrium.  In the case of non-LTE, the thermal relaxation time is dominated by the finite time for the exchange of energy between gas and dust via collisions \citep{malygin2017}.
}

\subsection{Protoplanetary Disk Model and Dust Grain Parameters}

For convenience, we introduce a protoplanetary disk model \citep{cuzzi93} and  express the radius from the star in units of astronomical units: $r_{au}\equiv r/(1~\mathrm{au})$.  In what follows, the subscript ``1'' indicates values at 1~au. Disk properties include: midplane gas temperature $T(r) = T_1r_{au}^{-0.50}$, $T_1=256$~K, gas surface density $\Sigma_g(r)=\Sigma_{g1}r_{au}^{-1.50}$, $\Sigma_{g1}=16{,}384$~kg/m$^{2}$, gas isothermal sound speed $c_s(r)=c_{s1}r_{au}^{-0.25}$, $c_{s1}=960$~m/s, and Toomre parameter $Q(r)=Q_1r_{au}^{-0.25}$, $Q_1=56$.  These values set the protostellar mass $M_{star}=1.01M_{\sun}$ and the disk gas mean molecular weight $\bar{m}=2.31$~amu.  Other properties include: disk aspect ratio $\delta(r) = \delta_1r_{au}^{+0.25}$, $\delta_1=0.0320$, isothermal gas scale height $H(r)=H_1r_{au}^{+1.25}$, $H_1=4.79$~million km, midplane gas density $\rho_{g}(r)=\rho_{g1}r_{au}^{-2.75}$, $\rho_{g1}=1.36\times10^{-6}$~kg/m$^3$, and molecular mean free path in the midplane $\ell_{g}(r)=\ell_{g1}r_{au}^{+2.75}$, $\ell_{g1}=7.79$~mm.

Grains are not monodisperse but are found in a broad distribution of sizes, initially imprinted from the interstellar medium out of which the protostellar nebula collapsed \citep{mathis1977}.  Within the early protoplanetary disk, grain growth is mediated by Brownian motion, vertical settling, radial drift, and turbulence \citep{weidenschilling77,weidenschilling80,weidenschilling84}.  \citet{dullemond2005} showed that small grains should grow so rapidly that in less than $10^4$ years there should be no grains smaller than 100~$\mu$m in protoplanetary disks, which is  inconsistent with infrared observations \citep{vanboekel2005}.  More rigorous modeling that includes fragmentation of grains in high speed collisions leads to power-law size distributions that achieve a quasi-steady state for millions of years \citep{dullemond2005,birnstiel2010b,garaud2013,estrada2016}.  From a theoretical point of view, power law size distributions are a natural outcome of self-similar collisional cascades \citep{dohnanyi1969}.

Henceforth, we will assume a power-law size distribution: the number density of dust grains with radii between $a$ and $a+da$ is given by $n(a)da=Ka^{-s}da$ within the size range $a_{min}$ to $a_{max}$; the normalization factor $K$ is set by the dust mass density.  A canonical value for the power law exponent is $s=3.5$, but can be made shallower, \eg $s=3.0$ for models in which the colliding particles have very little material strength, \ie ``rubble piles" or ``dust bunnies" \citep{pan2005}.  Millimeter observations of debris disks with the VLA, SMA, and ALMA show evidence of intermediate values of $s\approx3.25$ \citep{macgregor2016,wilner2018}.  Integrating over all sizes, one obtains the total number density of dust grains:
\begin{equation}
n_d \equiv \int_{a_{min}}^{a_{max}}n(a)~da = \frac{K}{(s-1)}(a_{min}^{1-s}-a_{max}^{1-s})\approx \frac{K}{(s-1)}a_{min}^{1-s},
\end{equation}
where the last approximation is valid for $a_{max}\gg a_{min}$.  Note that the total number density of dust grains is set by the size of the smallest grains and is relatively insensitive to the size of the larger grains.  Various moments of the size distribution (\ie averages of powers of $a$) often have useful physical interpretations.  The second moment is related to the total geometric cross-section of grains per unit volume:
\begin{equation}
\pi\langle a^2 \rangle n_d \equiv  \int_{a_{min}}^{a_{max}}n(a)\pi a^2~da = \frac{\pi K}{(s-3)}(a_{min}^{3-s}-a_{max}^{3-s})\approx \frac{\pi K}{(s-3)}a_{min}^{3-s}.
\end{equation}
The third moment is related to the total mass density of dust:
\begin{equation}
\rho_d = \tfrac{4}{3}\pi \langle a^3\rangle\rho_sn_d \equiv \int_{a_{min}}^{a_{max}}n(a)\tfrac{4}{3}\pi a^3\rho_s~da =  \frac{\tfrac{4}{3}\pi\rho_sK}{(4-s)}(a_{max}^{4-s}-a_{min}^{4-s})\approx \frac{\tfrac{4}{3}\pi\rho_sK}{(4-s)}a_{max}^{4-s},
\end{equation}
where $\rho_s$ is the solid density of a dust grain.  For the relevant range of the dust power law exponent, $3.0<s<3.5$, the total geometric area is sensitive to the size of the small grains, whereas the mass is in the larger grains.  As we will see in the following sections, the ratio of the third moment and second moment of the size distribution is a key parameter in collisional and radiative timescales:
\begin{equation}
a_S\equiv\frac{\langle a^3\rangle}{\langle a^2\rangle} = \left(\frac{s-3}{4-s}\right)\left(\frac{a_{max}^{4-s}-a_{min}^{4-s}}{a_{min}^{3-s}-a_{max}^{3-s}}\right).
\end{equation}
In fluid dynamics, this ratio (which has units of length) is sometimes called the Sauter mean radius \citep{sauter1926}.  A helpful interpretation for this quantity is that for a polydisperse distribution of grain sizes with a given total surface area and total volume, there is a monodisperse population of grains with the Sauter mean radius that has the same total surface area and total volume.  \citet{chin1986} have argued that this is the best single number to characterize the size distribution since it captures both surface area and volume effects. An interesting special case is for the power law exponent $s=3.5$ in which the Sauter mean radius turns out to simply be the geometric mean of the smallest and largest radii: $a_S\equiv \langle a^3\rangle/\langle a^2\rangle = \sqrt{a_{min}a_{max}}$.  For the case of $s=3.25$, $a_S\approx\tfrac{1}{3}\sqrt[4]{a_{max}^3a_{min}}$, and the Sauter mean radius shifts in weight toward larger particles.

In this work, we fix the minimum grain size $a_{min}=0.1~\mu$m.  We have confirmed that our results are insensitive to choosing smaller values; this makes sense because grains on the size scale of a few nanometers contribute insignificant mass and are very inefficient absorbers or radiators.  For the maximum size of grains, we consider $a_{max}$ in the range of 1~cm to 1~m, reflecting different levels of grain coagulation \citep{natta2007,perez2012,testi2014,carrasco2016,tazzari2016,perez2015}. For spherical grains, the surface area to volume ratio decreases with increasing grain size; the chief effect of grain growth is to decrease the opacity by locking up more mass into large particles that do not contribute significantly to the overall cross-section.  For $a_{min}=0.1~\mu$m and $s=3.5$, the Sauter mean radius is 30~$\mu$m, $100~\mu$m, and 300~$\mu$m for, respectively, $a_{max}=$ 1~cm, 10~cm, and 100~cm.

\subsection{Optically Thin Cooling Time \& Photon Mean Free Path -- Baroclinic Critical Layers Are Optically Thin}

\begin{figure}
\epsscale{0.5}
\plotone{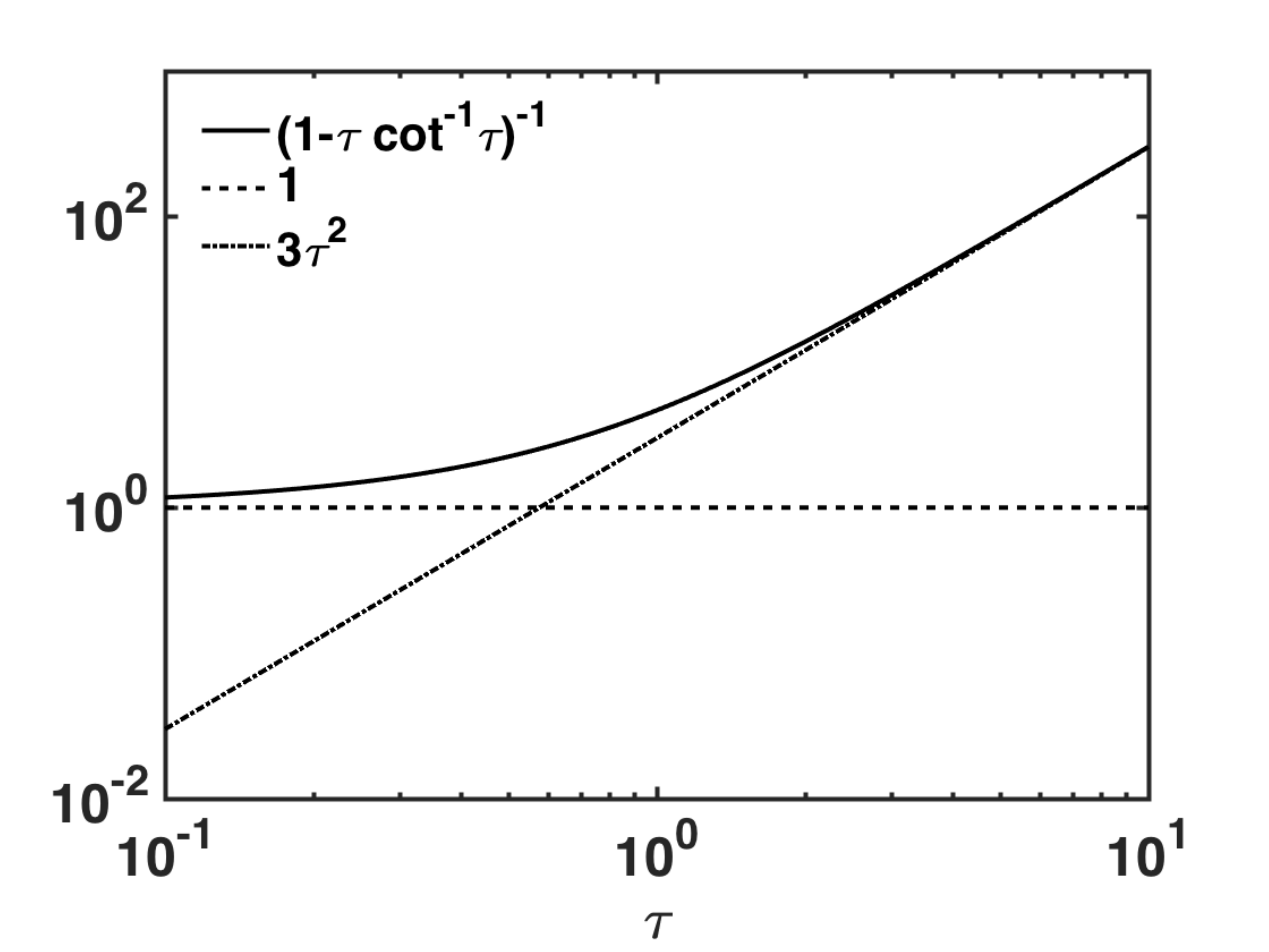}
\caption{{\bf Radiative damping as function of optical depth when the dust and gas temperatures are in local thermodynamic  equilibrium}.  The solid line represents the dependence of the radiative cooling time  $t_{rad}^{LTE}$ as a function of $\tau$. The dashed line is the optically thin limit and the dashed-dotted line is the optically thick limit.  Note that using the optically thin cooling time for large optical depth will overdamp optically thick features, while using the optically thick cooling time for small optical depth will overdamp optically thin features.}\label{fig:tau_graph}
\end{figure}

In this subsection we review the case of radiative cooling times when the temperatures of the dust and gas are tightly coupled and in local thermodynamic equilibrium (LTE). This subsection serves as a reference so that we can compare and contrast LTE results to the more general case when the gas and dust temperatures are not in LTE.  \citet{spiegel1957} derived the timescale $t_{rad}^{LTE}$ for the radiative damping of temperature fluctuations within a uniform gray atmosphere:
\begin{equation}
t_{rad}^{LTE} = \frac{c_p}{16\kappa\sigma T^3}\frac{1}{(1-\tau\cot^{-1}\tau)},\label{E:cooling_time1}
\end{equation}
where $c_p$ is the specific heat of the medium at constant pressure, $\kappa$ is the opacity of the medium, $\sigma$ is the Stefan-Boltzmann constant, $T$ is the temperature, $\rho$ is the mass density, and $\tau=\rho\kappa\ell$ is the optical depth through a perturbation of physical size $\ell$.  We note that \citet{spiegel1957} ignored compressibility effects and so had $c_v$ instead of $c_p$ in his original expressions.  A very good approximation valid for all optical depth is $[1-\tau\cot^{-1}\tau]^{-1}\approx 1+ 3\tau^2$ (see Fig.~\ref{fig:tau_graph}).  Optically thin and thick limits for the radiative timescale are:
\begin{subequations}
\begin{align}
t_{thin}^{LTE} &= c_p/(16\kappa\sigma T^3) \propto 1/\kappa &\mathrm{for}~\tau\ll 1,\label{E:cooling_time_thin}\\
t_{thick}^{LTE} &= 3c_p\rho^2\kappa\ell^2/(16\sigma T^3) \propto \kappa\ell^2 &\mathrm{for}~\tau\gg 1.\label{E:cooling_time_thick}
\end{align}
\end{subequations}
In the optically thick limit, photons diffuse through the region of interest, and so the radiative timescale scales with the number density of absorbers (opacity) and the square of the physical size, as expected for a random walk process. One can relate the optically thick cooling time to a radiative diffusivity: $\mathcal{K}_{rad}=\ell^2/t_{thick}^{LTE} =16\sigma T^3/(3c_p\rho^2\kappa)$.  As the size of the region of interest decreases, so does the cooling time, but only to the point at which photons freely stream out of the perturbation.  In this optically thin limit, the radiative timescale is independent of the physical size and inversely proportional to the number density of emitters. Figure~\ref{fig:tau_graph} shows the dependence of the radiative damping time as a function of optical depth.  Note that using the optically thin cooling time for large optical depth will overdamp optically thick features, while using the optically thick cooling time for small optical depth will overdamp optically thin features.  {\it With respect to numerical simulations, if one uses a diffusion term (a Laplacian operator), it is crucial that all relevant features in the flow are really optically thick, otherwise, optically thin small scale features may be overdamped}.  

Whether or not the dust and gas have the same temperature and are in local thermodynamic equilibrium, cool molecular hydrogen is very inefficient at absorbing or emitting electromagnetic radiation. Heat transfer in protoplanetary disks is primarily mediated via dust grains: gas molecules exchange energy with dust grains via collisions, and the dust grains absorb UV radiation from the protostar and emit IR (almost) blackbody radiation.  To first order, the dust opacity $\kappa_d$ is the total effective geometric cross-section of the grains per unit mass of gas and dust: $\rho\kappa_d\approx n_d\pi \langle a^2Q(a,T)\rangle$, where angle brackets indicate averaging over the grain size distribution, $Q(a,T)\approx \min[24ak_BT/hc,1]\approx\min[aT/(600~\mu\mathrm{m}\cdot\mathrm{K}),1]$ is a Planck-averaged emissivity for small grains to absorb infrared radiation, $h$ is Planck's constant, $c$ is the speed of light in vacuum, and $k_B$ is the Boltzmann constant \citep{natta2007,chiang1997}.\footnote{The wavelength dependent effective cross-section of a grain of radius $a$ can be expressed as $\pi a^2Q_\lambda(a)$ where $\lambda$ is wavelength of light and the emissivity is $Q_\lambda(a)\approx\min[2\pi a/\lambda, 1]$.  What we really need is the emissivity averaged over the Planck spectrum: $Q(a,T)\equiv(\pi/\sigma T^4)\int_0^\infty Q_{\lambda}(a)B_{\lambda}(T)d\lambda \approx (15/\pi^4)\{\int_0^bw^{-5}dw/[\exp(1/w)-1] + \int_b^\infty w^{-6}dw/[\exp(1/w)-1]\}$, where $B_\lambda(T)$ is the Planck intensity, $h$ is Planck's constant, $c$ is the speed of light in vacuum, $\sigma$ is the Stefan-Boltzmann constant, $k_B$ is the Boltzmann constant, $w\equiv\lambda k_BT/hc$, and $b\equiv 2\pi ak_BT/hc$.  For very large $b$, this yields $Q(a,T)\approx1$, whereas for very small $b$, $Q(a,T)\approx(360\zeta(5)/\pi^4)b= (720\zeta(5)/\pi^3)(ak_BT/hc)\approx 24ak_BT/hc$, where $\zeta(5)$ is the Riemann zeta function evaluated at 5. It should not be surprising that this is very close to $2\pi a/\lambda_{peak}$, where $\lambda_{peak}\approx hc/4k_BT$ is the wavelength of the peak of the spectral radiance per fractional bandwidth (\ie peak of $\lambda B_{\lambda}(T)$ vs.\ $\ln\lambda$).  See the classic papers \citet{draine1984} and \citet{draine2003}.}  Of special note is that for small grains ($aT<600~\mu\mathrm{m}\cdot\mathrm{K}$), the linear dependence of the emissivity on grain size combines with the area of grains to yield an opacity that depends only on the total mass of dust irrespective of the size of the grains.  For simplicity, we neglect other sources of opacity (\eg line cooling from vibro-rotational transitions of other molecules), but these can be included in a more sophisticated analysis \citep{semenov2003,cuzzi2014,malygin2017}.  The mean free path of an IR photon to small ($aT<600~\mu\mathrm{m}\cdot\mathrm{K}$) grains is thus:
\begin{align}
\ell_{IR} \equiv \frac{1}{\rho\kappa_d} \approx \frac{\rho_s}{\rho_d}\frac{(800~\mu\mathrm{m}\cdot\mathrm{K})}{T} = 1{,}600{,}00~\mathrm{km}\left(\frac{T}{100~\mathrm{K}}\right)^{-1}\left(\frac{\rho_d}{10^{-11}~\mathrm{kg}~\mathrm{m}^{-3}}\right)^{-1},
\end{align}
where $\rho_s\approx2000$~kg/m$^3$ is the solid density of the dust grains. 

\begin{figure}
\epsscale{1.0}
\plotone{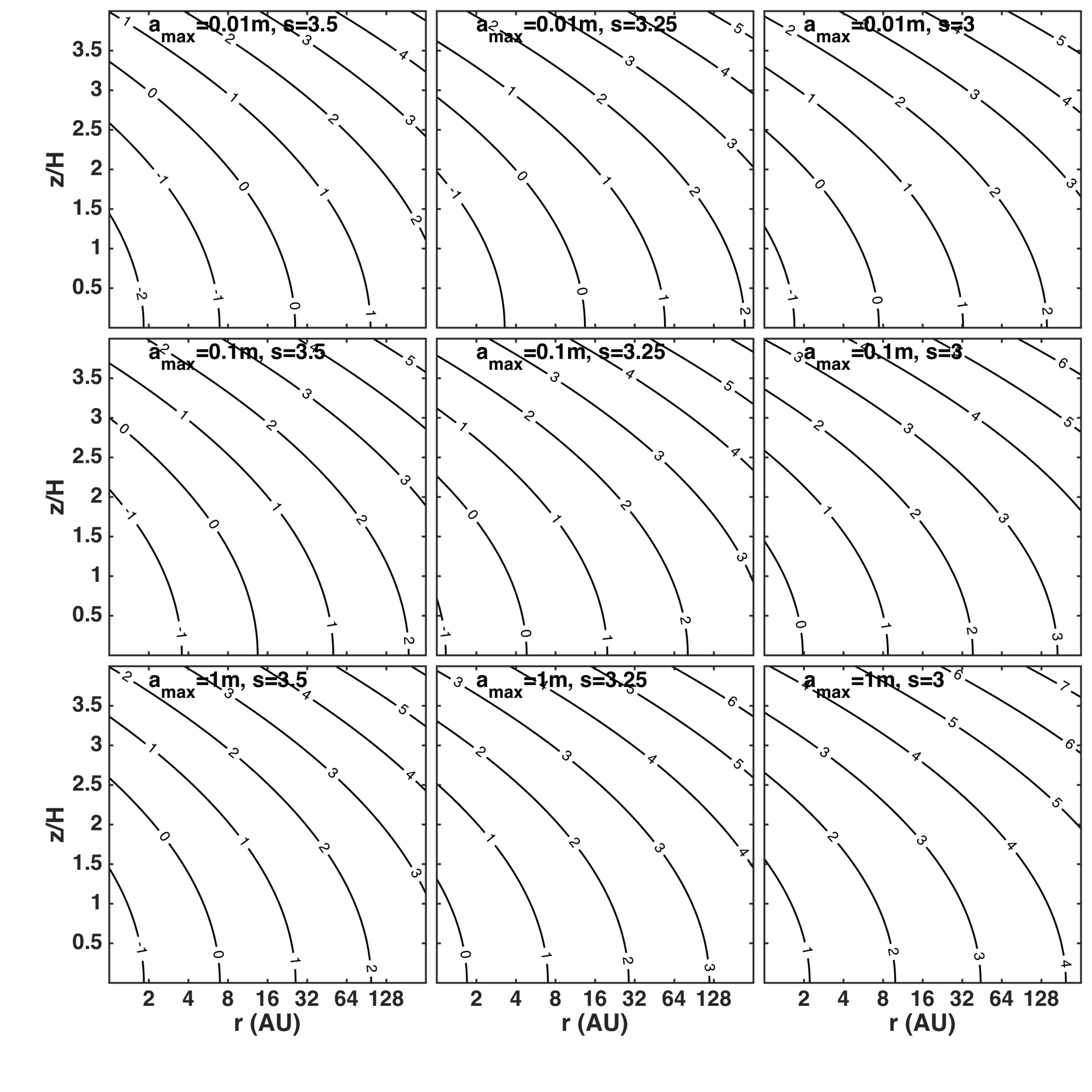}
\caption{{\bf Photon mean free path assuming no vertical settling or radial drift of dust grains.}  The photon mean free path $\ell_{IR}$ is everywhere much greater than the critical layer thickness $10^{-4}H$. Values on contours denote $\log_{10}(\ell_{IR}/H)$, so
  values greater than $-4$ indicate locations where the critical layer is optically thin.  Dust size distribution is assumed to be a power law where the number density of dust grains with radii between $a$ and $a+da$ is given by $n(a)da=Ka^{-s}da$ with a lower limit on grain size of $a_{min}=0.1~\mu$m, and an upper limit of $a_{max}$.  Across the rows, we vary the index $s$ from 3.50 on the left, 3.25 in the middle, and 3.00 on the right; down the columns, we vary the maximum grain size $a_{max}$ from 1 cm at the top, 10 cm  in the middle, and 1 m at the bottom. Dust-to-gas ratio is assumed to be uniform 0.01 throughout the entire protoplanetary disk.  If there is any settling of dust, then then $\ell_{IR}$ is larger than the values shown here, so $\ell_{IR}$ is greater than the thickness of the critical layer in all settling scenarios. Physical properties of protoplanetary disk are described in text. }\label{fig:ell_phot_no_settle}
\end{figure}

In Figure~\ref{fig:ell_phot_no_settle}, we graph the photon mean free path as a function of location within a protoplanetary disk, assuming a uniform dust-to-gas ratio of 0.01 (no vertical settling or radial migration of grains).  Values on contours denote $\log_{10}(\ell_{IR}/H)$.  The dust size distribution is assumed to be a power law with index $s$, with a lower limit on grain size of $a_{min}=0.1~\mu$m, and an upper limit of $a_{max}$.  Across the rows, we vary the index $s$ from 3.50 on the left, 3.25 in the middle, and 3.00 on the right; down the columns, we vary the maximum grain size $a_{max}$ from 1 cm at the top, 10 cm  in the middle, and 1 m at the bottom.  Of special note is that even without any dust settling or migration, the photon mean free path can vary by one or two orders of magnitude depending on the extent of grain growth (determined by $a_{max}$) or the slope of the grain size distribution. Also note, though, that irrespective of $a_{max}$ or $s$, the photon mean free path is many orders of magnitude larger than the width of the baroclinic critical layers $\delta_{CL}\sim10^{-4}H$. {\it Thus, baroclinic critical layers are optically thin.}  Including dust settling into the midplane will deplete grains in the regions above/below the midplane, making the photon mean free path even longer there. {\it Therefore, due to the facts that (1) it is the damping of the baroclinic critical layers that can potentially inhibit the triggering of ZVI, and (2)  $\ell_{IR}$ is always much greater than the thickness of the baroclinic critical layers, throughout that the remainder of this paper we shall consider radiative cooling in the optically thin limit.}

\subsection{Thermal Relaxation with Finite Gas-Dust Coupling Time}\label{sec:cooling_theory}

An implicit assumption for the validity of eq.~\eqref{E:cooling_time1} is that the gas molecules and dust grains are thermally coupled on a sufficiently short timescale so that they are in local thermodynamic equilibrium (\ie the kinetic temperature of the gas molecules is the same as the thermal temperature of the grains). This may not be the case in diffuse atmospheres where the energy exchange rate between gas and dust may be too slow \citep{fiocco1975,goldenson2008}.  In such cases, the gas and dust temperatures will not necessarily be the same, and we must explicitly follow the exchange of energy between gas and dust via collisions, and the absorption and emission of electromagnetic radiation from the the dust.    Consider a parcel of gas at temperature $T_g$ containing dust grains at temperature $T_d$, and let $\Lambda_{col}$ be the volumetric energy transfer rate from gas to dust via collisions, and let $\Lambda_{rad}^{net}$ be the net volumetric power radiated (emitted IR minus absorbed UV) by dust grains.  The evolution of the gas and dust temperatures within the parcel is governed by:
\begin{subequations}
\begin{align}
\rho_gc_p~dT_g/dt &= -\Lambda_{col},\\
\rho_dc_d~dT_d/dt &= +\Lambda_{col}-\Lambda_{rad}^{net},
\end{align}\label{E:dTdt1}
\end{subequations}
where $c_p$ and $c_d$ are the specific heats for gas at constant pressure and for dust grains, respectively.  The dust-gas collisional energy transfer rate is proportional to the difference between the gas and dust temperatures \citep{hollenbach1979,burke1983,glassgold2004}:
\begin{equation}
\begin{split}
\Lambda_{col} &\approx \int_{a_{min}}^{a_{max}} \pi a^2n(a)n_g\bar{v}_g(2\mathcal{A})k_B(T_g-T_d)~da= \left(\frac{3}{4\rho_s}\right)\left(\frac{\langle a^2\rangle}{\langle a^3\rangle}\right)\left(\frac{\rho_d}{\rho_g}\right)\left(\frac{\rho_g^2\bar{v}_g}{\bar{m}}\right)(2\mathcal{A})k_B(T_g-T_d)\\
{} &\approx 1.29\times 10^{-13}~\mathrm{W~m^{-3}~K^{-1}}\left(\frac{a_S}{0.1~\mathrm{mm}}\right)^{-1}\left(\frac{\rho_d/\rho_g}{0.01}\right)\left(\frac{\rho_{g}}{10^{-9}~\mathrm{kg~m^{-3}}}\right)^2\left(\frac{T_g}{100~\mathrm{K}}\right)^{1/2}(T_g-T_d), \label{E:lambda_col}
\end{split}
\end{equation}
where $n_g$ is the number density of gas molecules, $\bar{v}_g$ is the mean thermal speed of gas molecules, $k_B$ is the Boltzmann constant, and $\mathcal{A}$ is the accommodation coefficient, which measures the efficiency of heat transfer between a gas and a surface.  We have assumed the accommodation coefficient for molecular hydrogen impinging on amorphous carbon is $\mathcal{A}\sim 0.5$.  We can define response times for how the gas and dust temperatures respond to exchange of energy via collisions; these are of the form of a thermal inertia (\ie heat capacity) divided by rate of energy transport:
\begin{subequations}
\begin{align}
\begin{split}
t_{g}^{col} &\equiv \rho_gc_p/(\Lambda_{col}/|T_g-T_d|)\\
{}&\approx 7.0\times10^7~\mathrm{s}\left(\frac{a_S}{0.1~\mathrm{mm}}\right)\left(\frac{\rho_d/\rho_g}{0.01}\right)^{-1}\left(\frac{\rho_{g}}{10^{-9}~\mathrm{kg~m^{-3}}}\right)^{-1}\left(\frac{T_g}{100~\mathrm{K}}\right)^{-1/2},
\end{split}\label{E:tgcoll}\\
\begin{split}
t_{d}^{col} &\equiv \rho_dc_d/(\Lambda_{col}/|T_g-T_d|) = (\rho_d/\rho_g)(c_d/c_p)t_{g}^{col}\\
{}&\approx6.2\times10^4~\mathrm{s}\left(\frac{a_S}{0.1~\mathrm{mm}}\right)\left(\frac{\rho_{g}}{10^{-9}~\mathrm{kg~m^{-3}}}\right)^{-1}\left(\frac{T_g}{100~\mathrm{K}}\right)^{-1/2},
\end{split}\label{E:tdcoll}
\end{align}
\end{subequations}
where we assume the specific heats for gas (at constant pressure) and dust are $c_p=9000$~J/K/kg and $c_d=800$~J/K/kg.

Apart from energy exchange with gas molecules via collisions, grains absorb UV radiation from the protostar and emit IR thermal radiation, which is not quite blackbody because grains are inefficient at emitting electromagnetic radiation with wavelengths longer than the sizes of the grains. When radiative absorbtion and emission are balanced, the grains achieve an equilibrium temperature $T_{eq}$.  The net volumetric radiated power from the grains (IR emission minus UV absorption) is:
\begin{align}
\begin{split}
\Lambda_{rad}^{net} &\approx \int_{a_{min}}^{a_{max}}  4\pi a^2n(a)Q(a,T)\sigma(T_{d}^4-T_{eq}^4)~da\approx  12\left(\frac{\langle a^2Q\rangle}{\langle a^3\rangle}\right)\left(\frac{\rho_d}{\rho_s}\right)\sigma T_g^3(T_{d}-T_{eq})\\
{} &\approx 3.40\times 10^{-11}~\mathrm{W~m^{-3}~K^{-1}}\left(\frac{a_{SQ}}{0.1~\mathrm{mm}}\right)^{-1}\left(\frac{\rho_{d}}{10^{-11}~\mathrm{kg~m^{-3}}}\right)\left(\frac{T_g}{100~\mathrm{K}}\right)^{3}(T_d-T_{eq}),
\end{split}\label{E:lambda_rad}
\end{align}
{\color{black}where we have defined an emissivity-weighted Sauter mean radius:
\begin{equation}
a_{SQ}(T)\equiv\frac{\langle a^3\rangle}{\langle a^2Q(a,T)\rangle} \approx (s-3)a_{max}^{4-s}a_{Q=1}^{s-3} = (s-3)a_{max}^{4-s}\left(\frac{600~\mu\mathrm{m}\cdot\mathrm{K}}{T}\right)^{s-3},
\end{equation}
and where we define the smallest radius for a grain to have unity emissivity for a given temperature: $a_{Q=1} \equiv (600~\mu\mathrm{m}\cdot\mathrm{K})/T$.  The details of this calculation are left to Appendix B.  Unlike the previously introduced Sauter mean radius which is independent of temperature, the emissivity-weighted Sauter mean radius does indeed depend on temperature.   Note that the emissivity-weighted Sauter radius is insensitive to the size of the smallest grains (those with sizes much less than $a_{Q=1}$) because they are poor radiators and their effective surface area is much smaller than their geometric surface area.   Also note that for cool temperatures ($T<100$~K) in the outer disk, the emissivity-weighted Sauter mean radius can be 5-40 times larger than the Sauter mean radius with unity emissivity.}  Because it will be useful in the following analysis, we want to exploit the fact that the net radiated power is approximately linear in the temperature difference; so, we approximated $T_d^4-T_{eq}^4\approx 4T_g^3(T_d-T_{eq})$ and used the gas temperature instead of the dust temperature in the emissivity.  We also neglect the fact that different size grains have different equilibrium temperatures. These approximations are valid as long as $T_{eq}$, $T_d$, and $T_g$ do not deviate too much from each other.   Of course, one could relax these restrictions in a more sophisticated analysis.  As before, we can define a response time for how the dust temperature responds to emission and absorption of radiation:
\begin{align}
\begin{split}
t_{d}^{rad} &\equiv \rho_dc_d/(\Lambda_{rad}^{net}/|T_d-T_{eq}|)\\
{}&\approx 240~\mathrm{s}~\left(\frac{a_{SQ}}{0.1~\mathrm{mm}}\right)\left(\frac{T_g}{100~\mathrm{K}}\right)^{-3},
\end{split}\label{E:tdrad}
\end{align}

What is the relationship among $t_{d}^{rad}$, $t_g^{col}$, $t_d^{col}$ and the optically thin cooling time $t_{thin}^{LTE}$ when the dust and gas are in LTE?  In equation~\eqref{E:cooling_time_thin}, one should use the total heat capacity for the gas and dust mixture ($\rho c_{p,mix} = \rho_gc_p + \rho_dc_d$); for the opacity, simply use the dust opacity $\rho\kappa_d\approx n_d\pi \langle a^2Q(a,T)\rangle$.  One should see:
\begin{align}
\begin{split}
t_{thin}^{LTE} &= \frac{\rho_gc_p+\rho_dc_d}{16\pi n_d\langle a^2Q(a,T)\rangle\sigma T^3} = \left(1+\frac{\rho_g c_p}{\rho_d c_d}\right)t_{d}^{rad} \approx \left(\frac{t_g^{col}}{t_d^{col}}\right)t_d^{rad}\\
{}&\approx 2.7\times10^5~\mathrm{s}\left(\frac{a_{SQ}}{0.1~\mathrm{mm}}\right)\left(\frac{\rho_d/\rho_g}{0.01}\right)^{-1}\left(\frac{T_g}{100~\mathrm{K}}\right)^{-3}.
\end{split}\label{E:t_thin}
\end{align}
There is another way to get this result.  Consider the case that the gas and dust are perfectly coupled and always have the same temperature. Then one can add the two equations in \eqref{E:dTdt1} and form a timescale by dividing a heat capacity by an energy exchange rate.  Doing so, we obtain the exact same result as above.  This is consistent with the idea that the optically thin cooling time is valid in the limit that the gas-dust coupling happens nearly instantaneously.

\begin{figure}
\epsscale{1.0}
\plotone{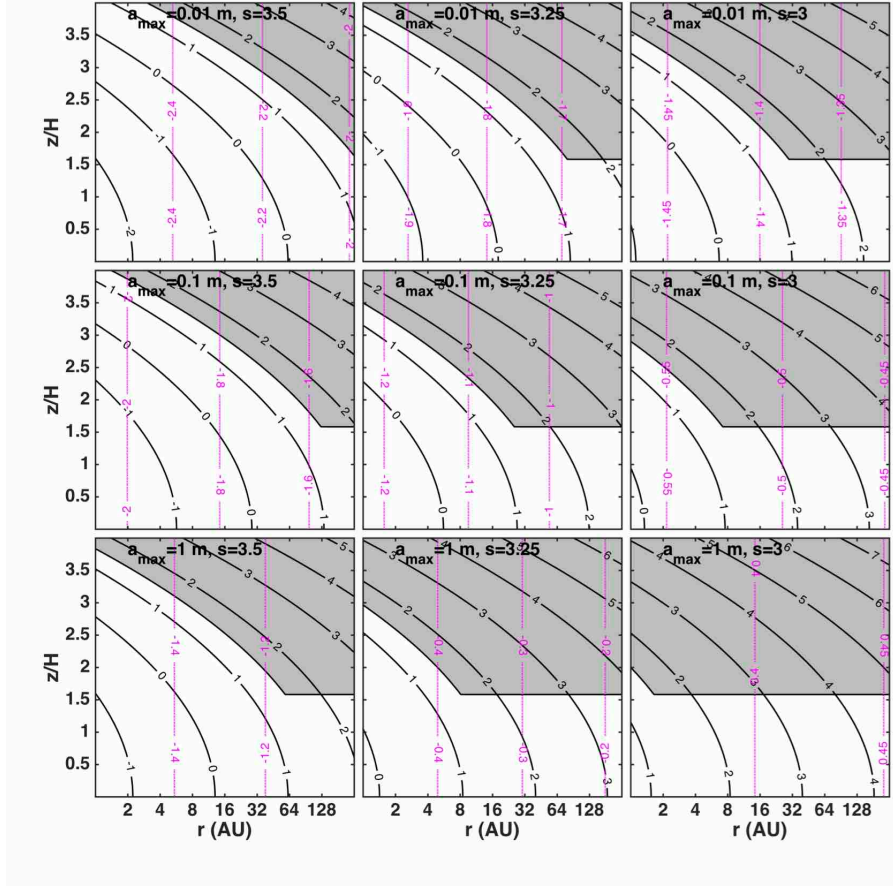}
\caption{{\bf Relaxation time $t_{thin}^{NLTE}$ assuming no vertical settling or radial drift of dust grains.}  Values on solid black contours denote $\log_{10}(\Omega t_{thin}^{NLTE})$.  Values on dotted magenta contours are for $\log_{10}(\Omega t_{thin}^{LTE})$.  {\color{black}The difference between these timescales is an indication of the magnitude of how far out of local thermodynamic equilibrium the gas and dust are with respect to each other.}  Note that throughout much of the disk, $t_{thin}^{NLTE}$ is many orders of magnitude longer than  $t_{thin}^{LTE}$.  Dust size distribution is assumed to be a power law where the number density of dust grains with radii between $a$ and $a+da$ is given by $n(a)da=Ka^{-s}da$ with a lower limit on grain size of $a_{min}=0.1~\mu$m, and an upper limit of $a_{max}$.  Across the rows, we vary the index $s$ from 3.50 on the left, 3.25 in the middle, and 3.00 on the right; down the columns, we vary the maximum grain size $a_{max}$ from 1 cm at the top, 10 cm  in the middle, and 1 m at the bottom. Dust-to-gas ratio is assumed to be a uniform 0.01 throughout the entire protoplanetary disk.  The regions shaded gray indicate where ZVI can be triggered, that is, where the gas is sufficiently stratified and the relaxation time is sufficiently long $\Omega t_{thin}^{NLTE}>10^{1.5}$.   Note that the boundaries of this region are only approximate; here we chose the stratification boundary for an isothermal background, but it could be lower depending on the exact stratification profile.}\label{fig:cooling_nosettle}
\end{figure}

\begin{figure}
\epsscale{1.0}
\plotone{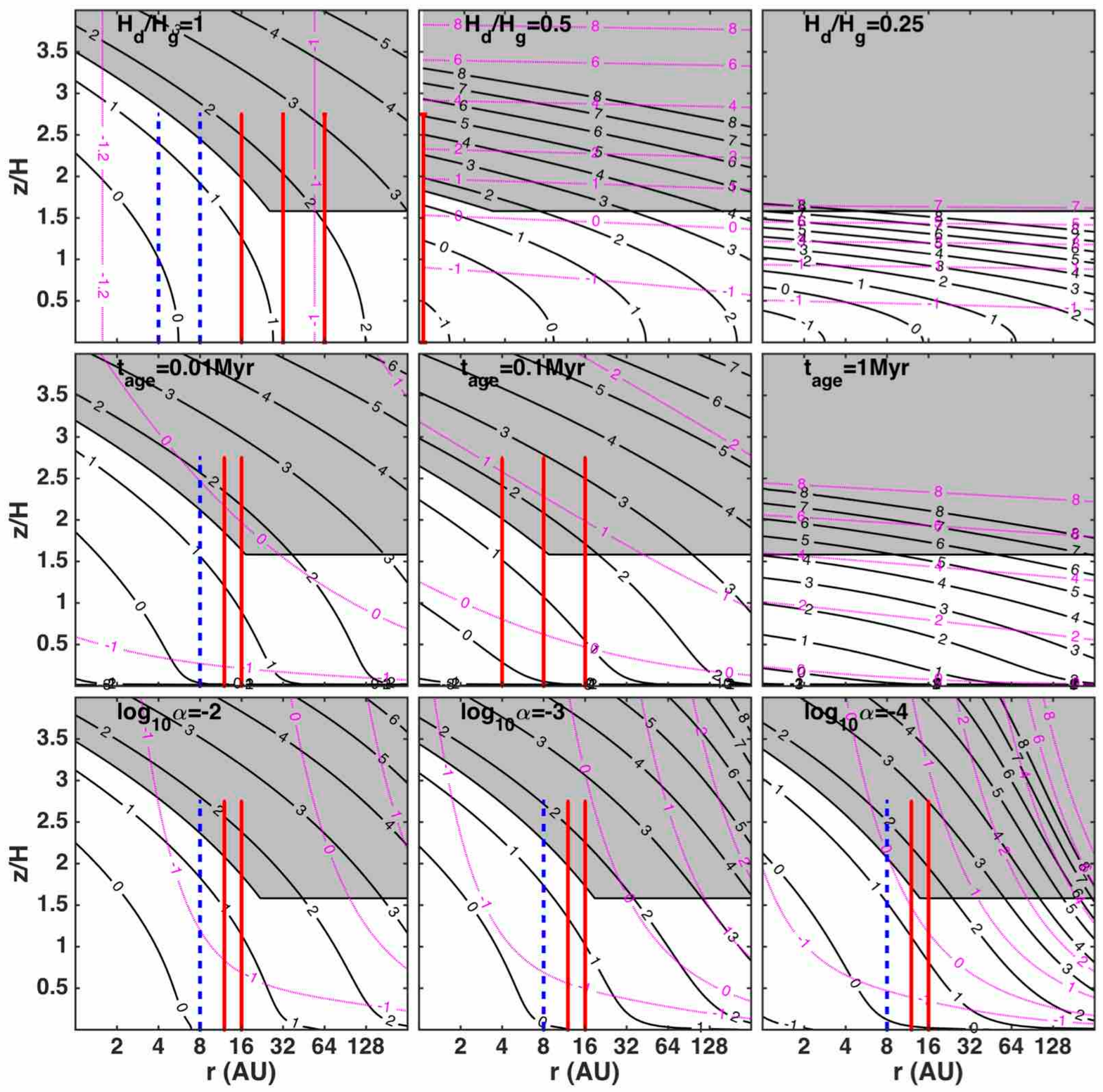}
\caption{{\bf Relaxation time $t_{thin}^{NLTE}$ for 3 different settling scenarios.}  Values on solid black contours denote $\log_{10}(\Omega t_{thin}^{NLTE})$.  Values on dotted magenta contours are for $\log_{10}(\Omega t_{thin}^{LTE})$.  {\color{black}The difference between these timescales is an indication of the magnitude of how far out of local thermodynamic equilibrium the gas and dust are with respect to each other.} We fix $s=3.25$ and $a_{max}=10$~cm for all cases.  Areas shaded in gray have sufficient stratification and long enough thermal relaxation times so as to be susceptible to ZVI.  {\bf Top row:} All particles settle to same dust scale height, irrespective of size.  Across the columns, $H_d/H_g$ = 1.0, 0.50, and 0.25. {\bf Middle row:} Particles are allowed to settle in a turbulent-free disk according to $H_d(a) = H_g\exp[-t_{age}/t_{sedi}(a)]$.  Particles of different sizes sediment at different rates: $t_{sedi}= (\Omega^2t_{stop})^{-1}$, where the stopping time is $t_{stop} = (\rho_s/\rho_g)(a/\bar{v}_g)$.  Across the columns, we vary the disk age: $t_{age}$ = 0.01~Myr, 0.1~Myr, 1~Myr.  {\bf Bottom row:} Particles reach a quasi-steady-state scale height in which downward settling is balanced by upward turbulent diffusion: $H_d(a)=H_g/\sqrt{1+\Omega t_{stop}/\alpha}$.  Across the columns, we vary the nondimensional turbulent viscosity: $\alpha=10^{-2}, 10^{-3}, 10^{-4}$.  {\bf Simulations with Radiative Damping:} We have simulated ZVI with the initial Run\_Brunt\_Step profile with vertical domains from $z=-3H$ to $z=+3H$ for 2000 orbits with nonuniform vertical cooling.  The simulations were local at one radius; that is, the radial extent of the computational domain was $6H\ll r$.  Because of additional damping near the vertical boundaries (for stability), the effective domain is $|z|\lesssim2.75H$.  Thick, blue vertical dashed lines indicate simulations that did not develop ZVI, whereas thick, red vertical solid lines indicate that ZVI was triggered and developed into full zombie turbulence.  The vertical extent of these lines indicates the effective size of the computational domain.  Note that when no part of the computational domain or only a small part of the computational domain extends into the gray region, ZVI is not triggered, while simulations in which the stratified regions of the computational domain have significant overlap with the gray regions do develop ZVI and turbulence.}\label{fig:cooling_settle}
\end{figure}

We will now show that the optically thin cooling time $t_{thin}^{NLTE}$ when the gas and dust are not in local thermodynamic equilibrium with each other can be significantly longer than $t_{thin}^{LTE}$.  Let's now rewrite the temperature evolution equations \eqref{E:dTdt1} in terms of deviations from the equilibrium temperature:
\begin{subequations}
\begin{align}
dT'_g/dt &= -(T'_g-T'_d)/t_{g}^{col},\\
dT'_d/dt &= +(T'_g-T'_d)/t_{d}^{col}-T'_d/t_{d}^{rad},
\end{align}
\end{subequations}
where we have defined the temperature deviations $T'_g\equiv T_g-T_{eq}$ and $T'_d\equiv T_d-T_{eq}$. This is a system of two coupled linear equations; we can look for eigenfunction solutions of the form $T'_g=\hat{T}_g\exp(-t/t_{thin}^{NLTE})$ and $T'_d=\hat{T}_d\exp(-t/t_{thin}^{NLTE})$, which yields a quadratic equation for the thermal relaxation time $t_{thin}^{NLTE}$.  One of the roots is:
\begin{align}
t_{thin}^{NLTE}=2t_{||}\left[1-\sqrt{1-4t_{||}^2/t_g^{col}t_d^{rad}}\right]^{-1}, \label{E:t_relax}
\end{align}
where we have defined $1/t_{||}\equiv1/t_g^{col}+1/t_d^{col}+1/t_d^{rad}$.  There are actually two eigenvalues, a slow (as above) and fast (switch the sign in front of the radical) timescale.  Associated with these two eigenvalues are two eigenvectors.  An arbitrary thermal perturbation will be a linear combination of the two eigenvectors; the component of the perturbation that corresponds to the fast eigenvalue will decay on a short timescale, while the component that corresponds to the slow eigenvalue will decay on a long timescale.  What ultimately matters for the lifetime of the thermal perturbation is the slow timescale.  Note that when the temperatures of the gas and dust are well-coupled and are in LTE, the timescale $t_{thin}^{NLTE}$ given in eq.~(\ref{E:t_relax}) reduces to  $t_{thin}^{LTE}$, so that our formula for $t_{thin}^{NLTE}$ can be used under all conditions. Furthermore, locations in the PPD where $t_{thin}^{LTE}$ and $t_{thin}^{NLTE}$ differ significantly (see Figs.~\ref{fig:cooling_nosettle} and~\ref{fig:cooling_settle}) indicate the locations where the dust and gas temperatures are not in local thermodynamic equilibrium.
We admit that an intuitive interpretation of form of the relaxation time in \eqref{E:t_relax} is not that obvious.  However, note that throughout the entire PPD, $t_{g}^{col}$ is always orders of magnitude longer than either $t_{d}^{col}$ or $t_{d}^{rad}$. One can expand \eqref{E:t_relax} in a Laurent series for large $t_{g}^{col}$ and obtain the approximation:
\begin{subequations}
\begin{align}
t_{thin}^{NLTE} &\approx t_g^{col}(1+t_d^{rad}/t_d^{col})\\
&\approx t_g^{col} + t_{thin}^{LTE}.
\end{align}
\end{subequations}
We emphasize that this is valid as long as $t_{g}^{col}\gg t_{d}^{col}, t_{d}^{rad}$, but is true whether $t_{g}^{col}$ is smaller or larger than $t_{thin}^{LTE}$.  As a rough analogy, one can say the cooling is a sequential two-stage process: energy must be transferred from the gas to the dust (on timescale $t_g^{col}$), and then radiated away via dust (on timescale $t_{thin}^{LTE}$), and the
actual cooling time $t_{thin}^{NLTE}$ is the sum of these two individual times.

We now present a series of graphs of the relaxation time $t_{thin}^{NLTE}$ for a particular protoplanetary disk model, for a variety of grain power-law distributions, and for the case of well-mixed gas and dust (no vertical settling or radial drift) and for various simple settling models.  These graphs are not meant to be  exhaustive, but rather to represent the wide range of possibilities for cooling times in disks.  First, in Figure~\ref{fig:cooling_nosettle}, we graph the radiative relaxation times  $t_{thin}^{NLTE}$ (normalized by the Keplerian frequency) for models with no radial drift or vertical settling of grains, that is, the dust and gas are well mixed with a uniform ratio throughout the disk. We fix the global dust-to-gas ratio $\Sigma_d/\Sigma_g=0.01$ and the minimum grain size $a_{min}=0.1~\mu$m.  Across the rows, we vary the power law index $s$ from 3.50 on the left, 3.25 in the middle, and 3.00 on the right; down the columns, we vary the maximum grain size $a_{max}$ from 1 cm at the top, 10 cm  in the middle, and 1 m at the bottom.  The dust grain size distribution is logarithmically binned, and collisional and radiative energy exchange rates $\Lambda_{col}$ and $\Lambda_{rad}^{net}$ are computed for each bin from equations \eqref{E:lambda_col} and \eqref{E:lambda_rad}.  The rates themselves are summed over all bins and the relevant timescales are computed.  For comparison, dotted magenta contours show the
optically thin cooling time $t_{thin}^{LTE}$ in \eqref{E:t_thin}.  In the case of uniform dust-to-gas ratio, $t_{thin}^{LTE}$ is independent of height off the midplane and only a function of distance from the protostar.  Note that $t_{thin}^{LTE}$ is everywhere smaller than $t_{g}^{col}$ and so the relaxation time is $t_{thin}^{NLTE} \approx t_{g}^{col}$.  We have shaded gray the region of the disk that is susceptible to ZVI, that is, the gas is sufficiently stratified ($N/\Omega\gtrsim1$) {\it and} the relaxation time is sufficiently long that $\Omega t_{thin}^{NLTE} >10^{1.5}$.  Even without any dust settling, ZVI can be triggered off the midplane throughout much of the outer disk.

Next, in Figure~\ref{fig:cooling_settle}, we graph the radiative relaxation times (normalized by the Keplerian frequency) for three different vertical dust settling scenarios.  We fix $s=3.25$ and $a_{max}=10$~cm. In the top row of plots, we assume that all grains settle to the same dust scale irrespective of grain size.  Across the columns, we choose $H_d/H_g$ = 1.0, 0.50, and 0.25.  In the middle row of plots, we assume a zero turbulence disk and allow dust grains to settle into the midplane according to $H_d(a) = H_g\exp[-t_{age}/t_{sedi}(a)]$.  Grains of different sizes settle with different sedimentation rates: $t_{sedi}= (\Omega^2t_{stop})^{-1}$, where the stopping time (the time for a particle to come to rest with respect to the gas due to aerodynamic drag) in the Epstein regime (particle size less than gas molecular mean free path) is $t_{stop} = (\rho_s/\rho_g)(a/\bar{v}_{g})$.  For a 1~mm particle at 1~au and $z\approx2H$, the stopping and sedimentation times (normalized by the Keplerian frequency) are $\Omega t_{stop}\approx0.0014$ and $\Omega t_{sedi}\approx700$.  In the middle row of graphs, we vary the disk age across the columns: $t_{age}$ = 0.01~Myr, 0.1~Myr, and 1~Myr.  In the last row of plots, we assume a turbulent disk in which the dust scale heights have reached a steady state between downward settling and upward turbulent diffusion: $H_d(a)=H_g/\sqrt{1+\Omega t_{stop}/\alpha}$, where $\alpha$ is the nondimensional turbulence viscosity $\alpha=\nu_{turb}/(c_sH)$ \citep{dubrulle1995,garaud2007}.  In the third row, we vary the turbulent viscosity across the columns: $\alpha=10^{-2}, 10^{-3}, 10^{-4}$. Note that we assume that there is enough time to settle to these steady-states, but that may not be the case for sub-micron grains.  The dotted magenta contours and the gray shaded region have the same meanings as in Figure~\ref{fig:cooling_nosettle}.  With even marginal dust settling, now large swathes of PPDs are susceptible to ZVI.

\subsection{Simulations of Zombie Vortex Instability with Radiative Damping}\label{sec:cooling_simulations}

If the photon mean free path is longer than the largest scale of interest within a flow, then all scales are optically thin and radiative transfer can be modeled with a size-scale independent Newton cooling (\ie radiated power proportional to temperature difference with respect to some equilibrium value).  On the other hand, if the photon mean free path is shorter than the smallest scale of interest within a flow, then all scales are optically thick and radiative transfer can be treated as simple diffusion.  A significant challenge is the case where, at the same location in space and time in the flow, small fluctuations are optically thin and large fluctuations are optically thick; using Newton cooling will overdamp large fluctuations, while using a diffusion operator will overdamp small fluctuations.  Monte Carlo techniques for radiative transfer are ideal for cases that do not fit safely within the optically thin or optically thick regimes, but these methods involve as substantial investment in additional coding and computational resources.   Our goal here is more modest: we want to determine how much radiative damping baroclinic critical layers can withstand and still support the development of ZVI.  In Appendix A, we report on simulations with uniform stratification that confirm that ZVI requires Newton cooling times longer than a few orbital periods.

In real protoplanetary disks, gravity and stratification are not uniform.  Not only that, the radiative relaxation time can vary in position within the disk and is especially sensitive to how much the dust has settled.  The baroclinic critical layers are most sensitive to radiative cooling;  their radial extent is much less than a gas scale height and are expected to be optically thin, thus we will focus only on Newton cooling.\footnote{We apply Newton cooling to the potential temperature: $-\mathcal{L}_{rad}\tilde{\theta} = -\tilde{\theta}/t_{thin}^{NLTE}$. Implicit in this approach is that the fractional change in potential temperature equals the fractional change in thermal temperature due to radiative cooling.  This can be shown to be exactly true in the limit of the anelastic approximation with uniform temperature backgrounds \citep{barranco06}, and is still a good approximation for the stratification profiles used here.}  This will likely overdamp the flow on the largest scales, but this should not be a concern for the onset of ZVI.  In this subsection, all simulations have a  domain size of $(6H)^3$ resolved with $192^3$ spectral modes.  The size of the time steps was adjusted every $\sim$0.1 orbits to keep the target CFL number $\lesssim0.125$.  We fix the background stratification to be the Run\_Brunt\_Step profile.  We do not assume the radiative relaxation time is spatially uniform, rather, we explore relaxation times that are functions of height and that correspond to different simple models of dust settling, as in Figure~\ref{fig:cooling_settle}.  We chose 21 cooling profiles to sample the parameter space, not to necessarily be exhaustive, but to show that the onset of ZVI in real protoplanetary disks is consistent with our understanding of where there is sufficient stratification and long enough cooling times.  Because of additional damping near the vertical boundaries (for stability), the effective domain is $|z|\lesssim2.75H$.  On Figure~\ref{fig:cooling_settle}, we plot thick, blue vertical dashed lines indicate simulations that did not develop ZVI after 2000 orbits, whereas thick, red vertical solid lines indicate that ZVI was triggered and developed into full zombie turbulence.   The vertical extent of these lines indicates the effective size of the computational domain.  Note that when no part of the computational domain or only a small part of the computational domain extends into the gray region, ZVI is not triggered, while simulations in which the stratified regions of the computational domain have significant overlap with the gray regions do develop ZVI and turbulence. 


\section{DISCUSSION \& OUTSTANDING ISSUES}\label{sec:discussion}

\paragraph{Summary of Results:} In our previous work, we investigated, via numerical simulation, ZVI with uniform stratification (uniform gravity, uniform background temperature, and uniform Brunt-V\"{a}is\"{a}l\"{a} frequency) and in the limit that the timescale associated with radiative damping was infinite \citep{MPJH13,MPJBHL15,MPJB16}.  In this work, we have relaxed both of these assumptions.  Here, we investigated ZVI with realistic vertical gravity and vertical stratification, with and without radiative damping. Our key results are:

(1) ZVI can occur with a variety of nonuniform stratification profiles.  In this work, we investigated the onset of ZVI and its nonlinear evolution for three profiles: (i) uniform temperature with a Brunt-V\"{a}is\"{a}l\"{a} frequency that increased linearly away from the midplane, (ii) a step profile in temperature which yielded a Brunt-V\"{a}is\"{a}l\"{a} frequency that had a local extremum near the step in temperature, and (iii) a step profile in the Brunt-V\"{a}is\"{a}l\"{a} frequency itself.  All showed robust development of ZVI.  {\color{black}The susceptibility to ZVI is local; that is, there needs to be a locality for which the stratification is sufficiently strong, the thermal relaxation time is sufficiently long, and the initial perturbations exceed some critical threshold.  Once ZVI is ignited at one location, it rapidly spreads and modifies the neighboring regions to push them towards susceptibility.}

(2) At late times, zombie turbulence resulted in vertical mixing. In Run\_Isothermal, the stratification was homogenized in the ZVI susceptible layer, resulting in a nearly uniform Brunt-V\"{a}is\"{a}l\"{a} frequency.  Similar behavior is seen in atmospheric and oceanic flows in which the breaking of internal gravity waves creates step-like or staircase patterns in stratification \citep{orlanski1969,phillips1972,pelegri1998}.  This also further justifies the initial profile for Run\_Brunt\_Step. 

(3) While the region in the immediate vicinity of the midplane lacks the requisite stratification for the excitation of baroclinic critical layers, we observe that zombie turbulence from the ZVI susceptible regions can penetrate into the midplane, albeit with a smaller magnitude.  The stratification profile Run\_Brunt\_Step showed the greatest penetration of zombie turbulence into the disk midplane.

(4) At late times, zombie turbulence resulted in the creation of azimuthal quasi-steady state zonal flows.  The zonal flows consisted of 5-6  pairs of dipolar vortex layers within a radial extent of $8H$.  The radial thickness of the cyclonic layers is approximately one third the thickness of the anticyclonic layers, but this is compensated by the fact that the cyclonic layers are roughly three times more intense as the anticyclonic layers.  We suspect that the width of the zones is imprinted from the initial separation of the baroclinic critical layers.  {\color{black} However, the stratification profiles are not uniform and the initial critical layer separations are functions of height.  Yet, the simulations seem to show that the widths of the zones are uniform in height.  How can this be consistent with the idea that the widths of the zones have memory of the initial critical layer thickness?  One possibility is that ZVI is first triggered at one specific height (because it has the combination of the most favorable stratification, cooling time, and strength of initial perturbations), and as ZVI rapidly develops at the initial location, it homogenizes the stratification around it, making the critical layer separations more uniform.}

(5) Fully-developed zombie turbulence shows intermittency where the flow cycles through near-laminar phases of zonal flow punctuated by chaotic bursts of new zombie vortices.  In some simulations, the bursting is quasi-periodic in time (with periods between 100-150 orbits), whereas in other cases, it appeared more stochastically. This phenomenon was not observed in any of our previous simulations of ZVI.

(6) The photon mean free path is always significantly longer than the width of baroclinic critical layers $\delta_{CL}\sim10^{-4}H$.  Baroclinic critical layers are optically thin structures.  In our spectral simulations, the baroclinic critical layers are resolved with only 3-5 collocation points.

{\color{black} (7) The hydrodynamics which leads to the creation of the baroclinic critical layers generates thermal anomalies in the gas (via pressure-volume work, buoyancy).  The temperature of the dust does not instantaneously match that of the gas, and the gas and dust become out of local thermodynamic equilibrium.  The gas responds to differences in temperature with the dust on the timescale $t_g^{col}$, and the dust responds to differences in temperature with the gas on the timescale $t_d^{col}$ (see equations \eqref{E:tgcoll} and \eqref{E:tdcoll}).  However, what ultimately matters for the lifetime of the critical layer is the time it takes for the gas to radiatively relax to the background equilibrium temperature.}

(8) The relevant radiative damping timescale in optically thin regions where the gas and dust are not in local thermodynamic equilibrium with each other is $t_{relax}\approx t_{g}^{col}+t_{thin}$.  That is, cooling is a sequential two-stage process: energy must be transferred from the gas to the dust (on timescale $t_g^{col}$), and the radiated away via dust (on timescale $t_{thin}$), and the total time is the sum of these individual times.  Throughout much of the off-midplane regions in PPDs, $t_{relax}$ can be orders of magnitude longer than $t_{thin}$ because the gas and dust are not in LTE (see Figs.~\ref{fig:cooling_nosettle} and \ref{fig:cooling_settle}).

(9) ZVI can still thrive in the presence of radiative damping.   As long as the thermal relaxation timescale is longer than a few orbital periods in the strongly stratified regions, then ZVI can locally be triggered. Without any dust settling, ZVI can be triggered in the outer regions of protoplanetary disks because the relaxation timescale is set by the rate at which the gas and dust exchange energy via collisions. With minimal dust settling, ZVI can operate throughout much of the planet-forming regions of protoplanetary disks. 

{\color{black} (10) If the thermal relaxation time is sufficiently long, ZVI can be triggered and the resultant late-time evolution of turbulence is virtually indistinguishable from the simulations without thermal relaxation.  That is, thermal relaxation can affect the initial development of the baroclinic critical layers, but once the critical layers give way to ZVI, the thermal relaxation has little effect on the subsequent evolution into zombie vortices and turbulence.}
{\color{black}
\paragraph{Where Within Protoplanetary Disks Is ZVI Potentially Relevant?} \citet{lesurlatter2016} showed that cooling times shorter than a few orbital periods can quench the onset of ZVI.  They also investigated the effect of radiative diffusion in optically thick regions, and showed that ZVI requires a large P\'{e}clet number $Pe\!\sim\!10^4$.  Our numerical results in Appendix A are broadly consistent.  However, we strongly disagree with their conclusion with respect to where within protoplanetary disks that ZVI may be operable.  They assumed opacities from \citet{semenov2003} that included virtually no grain growth (maximum grain sizes of only a few microns) nor any vertical dust settling, and thus erroneously concluded that the optically thin cooling time is so short throughout the bulk of the disk that the only place that ZVI can survive is in the very optically thick inner disk close to the protostar (which is likely to be active to the magnetorotational instability anyway).  However, the consensus of radio observations in cm, mm and sub-mm show that grain growth occurs early in the evolution of protoplanetary disks, and that the largest grains are likely to be centimeters or even larger \citep{natta2007,perez2012,testi2014,carrasco2016,tazzari2016,perez2015}.  Locking up mass in larger grains or having grains settle into the midplane reduces the opacity in off-midplane regions of protoplanetary disks and significantly lengthens the rate of thermal relaxation.

In this work, we explored various realistic grain growth and vertical settling scenarios, and found that in much of the off-midplane regions of protoplanetary disks, the gas and dust are not necessarily in local thermodynamic equilibrium with each other.  In such non-LTE cases, the relevant thermal relaxation timescale is actually the sum of the optically thin cooling timescale from dust infrared emission {\it plus} the timescale for the gas to respond to the collisional exchange of energy between the gas and dust: $t_{relax}\approx t_{thin} + t_g^{col}$.  Throughout much of the outer disk, $t_g^{col}\gg t_{orb}\gg t_{thin}$, and so ZVI is still an important mechanism for the generation of turbulence throughout large portions of protoplanetary disks. 

More recently \citet{malygin2017} did a comprehensive analysis of thermal relaxation in protoplanetary disks, including non-LTE effects such as the finite coupling time of gas and dust.  They also found that off-midplane regions of PPDs are not necessarily in LTE.  However, they did not correctly apply their results to ZVI and they did not do any numerical simulations of ZVI. They stated that ZVI would only be operable in a small region very close to the protostar, but this was based solely on the conclusion of \citet{lesurlatter2016} that ZVI required a large P\'{e}clet number (and thus assuming that the relevant timescale is optically thick diffusion).}

\paragraph{Outstanding Issues \& Future Work:} Not only ZVI, but also other purely hydrodynamic instabilities are very sensitive to the rate of thermal relaxation.  Vertical Shear Instability requires cooling times significantly shorter than an orbital period, Convective Overstability needs cooling times of order the orbital period, whereas ZVI is operable when the cooling time is longer than a few orbital periods.  The cooling timescales are very sensitive to the spatiotemporal density of dust as well as the size distribution of grains.  The size distribution, in turn, depends on the dynamic balance of agglomeration and fracturing processes, which are very sensitive the properties of turbulent motions which are set by the relevant hydrodynamic instabilities.  This is the classic chicken-and-egg problem: the dust density and size distribution depend on hydrodynamic turbulence and the hydrodynamic turbulence depends of the dust density and size distribution.  In future work, we plan to simulate ZVI explicitly with a dust phase.  We are interested in how ZVI mixes and/or concentrates dust grains, and how mass-loading affects zombie vortices and zonal flows.  This will also give us the opportunity to directly see how ZVI interacts with the Streaming Instability. 

{\color{black}
We did not include radial migration of dust grains in our models for computing thermal relaxation times in \S\ref{sec:cooling_theory}.   We want to remind the reader why we approached the grain growth and settling the way we did.  ZVI requires stratification and long thermal relaxation times.  The first condition means that ZVI is triggered only off the midplane, above approximately $1.5 H$.  To achieve longer cooling times, one needs only to have dust grains grow in size (locking a lot of mass in a fewer number of larger particles removes many small grains which contribute to the total surface area) and for grains to settle into the midplane.  We can treat grain growth and settling into the midplane in a  ``local" manner, that is at one radius in the disk.  Including radial migration effectively couples all radii of the disk.  One cannot simply vary, for example, $a_{max}$ as a function of radius without also varying the surface (2D, vertically integrated) dust to gas ratio as a function of radius, and probably also the size distribution power $s$ as a function of radius.  This will be time or age dependent as well.  These changes need to be done in a way that conserves dust mass (modulo dust lost to the protostar, dust vaporized at inner radii, new dust from residual infall).  All of these effects are very model dependent (on turbulence or ``viscous evolution") and cannot be boiled down to varying one parameter.  However, all of that is probably unnecessary for ZVI.  Radial migration affects larger grains (those with Stokes numbers near unity or larger) that are already in the midplane.  Particles of these sizes have a negligible effect on cooling.  Including radial migration is absolutely crucial to follow solid mass redistribution in the disk, collisional agglomeration, and the formation of planetesimals, but is not relevant for the cooling times away from the midplane because it is basically a redistribution of larger particles near the disk midplane.  (Though, we admit that changing the number and size of larger particles in the midplane could, as a second-order effect, create more smaller grains in the midplane via collisional fragmentation, that may then be lofted off the midplane via turbulent diffusion.)  Thus, vertical settling and grain growth is easy to do in a transparent way in a local simulation at one radius, and directly gets at what is necessary to trigger ZVI.  Including radial migration in a self-consistent way is not trivial, not well constrained, and probably not necessary for understanding the cooling times above $1.5 H$.  See \citet{estrada2016} and \citet{desch2017} for more details on radial migration of solids and its impact on opacity.
}

Because of the complexity of radiative transfer, it is common to work with simplified approaches that are appropriate for either optically thin or optically thick extremes.  However, it is a computational challenge when at the same point in space and time, there may coexist small features that are optically thin and large features that are optically thick.  Because the baroclinic critical layers are the most sensitive to thermal relaxation, and because they are optically thin, our approach in this paper was to employ a simple Newton cooling prescription.  Using a second-order diffusion operator allows different size scales to be damped at different rates, but with a constant radiative diffusivity, this will grossly overdamp the smallest scales.  What is needed is a radiative diffusivity that is itself scale dependent, \cf \citet{spiegel1957}.  It may be fruitful to explore Monte Carlo methods for radiative transfer for a more rigorous investigation of the radiative damping of the very thin baroclinic critical layers.

{\color{black} All of our previous studies of ZVI have used Cartesian domains in which the radial extent was no more than a few $H$. Future investigations will include annular domains in which the radial extent will be much larger.  The goal is not so much to check the effects of curvature (which we expect to be small), but to explore the extent to which turbulence from ZVI susceptible radii can penetrate into ZVI inactive regions (just like the midplane, which is not ZVI active but still can still harbor ZVI generated turbulence).}

\acknowledgments

JAB is supported by NSF grants AST-1010052 and AST-1510708.  PSM is supported by NSF grants AST-1009907 and AST-1510703 and by NASA PATM grants NNX10AB93G and NNX13AG56G.  This work used the Extreme Science and Engineering Discovery Environment (XSEDE), which is supported by National Science Foundation grant number ACI-1548562 \citep{xsede}.  Simulations were performed on Stampede at the Texas Advanced Computing Center (TACC) and Comet at the San Diego Supercomputer Center (SDSC) via allocation TG-AST020001S.  Graphics used the ``Balance'' colormap, which is a perceptually uniform diverging colormap \citep{thyng2016}.  Finally, the authors thank the following for useful discussions during the writing of this manuscript: Steven Beckwith, Eugene Chiang, Jeffrey Cuzzi, Paul Estrada, Jeffrey Fung, Alfred Glassgold, Daniel Lecoanet, Christopher McKee, Michael Shull, and Orkan Umurhan.

\facility{XSEDE}


\newpage
\appendix

\section{CRITICAL  VALUES OF NEWTON COOLING TIME \& P\'{E}CLET NUMBER}

In order to get the clearest sense for the critical magnitude of radiative damping, we will first study the case of uniform gravity, uniform thermal temperature background and uniform stratification, as we did in \citet{MPJBHL15} and \citet{MPJB16}.  For the simulations in this subsection alone, we solve equations 2a,b,c  in \citet{MPJB16} with a triply-periodic, anelastic code, but with Newton cooling or a diffusion operator added the the temperature equation:
\begin{align}
-\mathcal{L}_{rad}\tilde{T} =
\begin{cases}
-\tilde{T}/\tau_{relax}, & \text{optically thin}\\
-\mathcal{K}_{rad}\nabla^2\tilde{T}, & \text{optically thick},
\end{cases}
\end{align}
where $\tilde{T}\equiv T-T_0$ is the temperature deviation from the uniform background, and $\mathcal{K}_{rad}$ is the radiative diffusivity.  For the case of radiative diffusion, it is convenient to define the radiative P\'{e}clet number, which is the ratio of rate of heat transport via advection to the rate via diffusion:
\begin{equation}
Pe_{rad} \equiv \frac{|(\boldsymbol{v} \cdot \boldsymbol{\nabla})\tilde{T}|}{|\mathcal{K}_{rad}\nabla^2\tilde{T}|}\approx \frac{\Omega H^2}{\mathcal{K}_{rad}}.
\end{equation}
In this expression for the P\'{e}clet number, we had to make a choice on the scales of interest; here, we chose to define it with respect to a characteristic length scale of $H$ and a characteristic velocity $\Omega H$. 

\begin{figure}
\epsscale{1.0}
\plotone{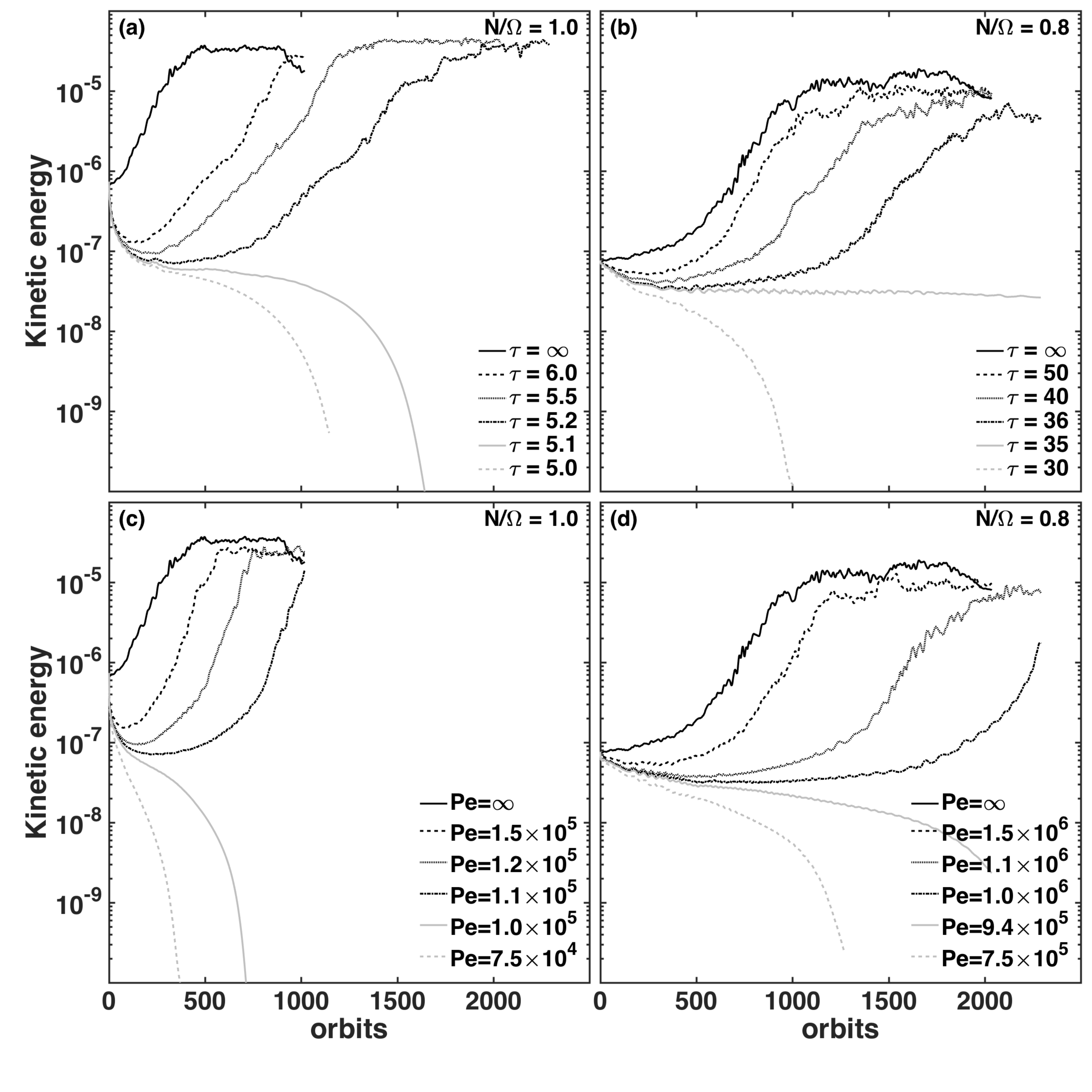}
\caption{{\bf Critical values for radiative damping time and P\'{e}clet number.}  Kinetic energy associated with the vertical component of the velocity $\int\bar{\rho}(z)v_z^2~dV$ (in units of $\rho_0H^5\Omega^2$) is on the vertical axes and time in units of orbital periods is on the horizontal axes. Simulations with different levels of optically thin cooling are shown in the top panels.  In panel (a), we see that for $N/\Omega=1.0$, the critical damping time is approximately $5$ orbits, whereas in panel (b) for $N/\Omega=0.8$, the critical damping time is $35$ orbits. Simulations with different magnitudes of radiative diffusivity are shown in the bottom row of plots. In panel (c), for $N/\Omega=1.0$, the critical P\'{e}clet number is around $10^5$, and in panel (d), for $N/\Omega=0.8$, the critical P\'{e}clet number is around $10^6$.}\label{fig:critical_cooling}
\end{figure}

Figure~\ref{fig:critical_cooling} shows the results of a series of simulations to ascertain the critical magnitude of thermal damping that halts the development of ZVI; kinetic energy associated with the vertical component of the velocity $\int\bar{\rho}(z)v_z^2~dV$ (in units of $\rho_0H^5\Omega^2$) is on the vertical axes and time (in units of orbital periods) is on the horizontal axes. Simulations with different levels of optically thin cooling are shown in the top panels.  ZVI depends strongly on the magnitude of stratification.  In panel (a), we see that for $N/\Omega=1.0$, the critical damping time is approximately $5$ orbits, whereas in panel (b) for $N/\Omega=0.8$, the critical damping time is $35$ orbits. ZVI is more robust for stronger stratification, and therefore can withstand shorter damping times. Simulations with different magnitudes of radiative diffusivity are shown in the bottom row of plots; panel (c) is for $N/\Omega=1.0$ and shows a critical P\'{e}clet number is around $10^5$, and panel (d) is for $N/\Omega=0.8$ and shows a critical P\'{e}clet number around $10^6$.  Again, we see that stronger stratification can withstand more radiative diffusion.  

\section{EMISSIVITY-WEIGHTED SAUTER MEAN RADIUS}

In this Appendix, we will derive an expression for the emissivity-weighted Sauter mean radius.  It will be convenient to define the smallest radius for a grain to have unity emissivity for a given temperature: $a_{Q=1} \equiv (600~\mu\mathrm{m}\cdot\mathrm{K})/T$.  For all realistic scenarios, we expect $a_{min}\ll a_{Q=1} \ll a_{max}$.  Next, let us compute the average square radius times emissivity:
\begin{align}
\begin{split}
\langle a^2Q(a,T)\rangle &\equiv \frac{1}{n_d}\int_{a_{min}}^{a_{max}}n(a) a^2Q(a,T)~da\\
{} &= \frac{K}{n_d}\left[\frac{1}{a_{Q=1}}\int_{a_{min}}^{a_{Q=1}} a^{3-s}~da + \int_{a_{Q=1}}^{a_{max}}a^{2-s}~da\right]\\
{} &= \frac{K}{n_d}\left[\frac{1}{a_{Q=1}}\frac{1}{(4-s)}\left(a_{Q=1}^{4-s}-a_{min}^{4-s}\right)+\frac{1}{(s-3)}\left(a_{Q=1}^{3-s}-a_{max}^{3-s}\right)\right].
\end{split}
\end{align}
Now taking the third moment of the distribution and dividing by the average square radius weighted by the emissivity, we obtain an expression for the emissivity-weighted Sauter radius:
\begin{align}
\begin{split}
a_{SQ}(T) &\equiv \frac{\langle a^3\rangle}{\langle a^2Q(a,T)\rangle} = \frac{a_{max}^{4-s}-a_{min}^{4-s}}{\frac{1}{a_{Q=1}}\left(a_{Q=1}^{4-s}-a_{min}^{4-s}\right)+\left(\frac{4-s}{s-3}\right)\left(a_{Q=1}^{3-s}-a_{max}^{3-s}\right)}\\
{} &\approx (s-3)a_{max}^{4-s}a_{Q=1}^{s-3} = (s-3)a_{max}^{4-s}\left(\frac{600~\mu\mathrm{m}\cdot\mathrm{K}}{T}\right)^{s-3}.
\end{split}
\end{align}
Note that the emissivity-weighted Sauter radius is insensitive to the size of the smallest grains (those with sizes much less than $a_{Q=1}$) because they are poor radiators and their effective surface area is much smaller than their geometric surface area.  For the special case $s=3.5$, then $a_{SQ}\approx 0.5\sqrt{a_{max}a_{Q=1}}$.  For the special case $s=3.25$, then $a_{SQ}\approx 0.25\sqrt[4]{a_{max}^3a_{Q=1}}$.


\newpage
\bibliography{zvi}

\begin{thebibliography}{}
\expandafter\ifx\csname natexlab\endcsname\relax\def\natexlab#1{#1}\fi
\providecommand{\url}[1]{\href{#1}{#1}}
\providecommand{\dodoi}[1]{doi:~\href{http://doi.org/#1}{\nolinkurl{#1}}}
\providecommand{\doeprint}[1]{\href{http://ascl.net/#1}{\nolinkurl{http://ascl.net/#1}}}
\providecommand{\doarXiv}[1]{\href{https://arxiv.org/abs/#1}{\nolinkurl{https://arxiv.org/abs/#1}}}

\bibitem[{{Aikawa} \& {Nomura}(2006)}]{aikawa2006}
{Aikawa}, Y., \& {Nomura}, H. 2006, \apj, 642, 1152, \dodoi{10.1086/501114}

\bibitem[{{Armitage}(2011)}]{armitage2011review}
{Armitage}, P.~J. 2011, \araa, 49, 195,
  \dodoi{10.1146/annurev-astro-081710-102521}

\bibitem[{{Bai}(2013)}]{bai2013}
{Bai}, X.-N. 2013, ApJ, 772, 96, \dodoi{10.1088/0004-637X/772/2/96}

\bibitem[{{Bai}(2014)}]{bai2014}
---. 2014, ApJ, 791, 137, \dodoi{10.1088/0004-637X/791/2/137}

\bibitem[{{Bai}(2015)}]{bai2015}
---. 2015, ApJ, 798, 84, \dodoi{10.1088/0004-637X/798/2/84}

\bibitem[{{Bai} \& {Stone}(2011)}]{baistone2011}
{Bai}, X.-N., \& {Stone}, J.~M. 2011, \apj, 736, 144,
  \dodoi{10.1088/0004-637X/736/2/144}

\bibitem[{{Bai} \& {Stone}(2013)}]{baistone2013b}
---. 2013, ApJ, 769, 76, \dodoi{10.1088/0004-637X/769/1/76}

\bibitem[{{Balbus} \& {Hawley}(1991)}]{balbus91}
{Balbus}, S.~A., \& {Hawley}, J.~F. 1991, \apj, 376, 214,
  \dodoi{10.1086/170270}

\bibitem[{{Bannon}(1996)}]{bannon96}
{Bannon}, P.~R. 1996, JAtS, 53, 3618,
  \dodoi{10.1175/1520-0469(1996)053<3618:OTAAFA>2.0.CO;2}

\bibitem[{{Barker} \& {Latter}(2015)}]{barker2015}
{Barker}, A.~J., \& {Latter}, H.~N. 2015, MNRAS, 450, 21,
  \dodoi{10.1093/mnras/stv640}

\bibitem[{{Barranco} {et~al.}(2000){Barranco}, {Marcus}, \&
  {Umurhan}}]{barranco00a}
{Barranco}, J., {Marcus}, P., \& {Umurhan}, O.~M. 2000, in Stanford Center for
  Turbulence Research -- Studying Turbulence Using Numerical Simulation
  Databases, 8. Proceedings of the 2000 Summer Program, 85--96

\bibitem[{{Barranco}(2009)}]{barranco09}
{Barranco}, J.~A. 2009, ApJ, 691, 907, \dodoi{10.1088/0004-637X/691/2/907}

\bibitem[{{Barranco} \& {Marcus}(2005)}]{barranco05}
{Barranco}, J.~A., \& {Marcus}, P.~S. 2005, ApJ, 623, 1157,
  \dodoi{10.1086/428639}

\bibitem[{{Barranco} \& {Marcus}(2006)}]{barranco06}
---. 2006, JCoPh, 219, 21, \dodoi{10.1016/j.jcp.2006.03.015}

\bibitem[{{Birnstiel} {et~al.}(2010){Birnstiel}, {Ricci}, {Trotta},
  {Dullemond}, {Natta}, {Testi}, {Dominik}, {Henning}, {Ormel}, \&
  {Zsom}}]{birnstiel2010b}
{Birnstiel}, T., {Ricci}, L., {Trotta}, F., {et~al.} 2010, \aap, 516, L14,
  \dodoi{10.1051/0004-6361/201014893}

\bibitem[{{Blaes} \& {Balbus}(1994)}]{blaes94}
{Blaes}, O.~M., \& {Balbus}, S.~A. 1994, \apj, 421, 163, \dodoi{10.1086/173634}

\bibitem[{{Blandford} \& {Payne}(1982)}]{blandfordpayne1982}
{Blandford}, R.~D., \& {Payne}, D.~G. 1982, MNRAS, 199, 883,
  \dodoi{10.1093/mnras/199.4.883}

\bibitem[{Boyd(2000)}]{boyd00}
Boyd, J. 2000, Chebyshev and Fourier Spectral Methods (Mineola, NY: Dover
  Publications, Inc.)

\bibitem[{{Brown} {et~al.}(2012){Brown}, {Vasil}, \& {Zweibel}}]{brown2012}
{Brown}, B.~P., {Vasil}, G.~M., \& {Zweibel}, E.~G. 2012, ApJ, 756, 109,
  \dodoi{10.1088/0004-637X/756/2/109}

\bibitem[{{Burke} \& {Hollenbach}(1983)}]{burke1983}
{Burke}, J.~R., \& {Hollenbach}, D.~J. 1983, ApJ, 265, 223,
  \dodoi{10.1086/160667}

\bibitem[{Canuto {et~al.}(1988)Canuto, Hussaini, Quarteroni, \&
  Zang}]{canuto88}
Canuto, C., Hussaini, M., Quarteroni, A., \& Zang, T. 1988, Spectral Methods in
  Fluid Dynamics (New York: Springer-Verlag), \dodoi{10.1007/978-3-642-84108-8}

\bibitem[{{Carr} {et~al.}(2004){Carr}, {Tokunaga}, \& {Najita}}]{carr2004}
{Carr}, J.~S., {Tokunaga}, A.~T., \& {Najita}, J. 2004, \apj, 603, 213,
  \dodoi{10.1086/381356}

\bibitem[{{Carrasco-Gonz{\'a}lez} {et~al.}(2016){Carrasco-Gonz{\'a}lez},
  {Henning}, {Chandler}, {Linz}, {P{\'e}rez}, {Rodr{\'{\i}}guez},
  {Galv{\'a}n-Madrid}, {Anglada}, {Birnstiel}, {van Boekel}, {Flock}, {Klahr},
  {Macias}, {Menten}, {Osorio}, {Testi}, {Torrelles}, \& {Zhu}}]{carrasco2016}
{Carrasco-Gonz{\'a}lez}, C., {Henning}, T., {Chandler}, C.~J., {et~al.} 2016,
  \apjl, 821, L16, \dodoi{10.3847/2041-8205/821/1/L16}

\bibitem[{{Chandrasekhar}(1960)}]{chandra60}
{Chandrasekhar}, S. 1960, PNAS, 46, 253, \dodoi{10.1073/pnas.46.2.253}

\bibitem[{{Chiang} \& {Goldreich}(1997)}]{chiang1997}
{Chiang}, E.~I., \& {Goldreich}, P. 1997, \apj, 490, 368,
  \dodoi{10.1086/304869}

\bibitem[{{Chin} \& {Lefebvre}(1986)}]{chin1986}
{Chin}, J.~S., \& {Lefebvre}, A.~H., eds. 1986, {Some comments on the
  characterization of drop-size distribution in sprays}, Vol.~2

\bibitem[{{Colovas} \& {Andereck}(1997)}]{colovas1997}
{Colovas}, P.~W., \& {Andereck}, C.~D. 1997, \pre, 55, 2736,
  \dodoi{10.1103/PhysRevE.55.2736}

\bibitem[{{Coughlin} \& {Marcus}(1996)}]{coughlin96}
{Coughlin}, K., \& {Marcus}, P.~S. 1996, Physical Review Letters, 77, 2214,
  \dodoi{10.1103/PhysRevLett.77.2214}

\bibitem[{Cuzzi {et~al.}(1993)Cuzzi, Dobrovolskis, \& Champney}]{cuzzi93}
Cuzzi, J., Dobrovolskis, A., \& Champney, J. 1993, Icarus, 106, 102

\bibitem[{{Cuzzi} {et~al.}(2014){Cuzzi}, {Estrada}, \& {Davis}}]{cuzzi2014}
{Cuzzi}, J.~N., {Estrada}, P.~R., \& {Davis}, S.~S. 2014, \apjs, 210, 21,
  \dodoi{10.1088/0067-0049/210/2/21}

\bibitem[{{D'Alessio} {et~al.}(2006){D'Alessio}, {Calvet}, {Hartmann},
  {Franco-Hern{\'a}ndez}, \& {Serv{\'{\i}}n}}]{dalessio2006}
{D'Alessio}, P., {Calvet}, N., {Hartmann}, L., {Franco-Hern{\'a}ndez}, R., \&
  {Serv{\'{\i}}n}, H. 2006, ApJ, 638, 314, \dodoi{10.1086/498861}

\bibitem[{{D'Alessio} {et~al.}(1998){D'Alessio}, {Cant{\"o}}, {Calvet}, \&
  {Lizano}}]{dalessio1998}
{D'Alessio}, P., {Cant{\"o}}, J., {Calvet}, N., \& {Lizano}, S. 1998, \apj,
  500, 411, \dodoi{10.1086/305702}

\bibitem[{{Desch} {et~al.}(2017){Desch}, {Estrada}, {Kalyaan}, \&
  {Cuzzi}}]{desch2017}
{Desch}, S.~J., {Estrada}, P.~R., {Kalyaan}, A., \& {Cuzzi}, J.~N. 2017, \apj,
  840, 86, \dodoi{10.3847/1538-4357/aa6bfb}

\bibitem[{{Dohnanyi}(1969)}]{dohnanyi1969}
{Dohnanyi}, J.~S. 1969, JGR, 74, 2531, \dodoi{10.1029/JB074i010p02531}

\bibitem[{{Draine}(2003)}]{draine2003}
{Draine}, B.~T. 2003, \araa, 41, 241,
  \dodoi{10.1146/annurev.astro.41.011802.094840}

\bibitem[{{Draine} \& {Lee}(1984)}]{draine1984}
{Draine}, B.~T., \& {Lee}, H.~M. 1984, \apj, 285, 89, \dodoi{10.1086/162480}

\bibitem[{Drazin \& Reid(2004)}]{drazinreid2004}
Drazin, P., \& Reid, W. 2004, Hydrodynamic Stability (Cambridge: Cambridge
  University Press), \dodoi{10.1017/CBO9780511616938}

\bibitem[{{Dubrulle} {et~al.}(1995){Dubrulle}, {Morfill}, \&
  {Sterzik}}]{dubrulle1995}
{Dubrulle}, B., {Morfill}, G., \& {Sterzik}, M. 1995, Icarus, 114, 237,
  \dodoi{10.1006/icar.1995.1058}

\bibitem[{{Dullemond} \& {Dominik}(2005)}]{dullemond2005}
{Dullemond}, C.~P., \& {Dominik}, C. 2005, \aap, 434, 971,
  \dodoi{10.1051/0004-6361:20042080}

\bibitem[{{Dullemond} {et~al.}(2002){Dullemond}, {van Zadelhoff}, \&
  {Natta}}]{dullemond2002}
{Dullemond}, C.~P., {van Zadelhoff}, G.~J., \& {Natta}, A. 2002, \aap, 389,
  464, \dodoi{10.1051/0004-6361:20020608}

\bibitem[{{Estrada} {et~al.}(2016){Estrada}, {Cuzzi}, \&
  {Morgan}}]{estrada2016}
{Estrada}, P.~R., {Cuzzi}, J.~N., \& {Morgan}, D.~A. 2016, \apj, 818, 200,
  \dodoi{10.3847/0004-637X/818/2/200}

\bibitem[{{Fiocco} {et~al.}(1975){Fiocco}, {Grams}, \& {Visconti}}]{fiocco1975}
{Fiocco}, G., {Grams}, G., \& {Visconti}, G. 1975, JATP, 37, 1327,
  \dodoi{10.1016/0021-9169(75)90125-7}

\bibitem[{{Flaherty} {et~al.}(2015){Flaherty}, {Hughes}, {Rosenfeld},
  {Andrews}, {Chiang}, {Simon}, {Kerzner}, \& {Wilner}}]{flaherty2015}
{Flaherty}, K.~M., {Hughes}, A.~M., {Rosenfeld}, K.~A., {et~al.} 2015, ApJ,
  813, 99, \dodoi{10.1088/0004-637X/813/2/99}

\bibitem[{{Flaherty} {et~al.}(2017){Flaherty}, {Hughes}, {Rose}, {Simon}, {Qi},
  {Andrews}, {K{\'o}sp{\'a}l}, {Wilner}, {Chiang}, {Armitage}, \&
  {Bai}}]{flaherty2017}
{Flaherty}, K.~M., {Hughes}, A.~M., {Rose}, S.~C., {et~al.} 2017, ApJ, 843,
  150, \dodoi{10.3847/1538-4357/aa79f9}

\bibitem[{{Gammie}(1996)}]{gammie96}
{Gammie}, C.~F. 1996, \apj, 457, 355, \dodoi{10.1086/176735}

\bibitem[{{Garaud}(2007)}]{garaud2007}
{Garaud}, P. 2007, \apj, 671, 2091, \dodoi{10.1086/523090}

\bibitem[{{Garaud} {et~al.}(2013){Garaud}, {Meru}, {Galvagni}, \&
  {Olczak}}]{garaud2013}
{Garaud}, P., {Meru}, F., {Galvagni}, M., \& {Olczak}, C. 2013, \apj, 764, 146,
  \dodoi{10.1088/0004-637X/764/2/146}

\bibitem[{Gilman \& Glatzmaier(1981)}]{gilman81}
Gilman, P., \& Glatzmaier, G. 1981, ApJS, 45, 335, \dodoi{10.1086/190714}

\bibitem[{{Glassgold} {et~al.}(2004){Glassgold}, {Najita}, \&
  {Igea}}]{glassgold2004}
{Glassgold}, A.~E., {Najita}, J., \& {Igea}, J. 2004, ApJ, 615, 972,
  \dodoi{10.1086/424509}

\bibitem[{Glatzmaier \& Gilman(1981{\natexlab{a}})}]{glatzmaier81a}
Glatzmaier, G., \& Gilman, P. 1981{\natexlab{a}}, ApJS, 45, 351,
  \dodoi{10.1086/190715}

\bibitem[{Glatzmaier \& Gilman(1981{\natexlab{b}})}]{glatzmaier81b}
---. 1981{\natexlab{b}}, ApJS, 45, 381, \dodoi{10.1086/190716}

\bibitem[{{Goldenson} {et~al.}(2008){Goldenson}, {Desch}, \&
  {Christensen}}]{goldenson2008}
{Goldenson}, N., {Desch}, S., \& {Christensen}, P. 2008, GeoRL, 35, L08813,
  \dodoi{10.1029/2007GL032907}

\bibitem[{Goldreich \& Lynden-Bell(1965{\natexlab{a}})}]{goldreich65b}
Goldreich, P., \& Lynden-Bell, D. 1965{\natexlab{a}}, MNRAS, 130, 125,
  \dodoi{10.1093/mnras/130.2.125}

\bibitem[{Goldreich \& Lynden-Bell(1965{\natexlab{b}})}]{goldreich65a}
---. 1965{\natexlab{b}}, MNRAS, 130, 97, \dodoi{10.1093/mnras/130.2.97}

\bibitem[{Gottlieb \& Orszag(1977)}]{gottlieborszag77}
Gottlieb, D., \& Orszag, S. 1977, Numerical Analysis of Spectral Methods:
  Theory and Applications (Philadelphia: Society for Industrial and Applied
  Mathematics), \dodoi{10.1137/1.9781611970425}

\bibitem[{{Gough}(1969)}]{gough69}
{Gough}, D.~O. 1969, JAtS, 26, 448,
  \dodoi{10.1175/1520-0469(1969)026<0448:TAAFTC>2.0.CO;2}

\bibitem[{{Guilloteau} {et~al.}(2012){Guilloteau}, {Dutrey}, {Wakelam},
  {Hersant}, {Semenov}, {Chapillon}, {Henning}, \&
  {Pi{\'e}tu}}]{guilloteau2012}
{Guilloteau}, S., {Dutrey}, A., {Wakelam}, V., {et~al.} 2012, A\&A, 548, A70,
  \dodoi{10.1051/0004-6361/201220331}

\bibitem[{{Hartmann} {et~al.}(2004){Hartmann}, {Hinkle}, \&
  {Calvet}}]{hartmann2004}
{Hartmann}, L., {Hinkle}, K., \& {Calvet}, N. 2004, \apj, 609, 906,
  \dodoi{10.1086/421317}

\bibitem[{Hill(1878)}]{hill1878}
Hill, G.~W. 1878, AmJM, 1, 5, \dodoi{10.2307/2369430}

\bibitem[{{Hollenbach} \& {McKee}(1979)}]{hollenbach1979}
{Hollenbach}, D., \& {McKee}, C.~F. 1979, ApJS, 41, 555, \dodoi{10.1086/190631}

\bibitem[{{Hughes} {et~al.}(2011){Hughes}, {Wilner}, {Andrews}, {Qi}, \&
  {Hogerheijde}}]{hughes2011}
{Hughes}, A.~M., {Wilner}, D.~J., {Andrews}, S.~M., {Qi}, C., \& {Hogerheijde},
  M.~R. 2011, ApJ, 727, 85, \dodoi{10.1088/0004-637X/727/2/85}

\bibitem[{Kelvin(1880)}]{kelvin1880}
Kelvin, L. 1880, Nature, 23, 45, \dodoi{10.1038/023045a0}

\bibitem[{{Klahr} \& {Hubbard}(2014)}]{klahrhubbard2014}
{Klahr}, H., \& {Hubbard}, A. 2014, ApJ, 788, 21,
  \dodoi{10.1088/0004-637X/788/1/21}

\bibitem[{Kundu(1990)}]{kundu90}
Kundu, P. 1990, Fluid Mechanics (San Diego: Academic Press, Inc.)

\bibitem[{{Kunz} \& {Lesur}(2013)}]{kunzlesur2013}
{Kunz}, M.~W., \& {Lesur}, G. 2013, \mnras, 434, 2295,
  \dodoi{10.1093/mnras/stt1171}

\bibitem[{{Lee} {et~al.}(2010{\natexlab{a}}){Lee}, {Chiang}, {Asay-Davis}, \&
  {Barranco}}]{lee2010a}
{Lee}, A.~T., {Chiang}, E., {Asay-Davis}, X., \& {Barranco}, J.
  2010{\natexlab{a}}, ApJ, 718, 1367, \dodoi{10.1088/0004-637X/718/2/1367}

\bibitem[{{Lee} {et~al.}(2010{\natexlab{b}}){Lee}, {Chiang}, {Asay-Davis}, \&
  {Barranco}}]{lee2010b}
---. 2010{\natexlab{b}}, ApJ, 725, 1938, \dodoi{10.1088/0004-637X/725/2/1938}

\bibitem[{{Lesur} {et~al.}(2014){Lesur}, {Kunz}, \& {Fromang}}]{lesur2014}
{Lesur}, G., {Kunz}, M.~W., \& {Fromang}, S. 2014, \aap, 566, A56,
  \dodoi{10.1051/0004-6361/201423660}

\bibitem[{{Lesur} \& {Latter}(2016)}]{lesurlatter2016}
{Lesur}, G.~R.~J., \& {Latter}, H. 2016, MNRAS, 462, 4549,
  \dodoi{10.1093/mnras/stw2172}

\bibitem[{{Lin} \& {Youdin}(2015)}]{lin2015}
{Lin}, M.-K., \& {Youdin}, A.~N. 2015, \apj, 811, 17,
  \dodoi{10.1088/0004-637X/811/1/17}

\bibitem[{{Lyra}(2014)}]{lyra2014}
{Lyra}, W. 2014, \apj, 789, 77, \dodoi{10.1088/0004-637X/789/1/77}

\bibitem[{{MacGregor} {et~al.}(2016){MacGregor}, {Wilner}, {Chandler}, {Ricci},
  {Maddison}, {Cranmer}, {Andrews}, {Hughes}, \& {Steele}}]{macgregor2016}
{MacGregor}, M.~A., {Wilner}, D.~J., {Chandler}, C., {et~al.} 2016, ApJ, 823,
  79, \dodoi{10.3847/0004-637X/823/2/79}

\bibitem[{{Malygin} {et~al.}(2017){Malygin}, {Klahr}, {Semenov}, {Henning}, \&
  {Dullemond}}]{malygin2017}
{Malygin}, M.~G., {Klahr}, H., {Semenov}, D., {Henning}, T., \& {Dullemond},
  C.~P. 2017, A\&A, 605, A30, \dodoi{10.1051/0004-6361/201629933}

\bibitem[{Marcus(1986)}]{marcus86a}
Marcus, P. 1986, in Proceedings of Astrophysical Radiation Hydrodynamics, ed.
  K.-H. Winkler \& M.~Norman (Springer-Verlag), 359--386

\bibitem[{{Marcus} {et~al.}(2016){Marcus}, {Pei}, {Jiang}, \&
  {Barranco}}]{MPJB16}
{Marcus}, P.~S., {Pei}, S., {Jiang}, C.-H., \& {Barranco}, J.~A. 2016, ApJ,
  833, 148, \dodoi{10.3847/1538-4357/833/2/148}

\bibitem[{{Marcus} {et~al.}(2015){Marcus}, {Pei}, {Jiang}, {Barranco},
  {Hassanzadeh}, \& {Lecoanet}}]{MPJBHL15}
{Marcus}, P.~S., {Pei}, S., {Jiang}, C.-H., {et~al.} 2015, ApJ, 808, 87,
  \dodoi{10.1088/0004-637X/808/1/87}

\bibitem[{{Marcus} {et~al.}(2013){Marcus}, {Pei}, {Jiang}, \&
  {Hassanzadeh}}]{MPJH13}
{Marcus}, P.~S., {Pei}, S., {Jiang}, C.-H., \& {Hassanzadeh}, P. 2013, PhRvL,
  111, 084501, \dodoi{10.1103/PhysRevLett.111.084501}

\bibitem[{{Marcus} \& {Press}(1977)}]{marcus77}
{Marcus}, P.~S., \& {Press}, W.~H. 1977, JFM, 79, 525,
  \dodoi{10.1017/S0022112077000305}

\bibitem[{{Maslowe}(1986)}]{maslowe86}
{Maslowe}, S.~A. 1986, AnRFM, 18, 405,
  \dodoi{10.1146/annurev.fl.18.010186.002201}

\bibitem[{{Mathis} {et~al.}(1977){Mathis}, {Rumpl}, \&
  {Nordsieck}}]{mathis1977}
{Mathis}, J.~S., {Rumpl}, W., \& {Nordsieck}, K.~H. 1977, ApJ, 217, 425,
  \dodoi{10.1086/155591}

\bibitem[{{Najita} {et~al.}(1996){Najita}, {Carr}, {Glassgold}, {Shu}, \&
  {Tokunaga}}]{najita1996b}
{Najita}, J., {Carr}, J.~S., {Glassgold}, A.~E., {Shu}, F.~H., \& {Tokunaga},
  A.~T. 1996, ApJ, 462, 919, \dodoi{10.1086/177205}

\bibitem[{{Najita} {et~al.}(2009){Najita}, {Doppmann}, {Carr}, {Graham}, \&
  {Eisner}}]{najita2009}
{Najita}, J.~R., {Doppmann}, G.~W., {Carr}, J.~S., {Graham}, J.~R., \&
  {Eisner}, J.~A. 2009, ApJ, 691, 738, \dodoi{10.1088/0004-637X/691/1/738}

\bibitem[{{Natta} {et~al.}(2007){Natta}, {Testi}, {Calvet}, {Henning},
  {Waters}, \& {Wilner}}]{natta2007}
{Natta}, A., {Testi}, L., {Calvet}, N., {et~al.} 2007, Protostars and Planets
  V, 767

\bibitem[{{Nelson} {et~al.}(2013){Nelson}, {Gressel}, \&
  {Umurhan}}]{nelson2013}
{Nelson}, R.~P., {Gressel}, O., \& {Umurhan}, O.~M. 2013, MNRAS, 435, 2610,
  \dodoi{10.1093/mnras/stt1475}

\bibitem[{{Ogura} \& {Phillips}(1962)}]{ogura62}
{Ogura}, Y., \& {Phillips}, N.~A. 1962, JAtS, 19, 173,
  \dodoi{10.1175/1520-0469(1962)019<0173:SAODAS>2.0.CO;2}

\bibitem[{{Orlanski} \& {Bryan}(1969)}]{orlanski1969}
{Orlanski}, I., \& {Bryan}, K. 1969, \jgr, 74, 6975,
  \dodoi{10.1029/JC074i028p06975}

\bibitem[{{Pan} \& {Sari}(2005)}]{pan2005}
{Pan}, M., \& {Sari}, R. 2005, Icarus, 173, 342,
  \dodoi{10.1016/j.icarus.2004.09.004}

\bibitem[{Pedlosky(1987)}]{pedlosky87}
Pedlosky, J. 1987, Geophysical Fluid Dynamics (New York: Springer-Verlag),
  \dodoi{10.1007/978-1-4684-0071-7}

\bibitem[{{Pelegr{\'{\i}}} \& {Sangr{\~a}}(1998)}]{pelegri1998}
{Pelegr{\'{\i}}}, J.~L., \& {Sangr{\~a}}, P. 1998, \jgr, 103, 30,
  \dodoi{10.1029/98JC01627}

\bibitem[{{Penev} {et~al.}(2009){Penev}, {Barranco}, \& {Sasselov}}]{penev09}
{Penev}, K., {Barranco}, J., \& {Sasselov}, D. 2009, ApJ, 705, 285,
  \dodoi{10.1088/0004-637X/705/1/285}

\bibitem[{{Penev} {et~al.}(2011){Penev}, {Barranco}, \& {Sasselov}}]{penev11}
---. 2011, ApJ, 734, 118, \dodoi{10.1088/0004-637X/734/2/118}

\bibitem[{{P{\'e}rez} {et~al.}(2012){P{\'e}rez}, {Carpenter}, {Chandler},
  {Isella}, {Andrews}, {Ricci}, {Calvet}, {Corder}, {Deller}, {Dullemond},
  {Greaves}, {Harris}, {Henning}, {Kwon}, {Lazio}, {Linz}, {Mundy}, {Sargent},
  {Storm}, {Testi}, \& {Wilner}}]{perez2012}
{P{\'e}rez}, L.~M., {Carpenter}, J.~M., {Chandler}, C.~J., {et~al.} 2012,
  \apjl, 760, L17, \dodoi{10.1088/2041-8205/760/1/L17}

\bibitem[{{P{\'e}rez} {et~al.}(2015){P{\'e}rez}, {Chandler}, {Isella},
  {Carpenter}, {Andrews}, {Calvet}, {Corder}, {Deller}, {Dullemond}, {Greaves},
  {Harris}, {Henning}, {Kwon}, {Lazio}, {Linz}, {Mundy}, {Ricci}, {Sargent},
  {Storm}, {Tazzari}, {Testi}, \& {Wilner}}]{perez2015}
{P{\'e}rez}, L.~M., {Chandler}, C.~J., {Isella}, A., {et~al.} 2015, \apj, 813,
  41, \dodoi{10.1088/0004-637X/813/1/41}

\bibitem[{{Phillips}(1972)}]{phillips1972}
{Phillips}, O.~M. 1972, Deep Sea Research and Oceanographic Abstracts, 19, 79,
  \dodoi{10.1016/0011-7471(72)90074-5}

\bibitem[{Rogallo(1981)}]{rogallo81}
Rogallo, R. 1981, Numerical experiments in homogeneous turbulence, Technical
  memorandum 81315, NASA

\bibitem[{Sauter(1926)}]{sauter1926}
Sauter, J. 1926, VDI-Forschungsheft, 279

\bibitem[{{Semenov} {et~al.}(2003){Semenov}, {Henning}, {Helling}, {Ilgner}, \&
  {Sedlmayr}}]{semenov2003}
{Semenov}, D., {Henning}, T., {Helling}, C., {Ilgner}, M., \& {Sedlmayr}, E.
  2003, \aap, 410, 611, \dodoi{10.1051/0004-6361:20031279}

\bibitem[{{Shakura} \& {Sunyaev}(1973)}]{shakura73}
{Shakura}, N.~I., \& {Sunyaev}, R.~A. 1973, \aap, 24, 337

\bibitem[{{Simon} {et~al.}(2018){Simon}, {Bai}, {Flaherty}, \&
  {Hughes}}]{simon2018}
{Simon}, J.~B., {Bai}, X.-N., {Flaherty}, K.~M., \& {Hughes}, A.~M. 2018, ArXiv
  e-prints.
\newblock \doarXiv{1711.04770}

\bibitem[{{Simon} {et~al.}(2015){Simon}, {Hughes}, {Flaherty}, {Bai}, \&
  {Armitage}}]{simon2015}
{Simon}, J.~B., {Hughes}, A.~M., {Flaherty}, K.~M., {Bai}, X.-N., \&
  {Armitage}, P.~J. 2015, ApJ, 808, 180, \dodoi{10.1088/0004-637X/808/2/180}

\bibitem[{{Spiegel}(1957)}]{spiegel1957}
{Spiegel}, E.~A. 1957, ApJ, 126, 202, \dodoi{10.1086/146386}

\bibitem[{{Stoll} \& {Kley}(2014)}]{stoll2014}
{Stoll}, M.~H.~R., \& {Kley}, W. 2014, \aap, 572, A77,
  \dodoi{10.1051/0004-6361/201424114}

\bibitem[{{Tazzari} {et~al.}(2016){Tazzari}, {Testi}, {Ercolano}, {Natta},
  {Isella}, {Chandler}, {P{\'e}rez}, {Andrews}, {Wilner}, {Ricci}, {Henning},
  {Linz}, {Kwon}, {Corder}, {Dullemond}, {Carpenter}, {Sargent}, {Mundy},
  {Storm}, {Calvet}, {Greaves}, {Lazio}, \& {Deller}}]{tazzari2016}
{Tazzari}, M., {Testi}, L., {Ercolano}, B., {et~al.} 2016, \aap, 588, A53,
  \dodoi{10.1051/0004-6361/201527423}

\bibitem[{{Teague} {et~al.}(2016){Teague}, {Guilloteau}, {Semenov}, {Henning},
  {Dutrey}, {Pi{\'e}tu}, {Birnstiel}, {Chapillon}, {Hollenbach}, \&
  {Gorti}}]{teague2016}
{Teague}, R., {Guilloteau}, S., {Semenov}, D., {et~al.} 2016, A\&A, 592, A49,
  \dodoi{10.1051/0004-6361/201628550}

\bibitem[{{Testi} {et~al.}(2014){Testi}, {Birnstiel}, {Ricci}, {Andrews},
  {Blum}, {Carpenter}, {Dominik}, {Isella}, {Natta}, {Williams}, \&
  {Wilner}}]{testi2014}
{Testi}, L., {Birnstiel}, T., {Ricci}, L., {et~al.} 2014, Protostars and
  Planets VI, 339, \dodoi{10.2458/azu_uapress_9780816531240-ch015}

\bibitem[{{Thyng} {et~al.}(2016){Thyng}, {Greene}, {Hetland}, {Zimmerle}, \&
  {DiMarco}}]{thyng2016}
{Thyng}, K.~M., {Greene}, C.~A., {Hetland}, R.~D., {Zimmerle}, H., \&
  {DiMarco}, S.~F. 2016, AGU Fall Meeting Abstracts, ED42B

\bibitem[{Towns {et~al.}(2014)Towns, Cockerill, Dahan, Foster, Gaither,
  Grimshaw, Hazlewood, Lathrop, Lifka, Peterson, Roskies, Scott, \&
  Wilkens-Diehr}]{xsede}
Towns, J., Cockerill, T., Dahan, M., {et~al.} 2014, Computing in Science and
  Engineering, 16, 62, \dodoi{10.1109/MCSE.2014.80}

\bibitem[{{Turner} \& {Drake}(2009)}]{turnerdrake2009}
{Turner}, N.~J., \& {Drake}, J.~F. 2009, \apj, 703, 2152,
  \dodoi{10.1088/0004-637X/703/2/2152}

\bibitem[{{Turner} {et~al.}(2014){Turner}, {Fromang}, {Gammie}, {Klahr},
  {Lesur}, {Wardle}, \& {Bai}}]{turner2014review}
{Turner}, N.~J., {Fromang}, S., {Gammie}, C., {et~al.} 2014, Protostars and
  Planets VI, 411, \dodoi{10.2458/azu_uapress_9780816531240-ch018}

\bibitem[{{Umurhan} {et~al.}(2016{\natexlab{a}}){Umurhan}, {Nelson}, \&
  {Gressel}}]{umurhan2016}
{Umurhan}, O.~M., {Nelson}, R.~P., \& {Gressel}, O. 2016{\natexlab{a}}, \aap,
  586, A33, \dodoi{10.1051/0004-6361/201526494}

\bibitem[{{Umurhan} {et~al.}(2016{\natexlab{b}}){Umurhan}, {Shariff}, \&
  {Cuzzi}}]{umurhanshariff2016}
{Umurhan}, O.~M., {Shariff}, K., \& {Cuzzi}, J.~N. 2016{\natexlab{b}}, ApJ,
  830, 95, \dodoi{10.3847/0004-637X/830/2/95}

\bibitem[{{Urpin}(2003)}]{urpin2003}
{Urpin}, V. 2003, \aap, 404, 397, \dodoi{10.1051/0004-6361:20030513}

\bibitem[{{van Boekel} {et~al.}(2005){van Boekel}, {Min}, {Waters}, {de Koter},
  {Dominik}, {van den Ancker}, \& {Bouwman}}]{vanboekel2005}
{van Boekel}, R., {Min}, M., {Waters}, L.~B.~F.~M., {et~al.} 2005, A\&A, 437,
  189, \dodoi{10.1051/0004-6361:20042339}

\bibitem[{{Vasil} {et~al.}(2013){Vasil}, {Lecoanet}, {Brown}, {Wood}, \&
  {Zweibel}}]{vasil2013}
{Vasil}, G.~M., {Lecoanet}, D., {Brown}, B.~P., {Wood}, T.~S., \& {Zweibel},
  E.~G. 2013, ApJ, 773, 169, \dodoi{10.1088/0004-637X/773/2/169}

\bibitem[{Velikhov(1959)}]{velikhov59}
Velikhov, E. 1959, JETP, 36, 1398

\bibitem[{Wang(2016)}]{wang2016}
Wang, M. 2016, PhD thesis, University of California, Berkeley, CA

\bibitem[{Weidenschilling(1984)}]{weidenschilling84}
Weidenschilling, S. 1984, Icarus, 60, 553, \dodoi{10.1016/0019-1035(84)90164-7}

\bibitem[{{Weidenschilling}(1977)}]{weidenschilling77}
{Weidenschilling}, S.~J. 1977, \mnras, 180, 57, \dodoi{10.1093/mnras/180.1.57}

\bibitem[{{Weidenschilling}(1980)}]{weidenschilling80}
---. 1980, \icarus, 44, 172, \dodoi{10.1016/0019-1035(80)90064-0}

\bibitem[{{Wilner} {et~al.}(2018){Wilner}, {MacGregor}, {Andrews}, {Hughes},
  {Matthews}, \& {Su}}]{wilner2018}
{Wilner}, D.~J., {MacGregor}, M.~A., {Andrews}, S.~M., {et~al.} 2018, ApJ, 855,
  56, \dodoi{10.3847/1538-4357/aaacd7}

\bibitem[{{Yu} {et~al.}(2017){Yu}, {Evans}, {Dodson-Robinson}, {Willacy}, \&
  {Turner}}]{yu2017}
{Yu}, M., {Evans}, II, N.~J., {Dodson-Robinson}, S.~E., {Willacy}, K., \&
  {Turner}, N.~J. 2017, ApJ, 850, 169, \dodoi{10.3847/1538-4357/aa9217}

\bibitem[{{Yu} {et~al.}(2016){Yu}, {Willacy}, {Dodson-Robinson}, {Turner}, \&
  {Evans}}]{yu2016}
{Yu}, M., {Willacy}, K., {Dodson-Robinson}, S.~E., {Turner}, N.~J., \& {Evans},
  II, N.~J. 2016, ApJ, 822, 53, \dodoi{10.3847/0004-637X/822/1/53}

\end{thebibliography}

\end{document}